\documentclass[useAMS,usenatbib]{mn2e}
\pdfoutput=1
\usepackage[letterpaper,total={17.8cm,24.0cm},centering]{geometry}

\usepackage{graphicx}
\usepackage{verbatim}
\usepackage{natbib}
\usepackage[fleqn]{amsmath}
\usepackage{amssymb}
\usepackage{color}
\usepackage{deluxetable}
\usepackage{enumerate}
\usepackage{comment}
\usepackage[bookmarks=False]{hyperref}
\newcommand{\dcoadd}{$\Delta_\mathrm{coadd}$}
\newcommand{\mr}{$M_r$}
\newcommand{\rfifty}{$R_{50}$}
\newcommand{\redshift}{$z$}
\newcommand{\bt}{$B/T$}

\title[Galaxy Zoo 2 data release]{Galaxy~Zoo~2: detailed morphological classifications for 304,122 galaxies from the Sloan Digital Sky Survey}
\author[Willett et al.]{
  \parbox[t]{16cm}{
  Kyle W. Willett$^{1}$\thanks{E-mail: willett@physics.umn.edu},
  Chris J. Lintott$^{2,3}$,
  Steven P. Bamford$^{4}$,
  Karen L. Masters$^{5,6}$,
  Brooke D. Simmons$^{2}$,
  Kevin R.V. Casteels$^{7}$,
  Edward M. Edmondson$^{5}$,
  Lucy F. Fortson$^{1}$,
  Sugata Kaviraj$^{2,8}$,
  William C. Keel$^{9}$,
  Thomas Melvin$^{5}$,
  Robert C. Nichol$^{5,6}$,
  M. Jordan Raddick$^{10}$,
  Kevin Schawinski$^{11}$,
  Robert J. Simpson$^{2}$,
  Ramin A. Skibba$^{12}$,
  Arfon M. Smith$^{3}$,
  Daniel Thomas$^{5,6}$
  \\
  }\\
$^{1}$School of Physics and Astronomy, University of Minnesota, 116 Church St. SE, Minneapolis, MN 55455, USA \\
$^{2}$Oxford Astrophysics, Denys Wilkinson Building, Keble Road, Oxford OX1 3RH, UK \\
$^{3}$Adler Planetarium, 1300 S. Lake Shore Drive, Chicago, IL 60605, USA \\
$^{4}$School of Physics and Astronomy, The University of Nottingham, University Park, Nottingham NG7 2RD, UK \\
$^{5}$Institute of Cosmology and Gravitation, University of Portsmouth, Dennis Sciama Building, Portsmouth PO1 3FX, UK \\
$^{6}$SEPnet, South East Physics Network, UK \\
$^{7}$Institut de Ci\`encies del Cosmos, Universitat de Barcelona (UB-IEEC), Mart\'i i Franqu\`es 1, E-08028 Barcelona, Spain \\
$^{8}$Centre for Astrophysics Research, University of Hertfordshire, College Lane, Hatfield AL10 9AB, UK \\
$^{9}$Department of Physics and Astronomy, University of Alabama, Box 870324, Tuscaloosa, AL 35487, USA \\
$^{10}$Department of Physics and Astronomy, The Johns Hopkins University, Homewood Campus, Baltimore, MD 21218, USA \\
$^{11}$Institute for Astronomy, Department of Physics, ETH Z\"urich, Wolfgang-Pauli-Strasse 16, CH-8093 Z\"urich, Switzerland \\
$^{12}$Center for Astrophysics and Space Sciences, University of California San Diego, 9500 Gilman Dr., San Diego, CA 92093, USA \\
}

\begin{document}

\date{Accepted 2 Aug 2013}

\pagerange{\pageref{firstpage}--\pageref{lastpage}} \pubyear{2013}

\maketitle

\label{firstpage}

\begin{abstract}
We present the data release for Galaxy~Zoo~2 (GZ2), a citizen science project with more than 16~million morphological classifications of 304,122 galaxies drawn from the Sloan Digital Sky Survey. Morphology is a powerful probe for quantifying a galaxy's dynamical history; however, automatic classifications of morphology (either by computer analysis of images or by using other physical parameters as proxies) still have drawbacks when compared to visual inspection. The large number of images available in current surveys makes visual inspection of each galaxy impractical for individual astronomers. GZ2 uses classifications from volunteer citizen scientists to measure morphologies for all galaxies in the DR7 Legacy survey with $m_r>17$, in addition to deeper images from SDSS Stripe~82. While the original Galaxy Zoo project identified galaxies as early-types, late-types, or mergers, GZ2 measures finer morphological features. These include bars, bulges, and the shapes of edge-on disks, as well as quantifying the relative strengths of galactic bulges and spiral arms. This paper presents the full public data release for the project, including measures of accuracy and bias. The majority ($\gtrsim90\%$) of GZ2 classifications agree with those made by professional astronomers, especially for morphological T-types, strong bars, and arm curvature. Both the raw and reduced data products can be obtained in electronic format at \url{http://data.galaxyzoo.org}.
\end{abstract}



\section{Introduction} \label{sec-intro}

The Galaxy~Zoo project \citep{lin08} was launched in 2007 to provide morphological classifications for nearly one million galaxies drawn from the Sloan Digital Sky Survey \citep[SDSS;][]{yor00} Main Galaxy Sample \citep{str02}. This scale of effort was made possible by combining classifications from hundreds of thousands of volunteers via a web-based interface. In order to keep the task at a manageable level of complexity, only the most basic morphological distinctions were requested, enabling the separation of systems into categories of elliptical (early-type), spiral (late-type) and mergers.\footnote{Galaxy Zoo is archived at \url{http://zoo1.galaxyzoo.org}.} Following the success of this project \citep{lin08,lin11}, the same methodology of asking for volunteer classifications was launched in 2009 with a more complex classification system. This paper presents data and results from this second incarnation of Galaxy Zoo, called Galaxy~Zoo~2 (GZ2). These data comprise detailed morphologies for more than 300,000 of the largest and brightest SDSS galaxies.\footnote{The Galaxy Zoo 2 site is archived at \url{http://zoo2.galaxyzoo.org}.}

While the morphological distinction used in the original Galaxy~Zoo (GZ1) -- that which divides spiral and elliptical systems -- is the most fundamental, the motivation for GZ2 was that galaxies demonstrate a much wider variety of morphological features. There is a long history of enhanced classifications (see \citealt{but13} for a historical review), but the most well-known approach \citep{hub26} included a division between barred and unbarred spirals, resulting in the famous `tuning fork' diagram. Further distinctions ordered ellipticals based on their apparent roundness and spirals on a combination of tightness and distinction of the arms and size of the central bulge. Along the late-type sequence, these traits are often correlated with physical parameters of the systems being studied \citep{rob94}, with spirals becoming (on average) redder, more massive, and less gas-rich for ``earlier'' locations in the sequence.

Morphological features can clearly provide insights into the physical processes that shape the evolution of galaxies. Most obviously, merger features reveal ongoing gravitational interactions, but even the presence of a central bulge in a disk galaxy is likely to indicate a history of mass assembly through significant mergers (\citealt{mar12} and references therein). On the other hand, galactic bars and rings reveal details of slower, secular evolution and stellar orbital resonances. For example, bars, are known to drive gas inwards and are related to the growth of a central bulge (reviews are given in \citealt{kor04,mas11c}). Careful classifications of morphological features are thus essential if the assembly and evolution history of galaxies is to be fully understood.

Traditional morphological classification relied on visual inspection of small numbers of images by experts \citep[e.g., ][]{san61,san94,dev91,but95,but02}. However, the sheer size of modern data sets (such as the SDSS Main Galaxy Sample) make this approach impractical. Detailed classifications of limited subsets of SDSS images have been made through huge efforts of a small number of experts. \citet{fuk07} and \citet{bai11} determined modified Hubble types for samples of 2253 and 4458 galaxies, respectively; the largest such effort to date is \citet{nai10}, who provide detailed classifications of 14,034 galaxies. Galaxy~Zoo~2 includes more than an order of magnitude more systems than any of these. Furthermore, each galaxy has a large number of independent inspections, which permits estimates of the classification likelihood (and in some cases the strength of the feature in question). The size of GZ2 allows for a more complete study of small-scale morphological features and their correlation with many other galaxy properties (e.g., mass, stellar and gas content, environment), while providing better statistics for the rarest objects.

The use of proxies for morphology --- such as colour, concentration index, spectral features, surface brightness profile, structural features, spectral energy distribution or some combination of these --- is a common practice in astronomy. However, proxies are not an adequate substitute for full morphological classification, as each has an unknown and likely biased relation with the features being studied. For example, most ellipticals are red and most spirals are blue; however, interesting subsets of both types have been found with the opposite colour \citep{sch09,mas10a}. With a sufficiently large set of galaxies, the diversity of the local population can be fully sampled and the relationship between morphology and proxies can be quantified.

Automated morphological classification is becoming much more sophisticated, driven in part by the availability of large training sets from the original Galaxy~Zoo \citep{ban10,hue11,dav13}. However, these methods do not yet provide an adequate substitute for classification by eye. In particular, as \citet{lin11} note, such efforts typically use proxies for morphology as their input (especially colour), meaning they suffer from the objections raised above. The release of the dataset Galaxy~Zoo~2 will be of interest to those developing such machine learning and computer vision systems. 

The GZ2 results were made possible by the participation of hundreds of thousands of volunteer `citizen scientists'. The original Galaxy~Zoo demonstrated the utility of this method in producing both large-scale catalogues as well as serendipitous discoveries of individual objects (see \citealt{lin11,for12} for reviews of Galaxy~Zoo~1 results). Since then, this method has been expanded beyond galaxy morphologies to include supernova identification \citep{smi11}, exoplanet discovery \citep{fis12,sch12} and a census of bubbles associated with star formation in the Milky~Way \citep{sim12a,ken12}, as well as a variety of ``big data'' problems outside of astronomy.\footnote{See \url{http://www.zooniverse.org/} for the full collection.} 

Several results based on early Galaxy Zoo 2 data have already been published. \citet{mas11c,mas12a} use GZ2 bar classifications to measure a clear increase in bar fraction for galaxies with redder colours, lower gas fractions, and more prominent bulges. \citet{hoy11} showed that the bars themselves are both redder and longer in redder disk galaxies. \citet{ski12} demonstrated that a significant correlation exists between barred and bulge-dominated galaxies at separations from $0.15$--$3$~Mpc. \citet{kav12a} used GZ2 to study early-type galaxies with visible dust lanes, while \citet{sim13} discovered a population of AGN host galaxies with no bulge, illustrating how black holes can grow and accrete via secular processes. Finally, \citet{cas13} quantify morphological signatures of interaction (including mergers, spiral arms, and bars) for galaxy pairs in the SDSS. This paper describes the data used in these studies, and goes further by quantifying and adjusting for classification biases and in comparing GZ2 classifications with other results. 

This paper is organised as follows. Section~\ref{sec-description} describes the sample selection and method for collecting morphological classifications. Section~\ref{sec-datareduction} outlines the data reduction and debiasing process, and Section~\ref{sec-catalogue} describes the tables that comprise the public data release. Section~\ref{sec-comparison} is a detailed comparison of GZ2 to four additional morphological catalogues that were created with SDSS imaging. We summarise our results in Section~\ref{sec-conclusion}. 

This paper uses the WMAP9 cosmological parameters of $H_0=71.8$~km/s/Mpc, $\Omega_m = 0.273$, and $\Omega_\Lambda = 0.727$ \citep{hin12}. 


\section{Project description} \label{sec-description}

\subsection{Sample selection} \label{ssec-sample}
The primary sample of objects used in Galaxy~Zoo~2 comprise approximately the brightest 25\% of the resolved galaxies in the SDSS North Galactic Cap region. The sample is generated from the SDSS Data Release~7 (DR7) `Legacy' catalogue \citep{aba09}, and therefore excludes observations made by SDSS for other purposes, such as the SEGUE survey. Spectroscopic targets come from the SDSS Main Galaxy Sample \citep{str02}.

Several cuts on the data were applied to the DR7 Legacy sample for selection in GZ2. The goal was to include only the nearest, brightest, and largest systems for which fine morphological features can be resolved and classified. GZ2 required a Petrosian half-light magnitude brighter than 17.0 in the $r$-band (after Galactic extinction correction was applied), along with a size limit of {\tt petroR90\_r}$>$3~arcsec ({\tt petroR90\_r} is the radius containing 90\% of the $r$-band Petrosian aperture flux). Galaxies which had a spectroscopic redshift in the DR7 catalogue outside the range $0.0005<z<0.25$ were removed; however, galaxies without reported redshifts were kept. Finally, objects which are flagged by the SDSS pipeline as \textsc{saturated}, \textsc{bright} or \textsc{blended} without an accompanying \textsc{nodeblend} flag were also excluded. The 245,609 galaxies satisfying all these criteria are referred to as the `original' sample.  

An error in the selection query meant that the `original' sample initially missed objects to which the SDSS photometric pipeline \citep{sto02} assigned both \textsc{blended} and \textsc{child} flags. These are objects that have been deblended from a larger blend (hence \textsc{child}), and have been identified as blended themselves (hence \textsc{blended}; due to containing multiple peaks). However, these are `final' objects, as the SDSS deblender doesn't attempt to further deblend already deblended objects.  These galaxies, which are typically slightly brighter, larger and bluer than the general population, were added to the GZ2 site on 2009-09-02. These additional 28,174 galaxies are referred to as the `extra' sample. 

In addition to galaxies from the DR7 Legacy, GZ2 also classified images from Stripe~82, a multiply-imaged section along the celestial equator in the Southern Galactic Cap. The selection criteria were the same as for the Legacy galaxies, with the exception of a fainter magnitude limit of $m_r < 17.77$. For the Stripe~82 sample only, GZ2 includes multiple images of individual galaxies: one set of images at single exposure depth, plus two sets of co-added images from multiple exposures. The coadded images combined 47 (south) or 55 (north) individual scans of the region, resulting in an object detection limit approximately two magnitudes lower than in normal imaging \citep{ann11}. 

The primary sample for GZ2 analysis consists of the combined `original', `extra', and Stripe~82 normal-depth images with $m_r\leq17.0$. We have verified that there are no significant differences in the classifications between these sub-samples (i.e., no significant bias is introduced by the fact that they were classified at different times) and thus can be reliably used as a single data set. This is hereafter referred to as the GZ2 {\bf main sample} (Table~\ref{tbl-sample}), and is used for the bulk of the analysis in this paper. Data from both the Stripe~82 normal-depth images with $m_r>17.0$ and the two sets of coadded images are separately included in this data release. 

\begin{table}
 \begin{tabular}{@{}lrcl}
 \hline
\multicolumn{1}{c}{Sample} &
\multicolumn{1}{c}{$N_\mathrm{gal}$} &
\multicolumn{1}{c}{$N_\mathrm{class}$} &
\multicolumn{1}{c}{$m_r$} 
\\ 
\multicolumn{1}{c}{} &
\multicolumn{1}{c}{} &
\multicolumn{1}{c}{median} &
\multicolumn{1}{c}{[mag]} 
\\ 
\hline
\hline						
original                       & 245,609 & 44  & 17.0   \\     
extra                          &  28,174 & 41  & 17.0   \\     
Stripe~82 normal               &  21,522 & 45  & 17.77  \\     
Stripe~82 normal ($m_r<17$)    &  10,188 & 45  & 17.0   \\     
Stripe~82 coadd 1              &  30,346 & 18  & 17.77  \\     
Stripe~82 coadd 2              &  30,339 & 21  & 17.77  \\     
\hline
main                           & 283,971 & 44  & 17.0   \\     
original + extra + S82 ($m_r<17$) & \\
\hline
 \end{tabular}
 \caption{Basic properties of the galaxy samples in GZ2, including the total number of galaxies ($N_\mathrm{gal}$), the median number of classifications per galaxy ($N_\mathrm{class}$), and the apparent magnitude limit. \label{tbl-sample}}
\end{table}

\subsection{Image creation}\label{ssec-imagecreation}

Images of galaxies for classification were generated from the SDSS ImgCutout web service \citep{nie04} from the Legacy and Stripe~82 normal depth surveys. Each image is a $gri$ colour composite $424\times424$ pixels in size, scaled to $(0.02\times${\tt petroR90\_r}) arcsec per pixel.

Coadded images from Stripe~82 were generated from the corrected SDSS FITS frames. Frames were combined using Montage \citep{jac10} and converted to a colour image using a slightly modified version of the SkyServer asinh stretch code \citep{lup04}, with parameters adjusted to replicate the normal SDSS colour balance. The parameterisation of the stretch function used is:

\begin{equation}
f(x)=\text{asinh}(\alpha Q x)/Q         
\label{eqn-imagegen}
\end{equation}

\noindent where $Q=3.5$ and $\alpha=0.06$. The colour scaling is [1.000,1.176,1.818] in $g$, $r$ and $i$, respectively. 

The first set of Stripe~82 coadded images were visually very different from the single-depth images. Changing the colour balance to maximise the visibility of faint features, however, resulted in more prominent background sky noise; since each pixel is typically dominated by a single band, the background is often brightly coloured by the \citet{lup04} algorithm. Due to concerns that this noise would be an obvious sign that the images were from deeper data (potentially biasing the classifications), we created a second set of coadd images in which the colour of background pixels was removed. This was achieved by reducing the colour saturation of pixels outside of a soft-edged object mask. 

The original and desaturated coadd image sets are labeled `stripe82\_coadd\_1' and `stripe82\_coadd\_2', respectively (Table~\ref{tbl-sample}). Subsequent analysis revealed very few differences between the classifications for the images using the two coadd methods (see \S\ref{ssec-s82}). 

\subsection{Decision tree}\label{ssec-decision_tree}

Morphological data for Galaxy~Zoo~2 were collected via a web-based interface. Volunteers needed to register with a username for their classifications to be recorded. Like Galaxy~Zoo~1, classification begins with the user being shown an SDSS colour composite image of a galaxy alongside a question and set of possible responses. More detailed data is then collected via a multi-step decision tree. In this paper, a {\it classification} is defined as the total amount of information collected about one galaxy by a single user completing the decision tree. Each individual step in the tree is a {\it task}, which consists of a {\it question} and a finite set of possible {\it responses}. The selection of a particular response is referred to as the user's {\it vote}.  

The first GZ2 task is a slightly modified version of GZ1, identifying whether the galaxy is either ``smooth'', has ``features or a disk'', or is a ``star or artifact''. The appearance of subsequent tasks in the interface depends on the user's previous responses. For example, if the user clicks on the ``smooth'' button, they are subsequently asked to classify the roundness of the galaxy; this task would not be shown if they had selected either of the other two responses. 

\begin{figure*}
\includegraphics[angle=0,width=7.0in]{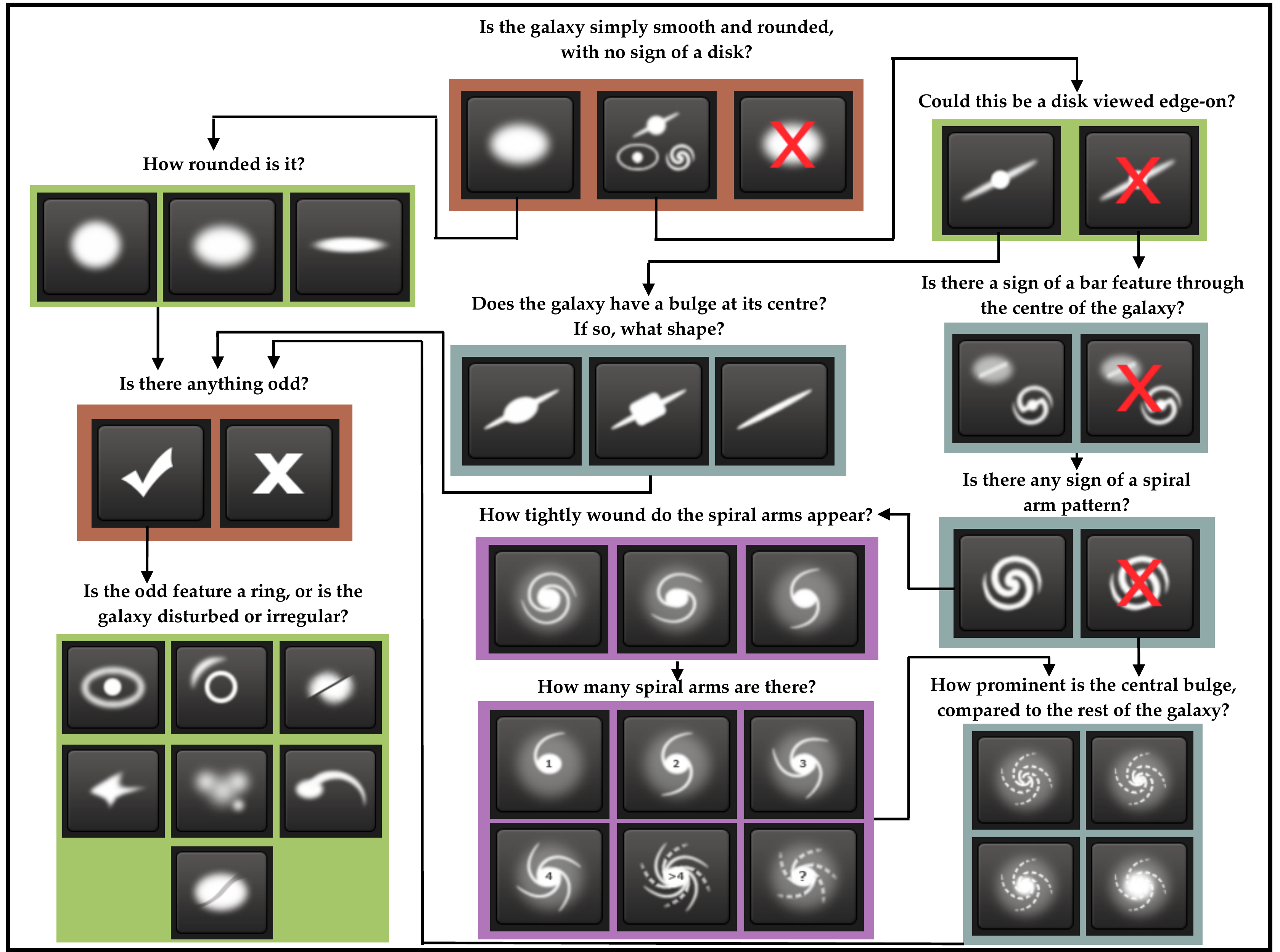}
\caption{Flowchart of the classification tasks for GZ2, beginning at the top centre. Tasks are colour-coded by their relative depths in the decision tree. Tasks outlined in brown are asked of every galaxy. Tasks outlined in green, blue, and purple are (respectively) one, two or three steps below branching points in the decision tree. Table~\ref{tbl-tree} describes the responses that correspond to the icons in this diagram.
\label{fig-flowchart}}
\end{figure*}

The GZ2 tree has 11 classification tasks with a total of 37 possible responses (Figure~\ref{fig-flowchart} and Table~\ref{tbl-tree}). A classifier selects only one response for each task, after which they are immediately taken to the next task in the tree. Tasks~01 and 06 are the only questions that are always answered for each and every classification. Once a classification is complete, an image of the next galaxy is automatically displayed and the user can begin classification of a new object. Importantly, in no case could a volunteer choose which galaxy to classify. 

\begin{table}
 \begin{tabular}{@{}cllr}
 \hline
\multicolumn{1}{l}{Task} &
\multicolumn{1}{c}{Question} &
\multicolumn{1}{c}{Responses} &
\multicolumn{1}{c}{Next} 
\\ 
\hline
\hline						
01    & {\it Is the galaxy simply smooth   }  & smooth           & 07 \\
      & {\it and rounded, with no sign of  }  & features or disk & 02 \\
      & {\it a disk?                       }  & star or artifact & {\bf end} \\
      \hline
02    & {\it Could this be a disk viewed   }  & yes              & 09 \\
      & {\it edge-on?                      }  & no               & 03 \\
      \hline
03    & {\it Is there a sign of a bar      }  & yes              & 04 \\
      & {\it feature through the centre    }  & no               & 04 \\
      & {\it of the galaxy?                }                                        \\
      \hline
04    & {\it Is there any sign of a        }  & yes              & 10 \\
      & {\it spiral arm pattern?           }  & no               & 05 \\
      \hline
05    & {\it How prominent is the          }  & no bulge         & 06 \\
      & {\it central bulge, compared       }  & just noticeable  & 06 \\
      & {\it with the rest of the galaxy?  }  & obvious          & 06 \\
      & {\it                               }  & dominant         & 06 \\
      \hline
06    & {\it Is there anything odd?        }  & yes              & 08 \\ 
      & {\it                               }  & no               & {\bf end}        \\
      \hline
07    & {\it How rounded is it?            }  & completely round & 06 \\
      & {\it                               }  & in between       & 06 \\
      & {\it                               }  & cigar-shaped     & 06 \\
      \hline
08    & {\it Is the odd feature a ring,    }  & ring             & {\bf end}        \\
      & {\it or is the galaxy disturbed    }  & lens or arc      & {\bf end}        \\
      & {\it or irregular?                 }  & disturbed        & {\bf end}        \\
      & {\it                               }  & irregular        & {\bf end}        \\  
      & {\it                               }  & other            & {\bf end}        \\  
      & {\it                               }  & merger           & {\bf end}        \\  
      & {\it                               }  & dust lane        & {\bf end}        \\  
      \hline
09    & {\it Does the galaxy have a        }  & rounded          & 06 \\
      & {\it bulge at its centre? If       }  & boxy             & 06 \\
      & {\it so, what shape?               }  & no bulge         & 06 \\
      \hline
10    & {\it How tightly wound do the      }  & tight            & 11 \\
      & {\it spiral arms appear?           }  & medium           & 11 \\
      & {\it                               }  & loose            & 11 \\    
      \hline
11    & {\it How many spiral arms          }  & 1                & 05 \\
      & {\it  are there?                   }  & 2                & 05 \\
      & {\it                               }  & 3                & 05 \\
      & {\it                               }  & 4                & 05 \\
      & {\it                               }  & more than four   & 05 \\
      & {\it                               }  & can't tell       & 05 \\
\hline
 \end{tabular}
 \caption{The GZ2 decision tree, comprising 11 tasks and 37 responses. The `Task' number is an abbreviation only and does {\em not} necessarily represent the order of the task within the decision tree. The text in `Question' and `Responses' are displayed to volunteers during classification, along with the icons in Figure~\ref{fig-flowchart}. `Next' gives the subsequent task for the chosen response. \label{tbl-tree}}
\end{table}

Data from the classifications were stored in a live Structured Query Language (SQL) database. In addition to the morphology classifications, the database also recorded a timestamp, user identifier, and image identifier for each classification. 

\subsection{Site history}\label{ssec-site_history}

Galaxy~Zoo~2 launched on 2009-02-16 with the `original' sample of 245,609 images. The `extra' galaxies from the Legacy survey were added on 2009-09-02. The normal-depth and the first set of coadded Stripe~82 images were mostly added on 2009-09-02, with an additional $\sim7700$ of coadded images added on 2010-09-24. Finally, the second version of the coadded images were added to the site on 2009-11-04. 

\begin{figure}
\includegraphics[angle=0,width=3.5in]{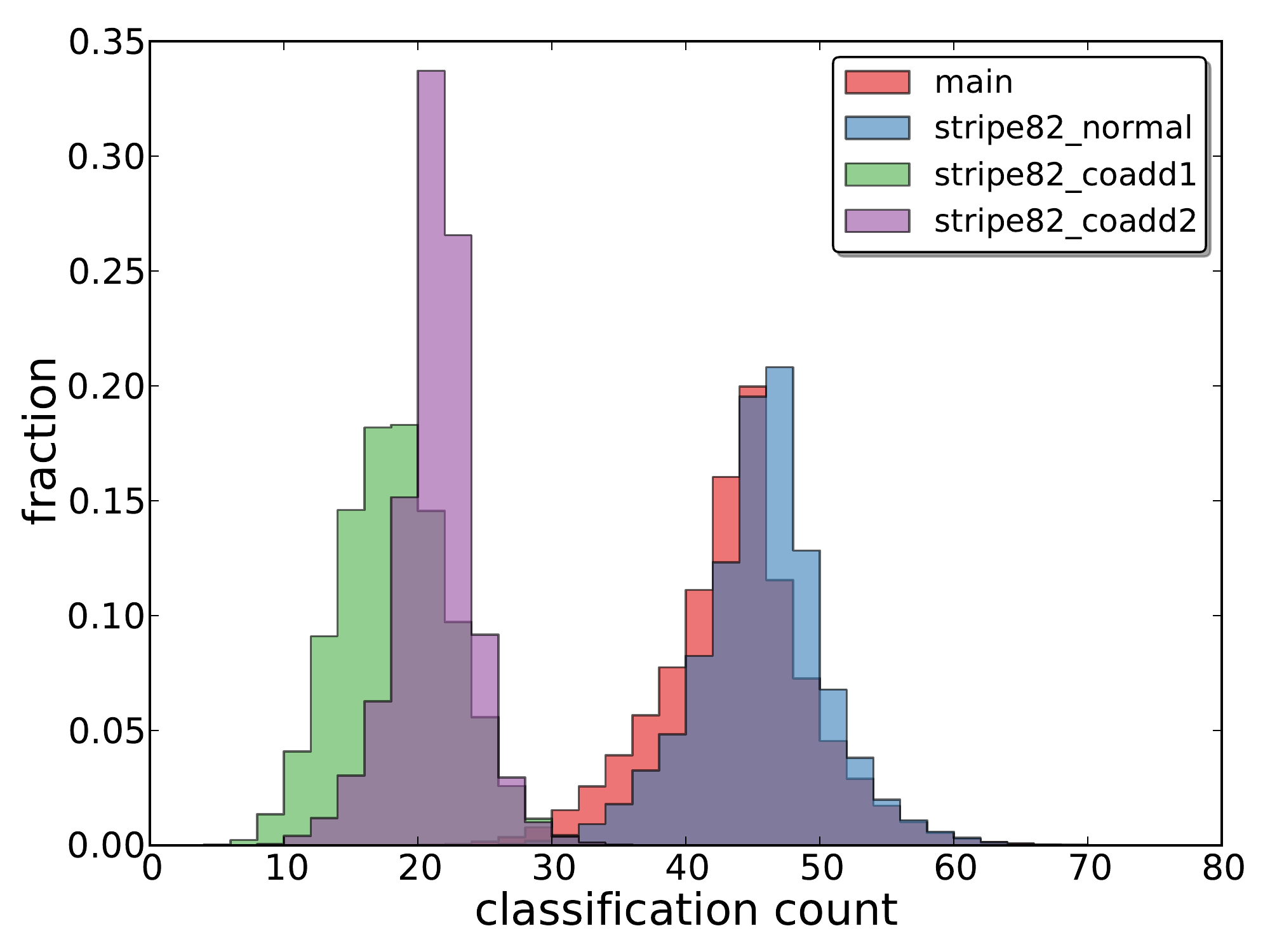}
\caption{Distribution of the number of classifications for the sub-samples within GZ2.
\label{fig-classification_histogram}}
\end{figure}

For most of the duration of GZ2, images shown to classifiers were randomly selected from the database. To ensure that each galaxy ultimately had enough responses to accurately characterize the likelihood of the classification, images with low numbers of classifications were shown at a higher rate toward the end of the project. The main sample galaxies finished with a median of 44 classifications; the minimum was 16, and $>99.9\%$ of the sample had at least 28~classifications. The `stripe82\_coadd\_2' galaxies had a median of 21 classifications and $>99.9\%$ had at least 10 (Figure~\ref{fig-classification_histogram}).

The last GZ2 classifications were collected on 2010-04-29, with the project spanning just over 14~months. The archived site continued to be maintained, but classifications were no longer recorded. The final dataset contains 16,340,298 classifications (comprising a total of 58,719,719 tasks) by 83,943 volunteers.


\section{Data reduction} \label{sec-datareduction}

\subsection{Multiple classifications}
In a small percentage of cases, individuals classified the same image more than once. In order to treat each vote as an independent measurement, classifications repeated by the same user were removed from the data, keeping only their votes from the last submission. Repeat classifications occurred for only $\sim1\%$ of all galaxies. The removal of the repeats only changed the morphological classifications for $\lesssim0.01\%$ of the sample.  

\subsection{Individual user weighting and combining classifications}\label{ssec-consistency}

The next step is to reduce the influence of potentially unreliable classifiers (whose classifications are consistent with random selection). To do so, an iterative weighting scheme (similar to that used for GZ1) is applied. First, we calculated the vote fraction ($f_\mathrm{r} = n_\mathrm{r}/n_\mathrm{t}$) for every response to every task for every galaxy, weighting each user's vote equally. Here, $n_\mathrm{r}$ is the number of votes for a given response and $n_\mathrm{t}$ is the total number of votes for that task. Each vote is compared to the vote fraction to calculate a user's consistency $\kappa$:

\begin{equation}
\kappa = \frac{1}{N_\mathrm{r}}\sum\limits_{i=1}^{N_\mathrm{r}}{\kappa_i},
\label{eqn-consistency}
\end{equation}

\noindent where $N_\mathrm{r}$ is the total number of possible responses for a task and

\begin{equation}
    \kappa_i = \left\{
    \begin{array}{l l}
      f_\mathrm{r}       & \text{ if vote corresponds to this response,} \\
      (1 - f_\mathrm{r}) & \text{ if vote does not correspond.}\\
    \end{array} \right.
    \label{eqn-consistency2}
 \end{equation}

For example, if a question has three possible responses, and the galaxy corresponds best to response $a$, then the vote fractions for responses $(a, b, c)$ might be $(0.7, 0.2, 0.1)$.
\begin{itemize}
\item If an individual votes for response $a$, then \\$\kappa = (0.7 + (1-0.2) + (1-0.1))/3 = 0.8$
\item If an individual votes for response $b$, then \\$\kappa = ((1-0.7) + 0.2 + (1-0.1))/3 = 0.467$
\item If an individual votes for response $c$, then \\$\kappa = ((1-0.7) + (1-0.2) + 0.1)/3 = 0.4$
\end{itemize}
\noindent Votes which agree with the majority thus have high values of consistency, whereas votes which disagree have low values.

\begin{figure}
\includegraphics[angle=0,width=3.3in]{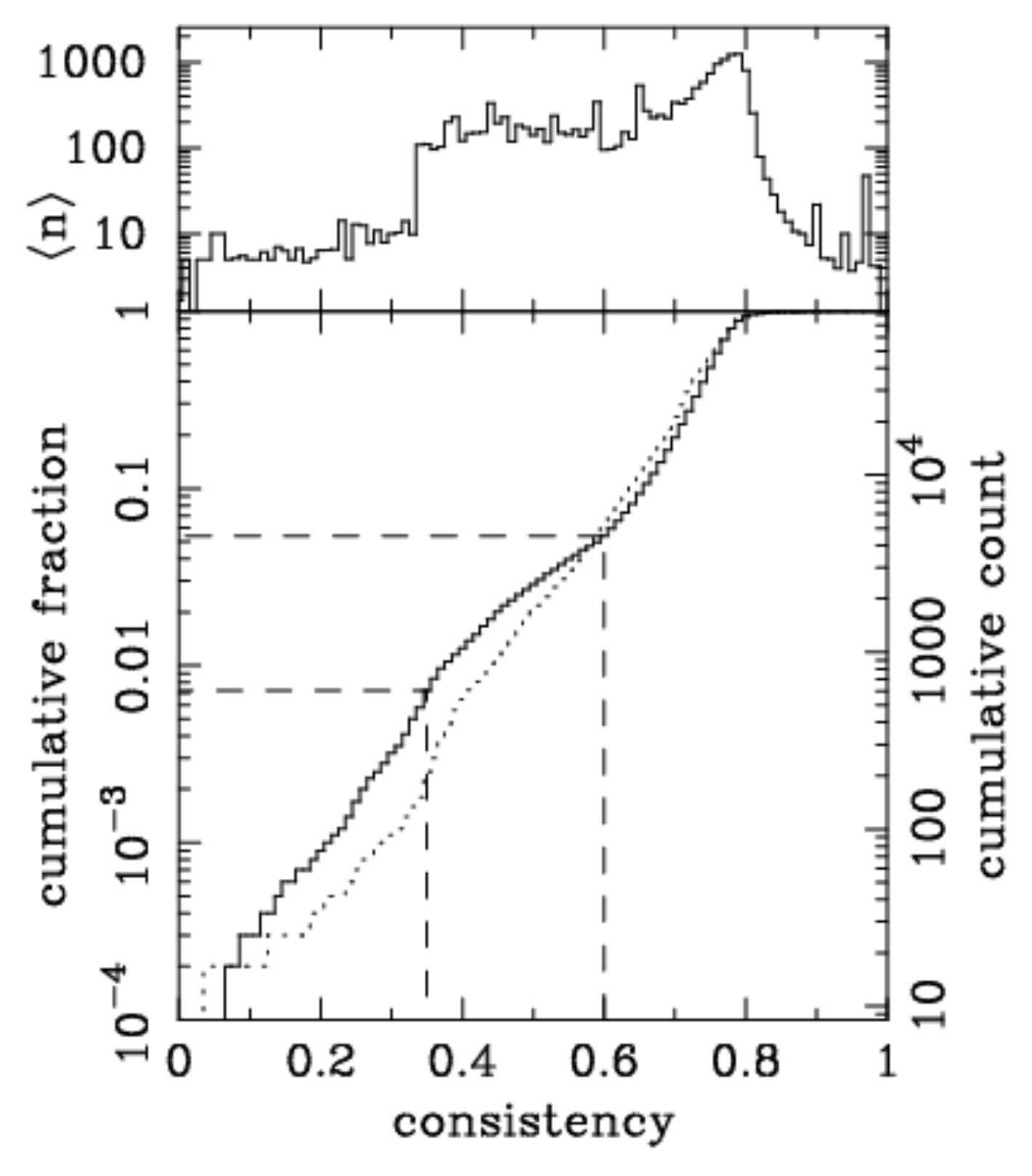}
\caption{Distribution of the user consistency $\kappa$. Top: mean number of galaxies classified per user as a function of their consistency. Bottom: Cumulative distribution of consistency. The dotted line shows the first iteration of weighting, and the solid line the third iteration. The second iteration is not shown, but is almost identical to the third. Dashed lines indicate where the user weighting function takes values of 0.01 and 1. 
\label{fig-consistency}}
\end{figure}

Each user was assigned an overall consistency ($\bar{\kappa}$) by taking the mean consistency of every response. From the distribution of results for the initial iteration (Figure~\ref{fig-consistency}), a weighting function is applied that down-weights classifiers in the tail of low consistency.

\begin{equation}
w = \min \left(1.0,(\bar{\kappa} / 0.6)^{8.5} \right)
\label{eqn-weight}
\end{equation}

\noindent For this function, $w=1$ for $\sim95\%$ of classifiers and $w<0.01$ for only $\sim1\%$ of classifiers. The vast majority of classifiers are thus treated equally; there is no up-weighting of the most consistent classifiers. The top panel of Figure~\ref{fig-consistency} also shows that the lowest-weighted classifiers completed only a handful ($<10$) of objects on average. This may demonstrate either that the volunteers are becoming more accurate as they classify more galaxies, or that inconsistent people are less likely to remain engaged with the project; further work on user behaviour is needed to distinguish between the two possibilities.

After computing $\kappa$, vote fractions were recalculated using the new user weights, and then repeated a third time to ensure convergence. For each task, individual responses are combined to produce the total vote count and a vote fraction for each task. The weighted votes and vote fractions generated by Equation~\ref{eqn-weight} are used exclusively hereafter when discussing GZ2 votes and vote fractions; for brevity, we typically drop the term ``weighted''. 

\subsection{Classification bias}\label{ssec-classificationbias}

The vote fractions are adjusted for what is termed {\it classification bias}. The overall effect of this bias is a change in observed morphology fractions as a function of redshift {\em independent of any true evolution in galaxy properties}, a trend also seen in the Galaxy~Zoo~1 data \citep{bam09}. The SDSS survey is expected to be shallow enough to justify an assumption of no evolution, and so the presumed cause is that more distant galaxies, on average, are both smaller and dimmer in the cutout images. As a result, finer morphological features are more difficult to identify. We note that this effect is not limited to crowd-sourced classifications; expert and automatic classifications must also suffer from bias to some degree, although smaller sample sizes make this difficult to quantify. 

\begin{figure*}
\includegraphics[angle=0,width=7.0in]{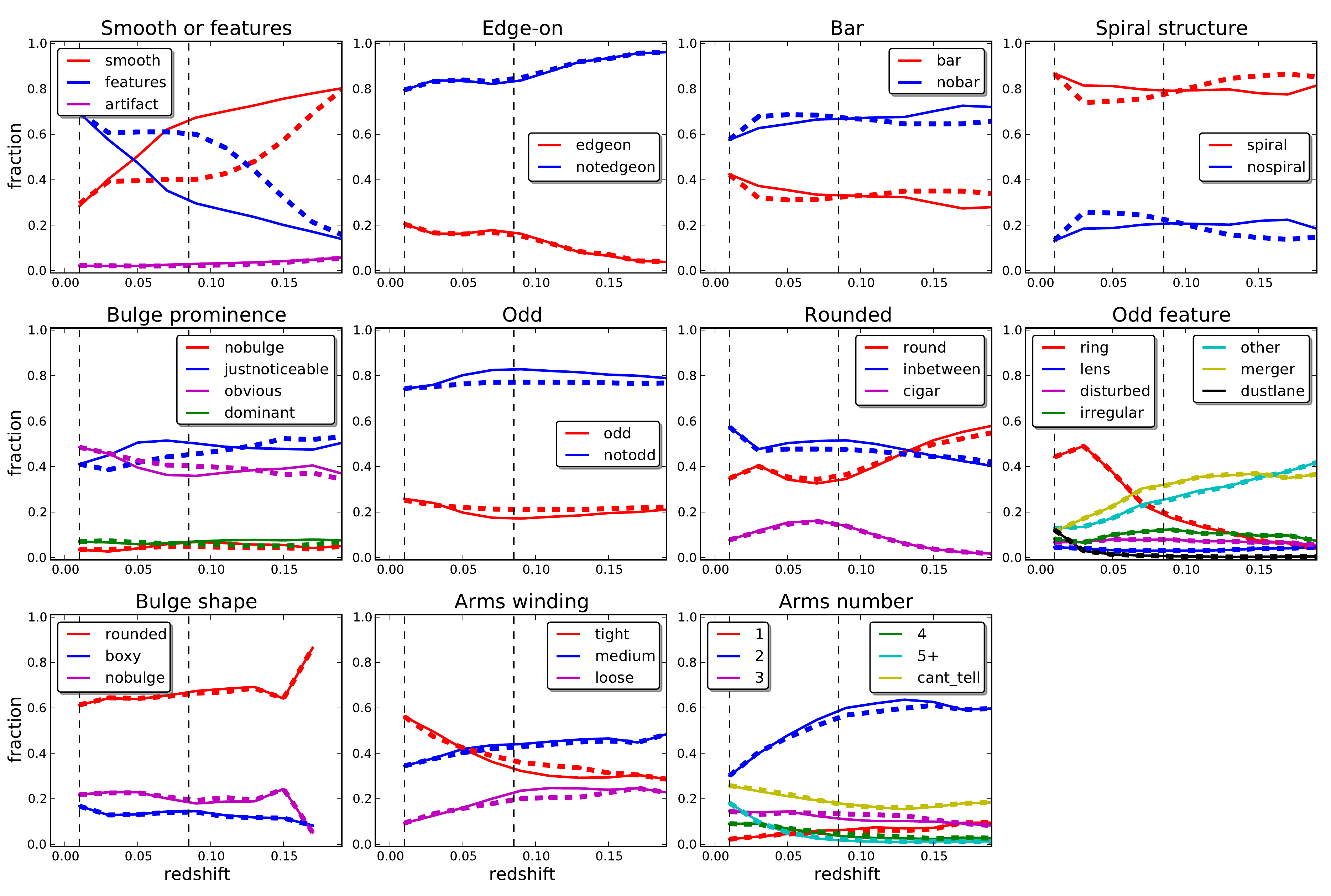}
\caption{Type fractions as a function of redshift for the classification tasks in GZ2. Solid (thin) lines show the vote fractions, while the thick (dashed) lines show the debiased vote fractions adjusted for classification bias. This is a luminosity-limited sample for \mr~$<-20.89$. The data for each task is plotted only for galaxies with enough votes to characterize the response distribution (Table~\ref{tbl-thresholds}). Vertical dashed lines show the redshift at $z=0.01$ (the lower limit of the correction) and $z=0.085$ (the redshift at which the absolute magnitude limit reaches the sensitivity of the SDSS). 
\label{fig-type_fractions}}
\end{figure*}

Figure~\ref{fig-type_fractions} demonstrates the effect of classification bias for the GZ2 tasks. The mean vote fraction for each response is shown as a function of redshift; the fraction of votes for finer morphological features (such as identification of disk galaxies, spiral structure, or galactic bars) decreases at higher redshift. The trend is strongest for the initial task of separating smooth and feature/disk galaxies, but almost all tasks exhibit some level of change. 

Part of the observed trends in type fractions at high redshifts is due to the nature of a magnitude-limited sample; high-redshift galaxies must be more luminous to be detected in the SDSS and are thus most likely to be giant red ellipticals. However, there is clear evidence of classification bias in GZ2 even in luminosity-limited samples. Since this bias contaminates any potential studies of galaxy demographics over the sample volume, it must be corrected to the fullest possible extent. 

\citet{bam09} corrected for classification bias in the GZ1 data for the elliptical and combined spiral classes. Their approach was to bin the galaxies as a function of absolute magnitude (\mr), the physical Petrosian half-light radius (\rfifty), and redshift. They then computed the average elliptical-to-spiral ratio for each (\mr,\rfifty) bin in the lowest redshift slice with significant numbers of galaxies; this yields a local baseline relation which gives the (presumably) unbiased morphology as a function of the galaxies' {\em physical}, rather than {\em observed} parameters. From the local relation, they derived a correction for each (\mr,\rfifty,$z$) bin and then adjusted the vote fractions for the individual galaxies in each bin. The validity of this approach is justified in part since debiased vote fractions result in a consistent morphology-density relation over a range of redshifts \citep{bam09}. We modify and extend this technique for the GZ2 classifications. 

There are two major differences between the GZ1 and GZ2 data. First, GZ2 has a decision tree, rather than a single question and response for each vote. This means that all tasks, with the exception of the first, depend on responses to previous tasks in the decision tree. For example, the bar question is only asked if the user classifies a galaxy as having ``features or disk'' and as ``not edge-on''. Thus, the value of the vote fraction for this example only addresses the total bar vote fraction {\em among galaxies that a user has classified as disks \underline{and} are not edge-on}, and not as a function of the total galaxy population (see \citealt{cas13} for further discussion). 

For a galaxy to be used in deriving a bias correction for a particular task, this method requires both a minimum weighted vote fraction for the preceding response(s) and a minimum number of votes for the task in question. The value of the threshold is determined by finding the minimum vote fraction for the preceding response for which $>99\%$ of galaxies with $N_\mathrm{votes}\geq N_\mathrm{crit}$ are preserved. We compute thresholds for both $N_\mathrm{crit}=10$ and $N_\mathrm{crit}=20$ (Table~\ref{tbl-thresholds}). The effect is to remove galaxies with high vote fractions but low (and potentially unreliable) numbers of total votes. 

Applying the thresholds to galaxies for deriving the bias correction does increase the number of bins with large variances; however, it is critical for reproducing accurate baseline measurements of individual morphologies. The correction derived from well-classified galaxies is then applied to the vote fractions for {\em all} galaxies in the sample. 

\begin{table}
\centering
 \begin{tabular}{@{}llcc}
 \hline
\multicolumn{1}{l}{Task} &
\multicolumn{1}{l}{Previous task} &
\multicolumn{1}{c}{Vote fraction} &
\multicolumn{1}{c}{Vote fraction}
\\ 
\multicolumn{1}{l}{} &
\multicolumn{1}{l}{} &
\multicolumn{1}{c}{$N_{task}\geq10$} &
\multicolumn{1}{c}{$N_{task}\geq20$}
\\ 
\hline					
01                      & --        & --        & --        \\
02                      & 01        & 0.227     & 0.430     \\
03                      & 01,02     & 0.519     & 0.715     \\
04                      & 01,02     & 0.519     & 0.715     \\
05                      & 01,02     & 0.519     & 0.715     \\
06                      & --        & --        & --        \\
07                      & 01        & 0.263     & 0.469     \\
08                      & 06        & 0.223     & 0.420     \\
09                      & 01,02     & 0.326     & 0.602     \\
10                      & 01,02,04  & 0.402     & 0.619     \\
11                      & 01,02,04  & 0.402     & 0.619     \\
\hline
 \end{tabular}
 \caption{Thresholds for determining well-sampled galaxies in GZ2. Thresholds depend on the number of votes for a classification task considered to be sufficient -- this table contains thresholds applied to previous task(s) for both 10 and 20 votes. As an example, to select galaxies that may or may not contain bars, cuts for $p_\mathrm{features/disk}>0.430$, $p_\mathrm{not~edgeon}>0.715$, and $N_\mathrm{not~edgeon}\geq20$ should be applied. No thresholds are given for Tasks 01 and 06, since these are answered for every classification in GZ2. \label{tbl-thresholds}}
\end{table}

The second major difference is that the adjustment of the GZ1 vote fractions assumed that the single task was essentially binary. Since almost every vote in GZ1 was for a response of either ``elliptical'' or ``spiral'' (either anticlockwise, clockwise, or edge-on), this ratio was employed as the sole metric. No systematic debiasing was done for the other GZ1 response options (``star/don't know'' or ``merger''), and the method of adjusting the vote fractions assumes that these other options do not significantly affect the classification bias for the most popular responses. This is not possible for GZ2: many tasks have more than two possible responses and represent a continuum of relative feature strength, rather than a binary choice.

The debiasing method relies on the assumption that for a galaxy of a given physical brightness and size, a sample of other galaxies with similar brightnesses and sizes will (statistically) share the same average mix of morphologies. This is quantified using the ratio of vote fractions $(f_i/f_j)$ for some pair of responses $i$ and $j$. We assume that the true (that is, unbiased) ratio of likelihoods for each task $(p_i/p_j)$ is related to the measured ratio via a single multiplicative constant $K_{j,i}$:

\begin{equation}
\frac{p_i}{p_j} = \frac{f_i}{f_j}\times K_{j,i}.
\label{eqn-kdef}
\end{equation}

\noindent The unbiased likelihood for a single task can trivially be written as:

\begin{equation}
p_i = \frac{1}{1/p_i},
\label{eqn-adjprob1}
\end{equation}

\noindent with the requirement that the sum of all likelihoods for a given task must be unity:
\begin{equation}
p_i + p_j + p_k + \dots = 1.
\label{eqn-sumprob}
\end{equation}

\noindent Multiplying (\ref{eqn-adjprob1}) by the inverse of (\ref{eqn-sumprob}) yields:

\begin{eqnarray}
p_i &=& \frac{1}{1/p_i} \times \frac{1}{p_i + p_j + p_k + \dots} \\
p_i &=& \frac{1}{p_i/p_i + p_j/p_i + p_k/p_i + \dots} \\
p_i &=& \frac{1}{1 + \sum\limits_{j\ne i}{(p_j/p_i)}} \\
p_i &=& \frac{1}{1 + \sum\limits_{j\ne i}{K_{j,i} (f_j/f_i)}}.
\label{eqnarray-adjprob2}
\end{eqnarray}

The corrections for each pair of tasks can be directly determined from the data. At the lowest redshift bin, $\frac{p_i}{p_j} = \frac{f_i}{f_j}$ and $K_{j,i}=1$. From Equation~\ref{eqn-kdef}:

\begin{eqnarray}
\left(\frac{f_i}{f_j}\right)_{z=0} &=& \left(\frac{f_i}{f_j}\right)_{z=z^\prime}\times K_{j,i} \\
K_{j,i} &=& \frac{\left(f_i/f_j\right)_{z=0} }{ \left(f_i/f_j\right)_{z=z^\prime}}
\label{eqnarray-adjprob3}
\end{eqnarray}

\noindent This can be simplified by defining $C_{j,i}\equiv\text{log}_{10}(K_{j,i})$ and substituting into (\ref{eqnarray-adjprob3}):

\begin{eqnarray}
C_{j,i} &=& \text{$log_{10}$}~\left(\frac{f_i}{f_j}\right)_{z=0} - \text{log$_{10}$}~\left(\frac{f_i}{f_j}\right)_{z=z^\prime}.
\label{eqnarray-adjprob4}
\end{eqnarray}

\noindent The correction $C_{j,i}$ for any bin is thus the difference between $f_i/f_j$ at the desired redshift and that of a local baseline, if the ratios between vote fractions are expressed as logarithms.  

Local morphology baselines and subsequent corrections for GZ2 are derived from the main sample data. Since determining the baseline ratio relies on absolute magnitude and physical size, only galaxies in the main sample with spectroscopic redshifts (86\%) are used. Corrections also use data only from galaxies with sufficient numbers of responses to determine their morphology. We apply the thresholds in Table~\ref{tbl-thresholds} for $N_\mathrm{task}\geq20$ to identify the well-answered galaxies for each task.

Bins for \mr~range from $-24$ to $-16$ in steps of $0.25$~mag, for \rfifty~from $0$ to $15$~kpc in steps of $0.5$~kpc, and for $z$ from $0.01$ to $0.26$ in steps of $0.01$. These bin ranges and step sizes are chosen to maximize the parameter space covered by the bias correction. Only bins with at least 20~galaxies are used in deriving a correction. 

Since each unique pair of responses to a question will have a different local baseline, there are $\binom{n}{2}$ correction terms for a task with $n$ responses. For $n=2$, this method is identical to that described in \citet{bam09}. 

\begin{figure*}
\includegraphics[angle=0,width=7.0in]{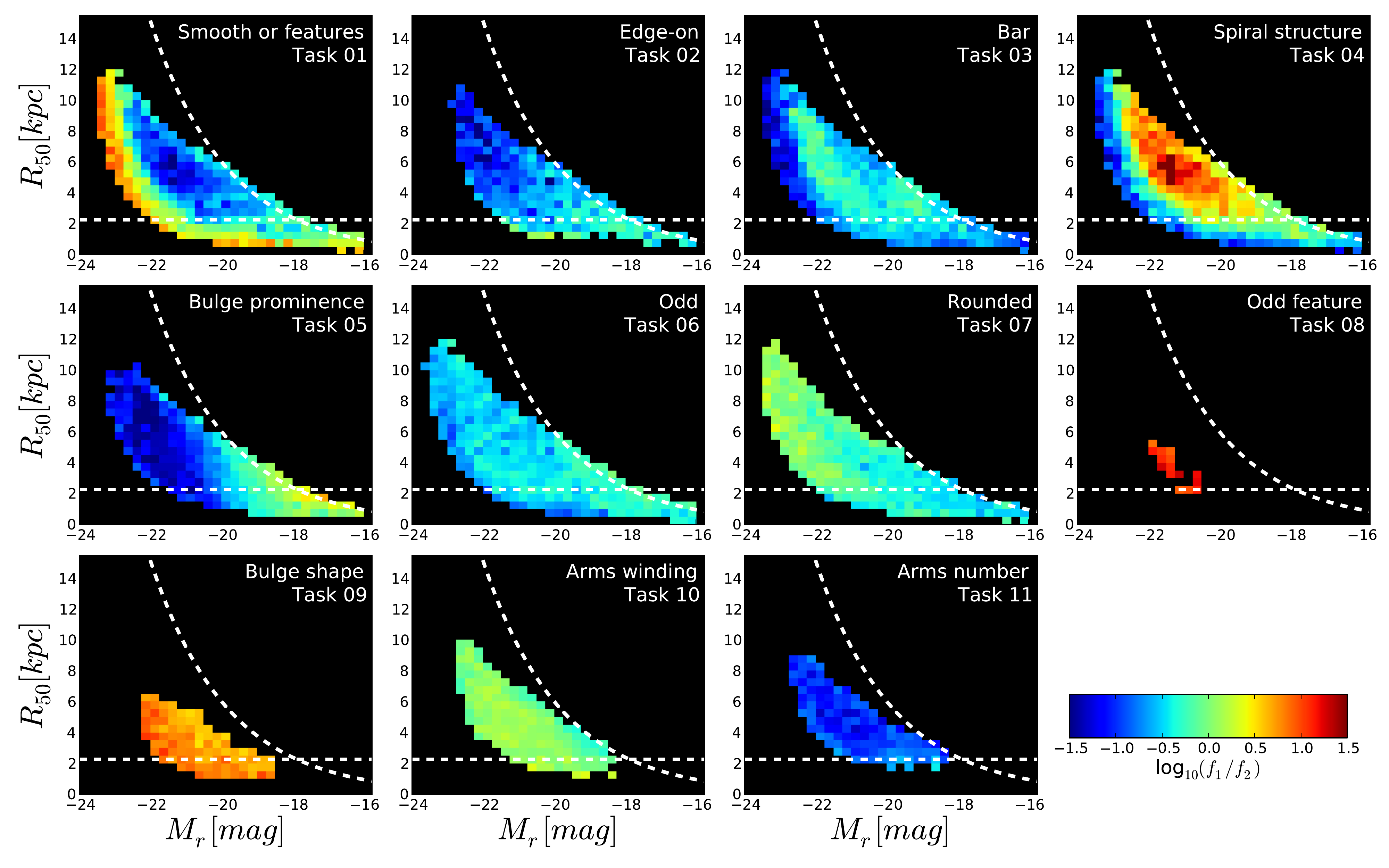}
\caption{Local morphology ratios for GZ2 classifications; these are used to derive the corrections that adjust data for classification bias (\S\ref{ssec-classificationbias}). The ratio of the binned vote fractions is for the first two responses in the decision tree (Table~\ref{tbl-tree}) for each task; there may be as many as 21 such pairs per task, depending on the number of unique responses. Dashed horizontal lines give the physical scale corresponding to $1\arcsec$, while the curved lines show a constant apparent surface brightness of $\mu_{50,r}=23.0$~mag~arcsec$^{-2}$.
\label{fig-baselines}}
\end{figure*}

The baseline morphology ratios for the GZ2 tasks are shown in Figure~\ref{fig-baselines} for the first two responses in each task. To derive a correction for bins not covered at low redshift, we attempted to fit each baseline ratio with an analytic, smoothly-varying function. The baseline ratio for the responses to Tasks~01 and 07 is functionally very similar to the GZ1 relation \citep[Figure~A5 in][]{bam09}. This ratio can be fit with an analytic function: 

\begin{equation}
\frac{f_j}{f_i}[R_{50},M_r] = \frac{s_6}{1 + exp[(\alpha - M_r)/\beta]} + s_7
\label{eqn-sb}
\end{equation}

\noindent where:

\begin{eqnarray}
\alpha &=& s_2\times\text{exp}[{-\left(s_1 + s_8{R_{50}}^{s_9}\right)}] + s_3, \\
\beta  &=& s_4 + s_5(x_0 - s_3),
\end{eqnarray}

\noindent where $\{s_1,s_2,s_3,s_4,s_5,s_6,s_7,s_8,s_9\}$ are minimised to fit the data. 

None of the other tasks are well-fit by a function of the form in Equation~\ref{eqn-sb}. For these, a simpler function is used where both \mr~and \rfifty~vary linearly:

\begin{equation}
\frac{f_j}{f_i}[R_{50},M_r] = t_1(R_{50} - t_2) + t_3(M_r - t_4) + t_5,
\label{eqn-tilt}
\end{equation}

\noindent where $\{t_1,t_2,t_3,t_4,t_5\}$ are the parameters to be minimized. Equation~\ref{eqn-tilt} is fit to all other tasks where enough non-zero bins exist to get a good fit. Finally, for pairs of responses with only a few sampled bins, we instead directly measured the difference bin-by-bin between the local ratio and the measured ratio at higher redshift. Galaxies falling in bins that are not well-sampled are assigned a correction of $C_{i,j}=0$ for that term; this is necessary to avoid overfitting based on only a few noisy bins. 



This method succeeds for most GZ2 tasks and responses. Figure~\ref{fig-type_fractions} illustrates the comparison between the mean raw and debiased vote fractions as a function of redshift. The debiased results ({\it thick lines}) are flat over $0.01<z<0.085$, where $L^\star$ galaxies \citep[$M_r\sim-20.44$;][]{bla03a} are within the detection limit of the survey and there are fewer empty bins. The debiased early- and late-type fractions of 0.45 and 0.55 agree with the GZ1 type fractions derived by \citet{bam09} for the same selection criteria. The bar fraction in disk galaxies is approximately 0.35, slightly higher than the value found by using thresholded GZ2 data in \citet{mas11c}.

\subsection{Angular separation bias}\label{ang-bias}

The vote fractions also suffer from a bias which depends on the angular separation between galaxies. For some classifications, participants perceive a galaxy's morphology differently when it has a close apparent companion. \citet{cas13} found that this bias is particularly strong for Task~08 (``odd features'') and its ``merger'' classification. The mean merger vote fractions of both physically close galaxies with similar redshifts and projected pairs with very different redshifts increase strongly as a function of decreasing angular separation. This results in projected pairs of non-interacting galaxies being classified as mergers. To determine an unbiased estimate of the mean probability for a given classification, \citet{cas13} subtracted the mean probabilities of projected pairs (with very different redshifts) from physically close pairs (with similar redshifts) for each projected separation bin. This results in a residual probability which is considered to represent the true change in morphology due to strong tidal interaction. While such a correction can be applied in a statistical way to the mean vote fractions for a given classification, applying such a correction to the individual vote fractions is not as straightforward.

The vote fractions presented in this data release have not been corrected for angular separation bias and readers using GZ2 data to study very close pairs are advised to keep this in mind, particularly for Task~08. Fortunately, the angular separation bias is minimal for the rest of the classifications and can usually be ignored. A detailed discussion of the angular separation bias (and how it affects individual classifications) is given in \citet{cas13}.


\section{The catalogue} \label{sec-catalogue}

The data release for GZ2 includes the vote counts and fractions (raw, weighted, and debiased) for each task in the classification tree for each galaxy. Data for the five subsamples described below can be accessed at \url{http://data.galaxyzoo.org}, and are available on CasJobs\footnote{\url{http://skyserver.sdss3.org/casjobs/}} in SDSS Data Release~10 \citep{ahn13}. Abridged portions of each data table are included in this paper (Tables~\ref{tbl-mainclass}--\ref{tbl-stripe82_coadd2}).

\subsection{Main sample}\label{ssec-catalogue_main}

Table~\ref{tbl-mainclass} contains classification data for the 243,500 galaxies in the main sample with spectroscopic redshifts. Each galaxy is identified by its unique SDSS DR7 object ID, as well as its J2000.0 coordinates and GZ2 subsample (either original, extra or Stripe~82 normal-depth). $N_\mathrm{class}$ is the total number of volunteers who classified the galaxy, while $N_\mathrm{votes}$ gives the total number of votes summed over all classifications and all responses. For each of the 37 morphological classes, six parameters are given: the raw number of votes for that response (e.g., {\tt t01\_smooth\_or\_features\_a01\_smooth\_count}), the number of votes weighted for consistency ({\tt $^\star$\_weight}), the fraction of votes for the task ({\tt $^\star$\_fraction}), the vote fraction weighted for consistency ({\tt $^\star$\_weighted\_fraction}), the debiased likelihood ({\tt $^\star$\_debiased}), which is the weighted vote fraction adjusted for classification bias (see Section~\ref{ssec-classificationbias}), and a Boolean flag ({\tt $^\star$\_flag}) that is set if the galaxy is included in a clean, debiased sample.

Flags for each morphological parameter are determined by applying three criteria. First, the vote fraction for the preceding task(s) must exceed some threshold (Table~\ref{tbl-thresholds}) to ensure that the question is well-answered. For example, selecting galaxies from which a clean barred sample can be identified requires both $p_\mathrm{features/disk}\geq0.430$ and $p_\mathrm{not edge-on}\geq0.715$. Secondly, the task must exceed a minimum number of votes (10 for Stripe~82, 20 for the main sample) in order to eliminate variance due to small-number statistics. Finally, the debiased vote fraction itself must exceed a given threshold of 0.8 for all tasks. We note this is a highly conservative selection -- each of the above parameters may be adjusted to provide different clean thresholds, depending on the use case for the data.

Table~\ref{tbl-mainclass} also includes an abbreviated version of the classification designated as $gz2\_class$. It is intended to serve as a quick reference for the consensus GZ2 classification; any quantitative analyses, however, should use the vote fractions instead. A description of how the string is generated is given in Appendix~\ref{app-gzstring}. 

Table~\ref{tbl-mainclass_photoz} gives the GZ2 classifications for the 42,462 main sample galaxies without spectroscopic redshifts. To compute the debiased likelihoods, we used the morphology corrections obtained for galaxies in the spectroscopic main sample. SDSS photometric redshifts \citep{csa03} are used to derive \mr~and \rfifty~ for each galaxy in the photometric sample and select the appropriate correction bin. The mean error in the redshift of the photometric sample (from the SDSS photo-$z$) is $\Delta~z=0.021$ (a fractional uncertainty of 27\%), compared to the spectroscopic accuracy of $\Delta~z=0.00016$ (0.3\%). Since the size of the redshift bins in $C_{j,i}$ is 0.01, a shift of several bins can potentially produce a very large change in the debiased vote fractions. 

Since the redshift can have a strong effect on classification bias, galaxies with spectroscopic and photometric redshifts from the SDSS are separated; we do not recommend that their debiased data be combined for analysis. For science cases where the main driver is the number of galaxies, however, it may be possible to combine the \underline{raw} vote fractions for the two samples.

\subsection{Stripe~82}\label{ssec-s82}

Data for Stripe~82 is reduced separately from the GZ2 main sample. This is due to the deeper magnitude limit of the samples (both normal and coadded) as well as the improved seeing in the latter. Since different image qualities potentially affect the debiasing, all three Stripe~82 samples are individually adjusted for classification bias. The method is the same as that used for the spectroscopic main sample galaxies -- the only difference is that the threshold for classification in the coadded sample is lowered from 20 to 10 votes. 

Table~\ref{tbl-stripe82_normal} gives classifications for the Stripe~82 normal-depth images with spectroscopic redshifts. Galaxies in this table with $m_r < 17.0$ also appear in Table~\ref{tbl-mainclass}; however, the corrections for classification bias here are derived based only on Stripe~82 data, and so debiased likelihoods and flags are slightly different. Classifications for galaxies with photometric redshifts only are not included.

Tables~\ref{tbl-stripe82_coadd1} and \ref{tbl-stripe82_coadd2} contain classifications for the first and second sets of coadded Stripe~82 galaxies with spectroscopic redshifts. Since both the number of galaxies and the average number of classifications per galaxy are a small fraction of that in the main sample, though, the corrections encompass a smaller range of tasks and phase space in (\mr,\rfifty,\redshift). The increased exposure time and improved seeing, however, means that the effect of classification bias is lessened at lower redshifts; the raw vote fractions may thus be more suitable for some science cases that require deeper imaging. 

\begin{figure*}
\includegraphics[angle=0,width=7.0in]{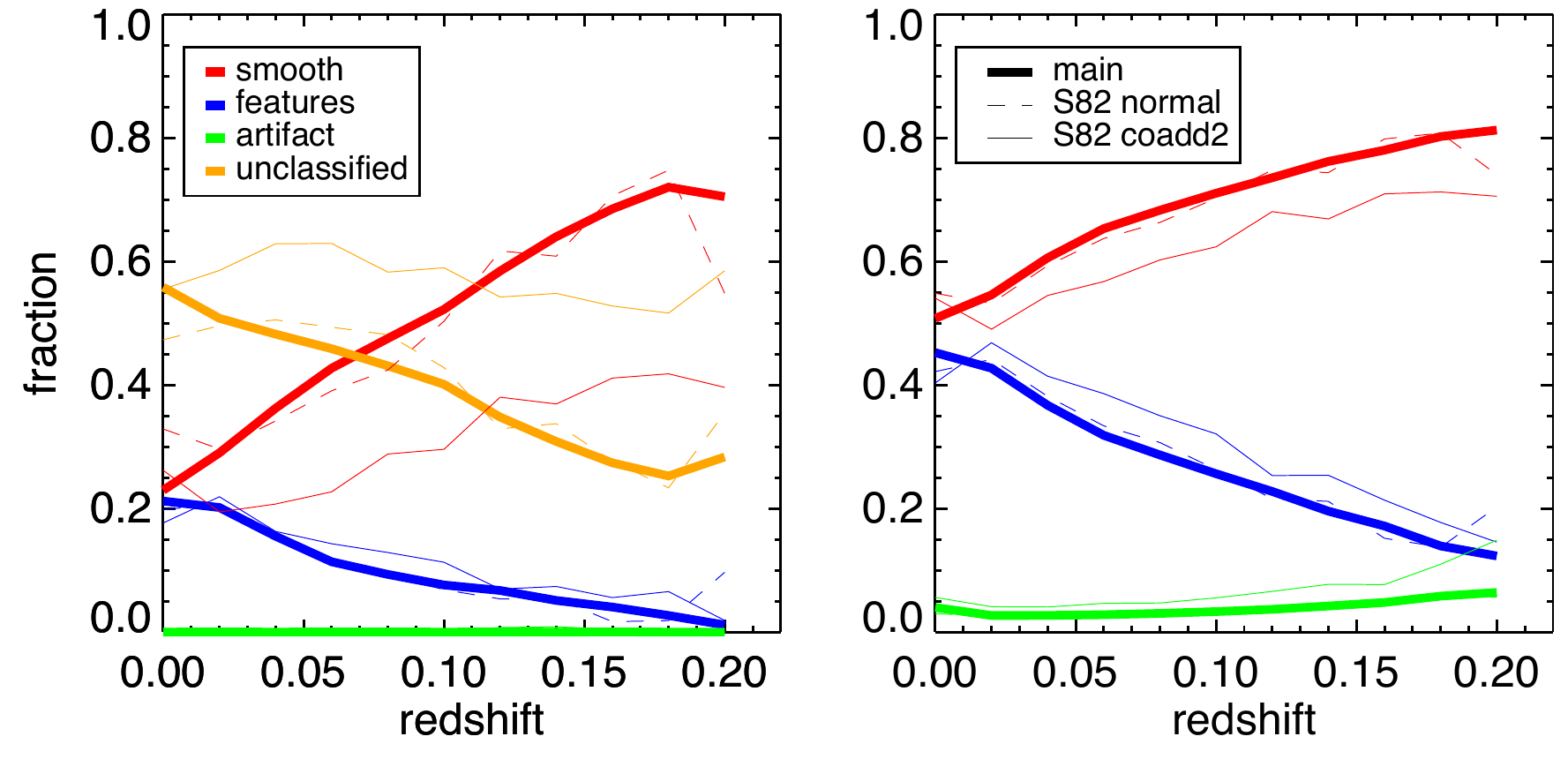}
\caption{GZ2 vote fractions for Task~01 ({\it smooth, features/disk, or star/artifact?}) as a function of spectroscopic redshift. {\it Left:} fraction of galaxies for which the GZ2 vote fraction exceeded 0.8. Galaxies with no responses above 0.8 are labeled as ``unclassified''. {\it Right:} mean GZ2 vote fractions weighted by the total number of responses per galaxy. Data are shown for the GZ2 original + extra (thick solid), Stripe~82 normal-depth (thin dotted), and Stripe~82 co-add depth (thin solid) samples with a magnitude limit of $m_r < 17.0$.
\label{fig-task01}}
\end{figure*}

Figure~\ref{fig-task01} compares the results of the Task~01 classifications for the GZ2 main and Stripe~82 samples. The distributions of the responses for both the main sample and Stripe~82 normal-depth show similar behavior as a function of redshift. This applies both when using thresholded vote fractions and the raw likelihoods. The type fractions for the coadded data, however, are significantly different -- there is a significant increase at all redshifts in the fraction of responses for ``features or disk''. This increases the fraction of unclassified galaxies (and subsequently decreases the fraction of smooth galaxies) when using thresholds, and a similar shift of vote fractions from smooth to feature/disk when using the raw likelihoods. 

This difference demonstrates why the main sample corrections cannot be applied to the coadded images. The likely cause is that the coadded data allows classifiers to better distinguish faint features and/or disks, due to both improved seeing \citep[from $1.4\arcsec$ to $1.1\arcsec$;][]{ann11} and higher signal-to-noise ratio.

\begin{figure*}
\includegraphics[angle=0,width=7.0in]{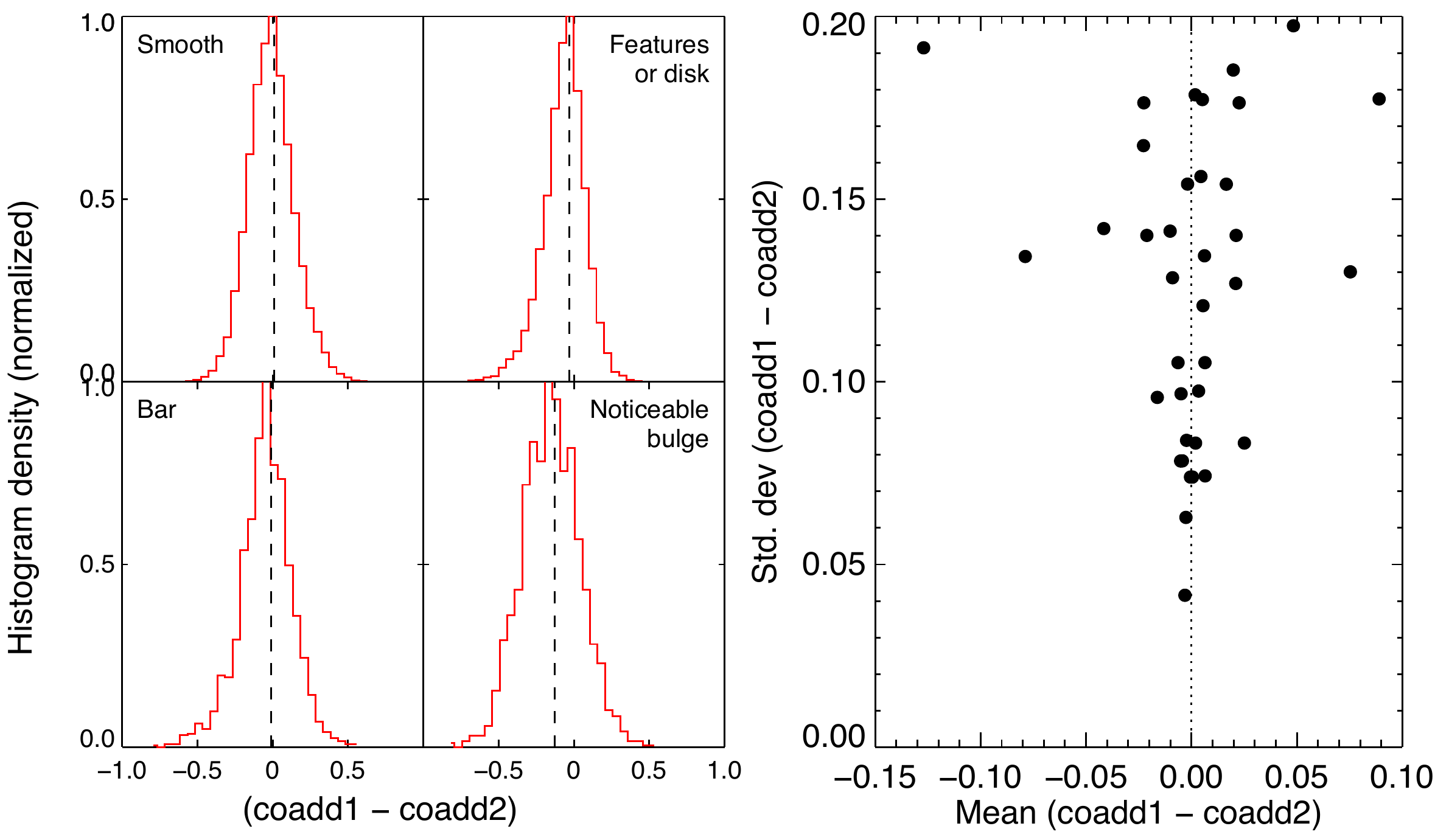}
\caption{Comparison of GZ2 classifications for the coadded images of Stripe~82. Left: Distribution of the difference in vote fractions (\dcoadd) for galaxies that appear in both the coadd1 and coadd2 samples. Four example tasks are shown, including only galaxies with at least 10 responses per task. The dashed line shows the median of each distribution. The response ``noticeable bulge'' for Task~05 was the only example for which the mean $|\Delta_\mathrm{coadd}| > 0.1$. Right: mean values \dcoadd~for every response in the GZ2 tree. 
\label{fig-stripe82_compare}}
\end{figure*}

Classifications for the two sets of coadded Stripe~82 images show no systematic differences for the majority of the GZ2 tasks. Figure~\ref{fig-stripe82_compare} shows the difference between the two vote fractions ($\Delta_\mathrm{coadd} = p_\mathrm{coadd1} - p_\mathrm{coadd2}$) for four examples. A non-zero mean value of \dcoadd~would indicate a systematic bias in classification, possibly due to the differences in image processing. In GZ2, 33/37~tasks have $|\Delta_\mathrm{coadd}| < 0.05$ for galaxies with at least 10 responses to the task. 

The biggest systematic difference is for the response to Task~05 (bulge prominence) of the bulge being ``just noticeable''. The mean fraction in coadd2 is 35\% higher than that in coadd1. This effect is opposite (but not equal) to that for an ``obvious'' bulge, for which coadd1 is 13\% higher; this may indicate a general shift in votes toward a more prominent bulge. A similar but smaller effect is seen in classification of bulge shapes for edge-on disks (Task~09), where votes for ``no bulge'' in coadd1 data go to ``rounded bulge'' in coadd2. The specific cause for these effects as it relates to the image quality is not investigated further in this paper. 

For most morphological questions, the two versions of coadded images showed no significant difference. While either set of coadded data can likely be used for science, we recommend using coadd2 if choosing between them. The overall consistency indicates that the votes for both could potentially be combined and treated as a single data set; this could be useful for increasing the classification accuracy for deeper responses (such as spiral arm properties) within the GZ2 tree.

\subsection{Using the classifications}\label{ssec-usingdata}

Since GZ2 is intended to be a public catalogue for use by the community, we present two examples of how classifications can be selected. Actual use will depend on the individual science case, and additional cuts (e.g., making a mass or volume-limited sample) may be required to define the parameters more appropriately.

The first use case suggested for the GZ2 data is the selection of pure samples matching a specific morphology category. This is appropriate for when some finite number of objects with clear morphological classifications is required (perhaps for individual study or an observing proposal), but there is no requirement to have a complete sample. An example would be the selection of three-armed spirals. The simplest way is to search for galaxies in the GZ2 spectroscopic main sample (Table~\ref{tbl-mainclass}) with {\tt t11\_arms\_number\_a33\_3\_flag}$ = 1$, which returns 308~galaxies. Inspection of the flagged images shows that they are all in fact disk galaxies with three spiral arms, with no object that is a clear false positive. 

Alternatively, those making use of the catalogue can set their own thresholds for the debiased likelihoods to change the strength of the selection criteria. This flag is currently set via the combination of $p_\mathrm{features/disk}>0.430$, $p_\mathrm{edge-on,no}>0.715$, $p_\mathrm{spiral,yes}>0.619$, $N_\mathrm{spiral,yes}>20$, and $p_\mathrm{3~arms}>0.8$ (Table~\ref{tbl-thresholds}). These cuts are generally regarded as conservative, and more genuine three-armed spirals might be discovered by, for example, lowering the threshold on $p_\mathrm{3~arms}$. If the number of objects returned by such a query is of a manageable size, we suggest that images be individually examined -- this is the only way to differentiate spirals with true radial symmetry from 2-armed spirals with an additional tidal tail, for example. Galaxy Zoo~2 papers that employ similar methods of selecting specific morphologies include \citet{mas11c}, \citet{kav12a}, \citet{sim13}, and \citet{cas13}. 

The second common use case for the morphologies is the direct use of the likelihoods. While thresholds on the likelihoods are appropriate for some studies, many classifications do not exceed $p>0.8$ for any available response, especially when more than two responses are available. These intermediate classifications are a combination of genuine physical attributes of the galaxy (vote fractions of $p_\mathrm{smooth}=0.5,p_\mathrm{features/disk}=0.5$ may accurately characterize a galaxy with both strong bulge and disk components) in addition to limitations in accuracy from the image quality and variance among individual classifiers. The problem is that thresholding only samples a small portion of the vote distributions.

In order to use data for the {\it entire} sample, the debiased likelihoods for each response can be treated as probabilistic weights. As an example, consider the type fractions from Task~01 shown in Figure~\ref{fig-task01}. The left hand side shows the average fraction of morphological classes at each redshift only defining a ``class'' as exceeding some vote fraction threshold; as a result, more than half the galaxies are left ``unclassified'', with no strong majority. The panel on the right in Figure~\ref{fig-task01} uses the likelihoods directly. A galaxy with $p_\mathrm{smooth}=0.6$, $p_\mathrm{features/disk}=0.3$, and $p_\mathrm{star/artifact}=0.1$ contributes $0.6$ of a ``vote'' to smooth, $0.3$ to features/disk, and $0.1$ to star/artifact. This approach is generally suitable for studying morphology dependence on global variables, such as environment or colour. Further examples of using the likelihoods as weights can be found in \citet{bam09}, \citet{ski12}, and \citet{cas13}.


\section{Comparison of GZ2 to other classification methods}\label{sec-comparison}

To assess both the scope and potential accuracy of the GZ2 classifications, we have compared our results to four morphological galaxy catalogues (including the previous version of Galaxy~Zoo). All four catalogues contain classifications based on optical SDSS images and have significant overlaps with the galaxies in GZ2. 

\begin{itemize}
	\item Galaxy~Zoo~1 \citep{lin11}: Citizen science
	\item \citet{nai10} : Expert visual classification
	\item EFIGI \citep{bai11} : Expert visual classification
	\item \citet{hue11} : Automatic classification
\end{itemize}

\begin{table}
\centering
 \begin{tabular}{@{}lrcrcrcrc}
 \hline
\multicolumn{1}{l}{\underline{GZ2}} &
\multicolumn{2}{c}{\underline{GZ1}} &
\multicolumn{2}{c}{\underline{HC11}} &
\multicolumn{2}{c}{\underline{NA10}} &
\multicolumn{2}{c}{\underline{EFIGI}}
\\ 
\multicolumn{1}{l}{} &
\multicolumn{1}{c}{$N$} &
\multicolumn{1}{c}{\%} &
\multicolumn{1}{c}{$N$} &
\multicolumn{1}{c}{\%} &
\multicolumn{1}{c}{$N$} &
\multicolumn{1}{c}{\%} &
\multicolumn{1}{c}{$N$} &
\multicolumn{1}{c}{\%}
\\ 
\hline					
early-type     & 79214    & 86.2     & 26732    & 82.1   & 1995    & 96.7    & 214     & 84.6       \\
late-type      & 26314    & 97.9     & 79277    & 88.6   & 5481    & 94.9    & 1675    & 98.2       \\
bar            & --       &  --      & --       &  --    & 651     & 94.9    & 238     & 98.7       \\
ring           & --       &  --      & --       &  --    & 438     & 91.6    & 110     & 83.6       \\
merger         & 526      & 63.3     & --       &  --    & 43      & 100.    & 6       & 100.       \\
\hline
 \end{tabular}
 \caption{Comparison of the agreement in morphology between the GZ2 main sample and other catalogues. For each category, a galaxy is considered to ``agree'' if it has a likelihood of at least $0.8$ (``clean'') in GZ2 and at least $0.5$ (``majority'') in the other catalogue (or for NA10, inclusion in the relevant category). This table gives both the total number of overlapping galaxies and the fraction that agreed in the corresponding catalogue when matched to GZ2. \label{tbl-compare}}
\end{table}

A summary of the agreement between GZ2 and other catalogues is given in Table~\ref{tbl-compare}; the remainder of this section discusses the results in more detail. 

\subsection{Galaxy~Zoo~1 and Galaxy~Zoo~2}\label{ssec-gz1gz2}

The galaxies in GZ2 are a subset of GZ1, with $248,883$~in both catalogues. The similarities between GZ1 and Task~01~in GZ2 allow their results to be compared in detail. We analyzed vote fractions for the GZ1 ``elliptical'' category as compared to GZ2 ``smooth'' galaxies, and combined responses for all three GZ1 spiral categories to the GZ2 ``features or disk'' response. 

The matched GZ1-GZ2 catalogue contains $33,833$ galaxies identified as ellipticals based on their debiased GZ1 likelihoods \citep{lin11}. Using the GZ2 debiased likelihoods, $50.4\%$ of galaxies have vote fractions exceeding $0.8$ in both samples, while $97.6\%$ have vote fractions exceeding $0.5$. There are only $\sim1200$ ellipticals identified in GZ1 that have $p_\mathrm{features/disk}>0.5$ in GZ2. Of these, roughly $40\%$ are barred galaxies, and almost all show obvious bulges, the likely cause of their identification as early-type in GZ1. 

The GZ2 main sample contains $83,956$ galaxies identified as spirals by GZ1. The agreement with the ``features or disk'' response in GZ2 is significantly lower than that of ellipticals. Only $31.6\%$ of the GZ1 clean spirals had vote fractions greater than in GZ2, with $59.2\%$ having a vote fraction greater than $0.5$. The GZ2 debiased likelihoods for the same galaxies agree at $38.1\%$ (for $0.8$) and $78.2\%$ (for $0.5$). Of the $\sim1600$ spirals in GZ1 with $p_\mathrm{features/disk}<0.4$, visual inspection shows these to be almost entirely inclined disks or lenticular galaxies without spiral arms. These galaxies are slightly bluer than the other early-types in GZ1; however, we emphasize that most elliptical galaxies with this colour \citep{sch09} were correctly classified in the initial GZ1 project. 

\begin{figure}
\includegraphics[angle=0,width=3.3in]{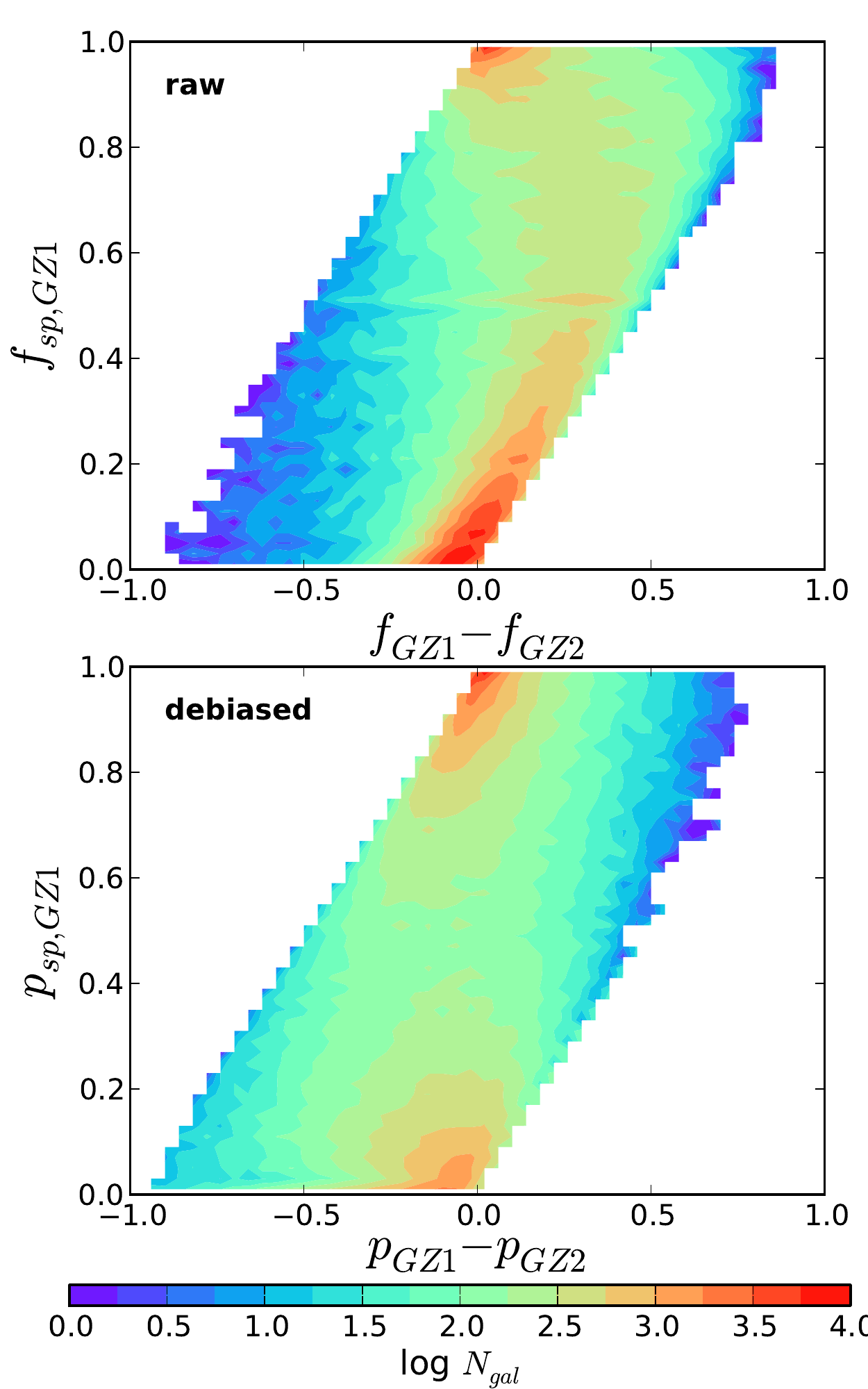}
\caption{Comparison of spiral galaxies using classifications for ``combined spiral'' (GZ1) and ``features or disk'' (GZ2). {\it Top}: raw vote fractions. At intermediate values ($f_{sp}\sim0.5$), GZ1 classifiers are more likely to identify galaxies as spiral compared to GZ2. {\it Bottom}: debiased vote fractions. At intermediate values, GZ1 and GZ2 classifications are consistent with each other; however, there is an increased scatter in the vote fractions near $p_\mathrm{sp}\simeq0$ and $p_\mathrm{sp}\simeq1$. 
\label{fig-trumpet}}
\end{figure}

Figure~\ref{fig-trumpet} shows the difference between the vote fractions for the spiral classifications in GZ1 and features/disk classifications in GZ2 for all galaxies that appear in both catalogues. The vote fractions show a tight correlation at both very low and very high values of the GZ1 vote fraction for combined spiral ($f_{sp}$), indicating that both projects agree on the strongest spirals (and corresponding ellipticals). At intermediate ($0.2--0.8$) values of $f_{sp}$, however, the GZ1 vote fractions are consistently higher than those in GZ2, differing by up to $0.25$. When using debiased likelihoods in place of the vote fractions, this effect decreases dramatically; however, the tightness of the correlation correspondingly drops at low and high $p_\mathrm{sp}$. 

\begin{figure}
\includegraphics[angle=0,width=3.5in]{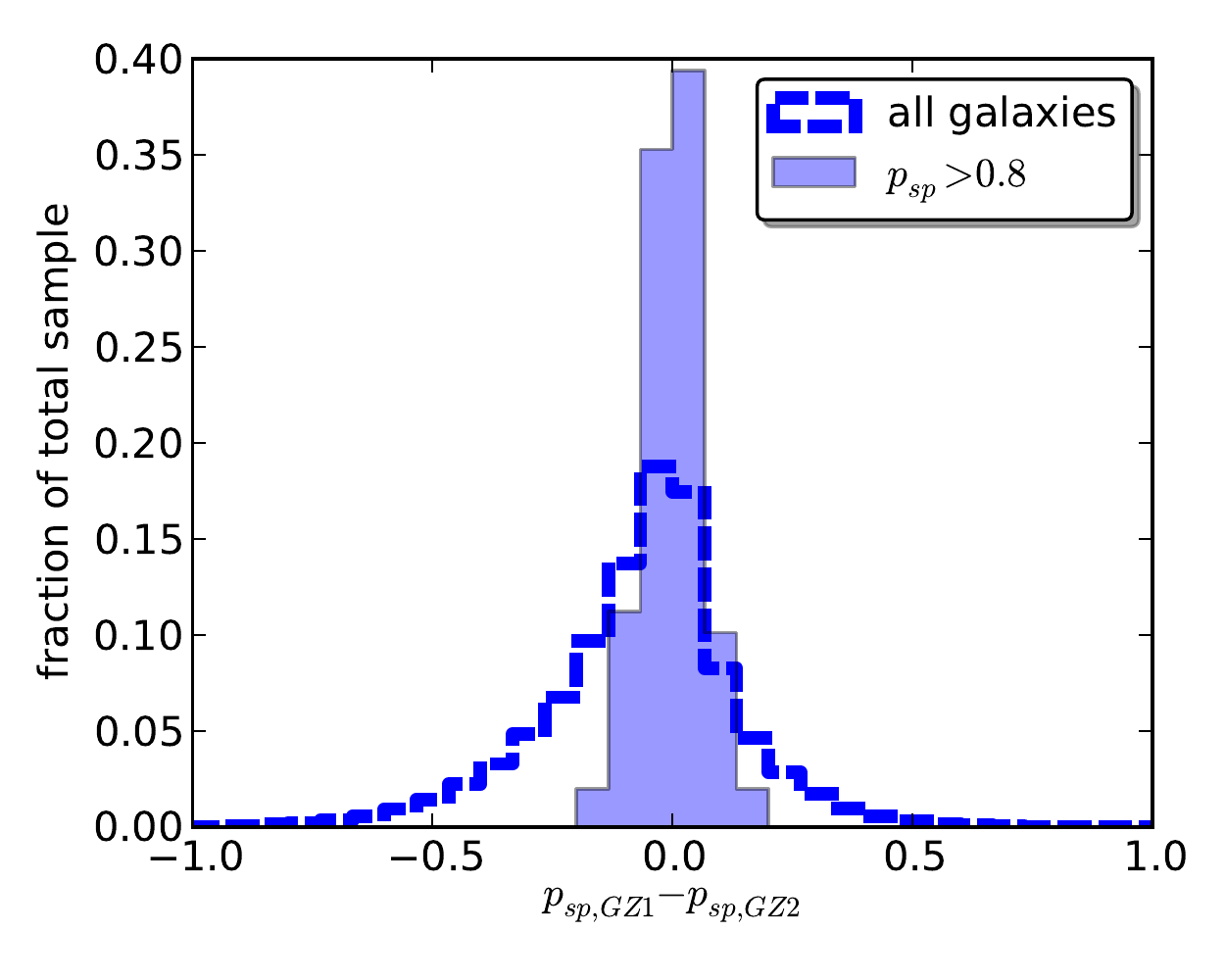}
\caption{Comparison of the spiral feature vote fractions for objects in Galaxy~Zoo~1 (GZ1) and Galaxy~Zoo~2 (GZ2). The dashed line shows the difference between $p_\mathrm{combined~spiral}$ for GZ1 and $p_\mathrm{features~or~disk}$ for GZ2 for the $240,140$ galaxies in both samples. The filled histogram shows the same metric for the $57,994$ galaxies classified as ``clean'' spirals in both GZ1 and GZ2. 
\label{fig-gz1_gz2}}
\end{figure}

Galaxies are slightly more likely to be identified as a spiral in GZ2 than in GZ1. Figure~\ref{fig-gz1_gz2} shows the distribution of the difference between spiral classifications, using the debiased likelihoods for combined spirals for GZ1 and ``features or disk'' galaxies in GZ2. The slight leftward skew indicates that a galaxy is more likely to be identified as a spiral in GZ2 compared to GZ1. When restricted only to galaxies in the joint clean samples ($p>0.8$), the spread is greatly reduced and the distribution is centered around a difference of zero, indicating that the two agree very well for classifications with high levels of confidence. 

Based on classifications from galaxies in both projects, GZ2 is more conservative than GZ1 at identifying spiral structure. A possible explanation is that this is a bias from classifiers who are anticipating subsequent questions about the details of any visible structures. An experienced classifier, for example, would know that selecting ``features or disk'' is followed by additional questions, none of which offer options for an uncertain classification. If the classifier is less confident in identifying a feature, it is possible they would avoid this by clicking ``smooth'' instead. 



The GZ1 interface had one option to classify merging galaxies. This was a rare response, comprising less than $1$~percent of the total type fraction at all redshifts in GZ1 \citep{bam09}. \citet{dar10a} found that a vote fraction of $f_{mg} > 0.6$ robustly identified merging systems in GZ1. Of the $1632$ such systems classified in GZ2, more than $99\%$ were identified as ``odd'' galaxies, and $77.7\%$ had $p_\mathrm{mg}>0.5$ in GZ2. This is partly due to early-stage merging spirals avoiding the ``merger'' classification, with only late-state mergers with extremely disturbed morphologies recording high vote fractions for the merger question. 

In addition to the angular separation bias discussed in \S\ref{ang-bias}, GZ2 responses to Task~08 (``odd feature'') also suffer from crosstalk. This is the result of more than one response being applicable for some galaxies, which forces the participant to choose the one they consider most relevant. For example, a merging galaxy may display a strong dust lane, be highly irregular in shape, {\em and} have a disturbed appearance. While ``merger'', ``dust lane'', ``irregular'' and ``disturbed'' are all possible classifications, the participant will usually choose the ``merger'' classification and information about the other morphological features is lost. For close pairs, this crosstalk is a function of angular separation -- the fraction of galaxies classified as mergers increases with decreasing separation, while the other ``odd feature'' classifications lose votes correspondingly \citep{cas13}. We note that in later incarnations of Galaxy Zoo\footnote{\url{www.galaxyzoo.org}} it is possible to select multiple classifications from the ``odd feature'' task.


To summarize, GZ1 and GZ2 share nearly 250,000 galaxies that have been classified in both samples. The separation of early and late-type galaxies from the two projects is mostly consistent, especially for high-confidence ($p>0.8$) galaxies. GZ2 classifications are more conservative than GZ1 at identifying spiral structure for intermediate vote fractions. Mergers identified in GZ1 appear at a very high rate in GZ2 as ``odd'' galaxies, although classification as a merger is complicated by cross-talk between other GZ2 responses to Task~08. 

\subsection{Expert visual classifications}

The standard for detailed morphological classifications for many years has come from visual identifications by individual expert astronomers. We compare the GZ2 classifications to two SDSS morphological catalogues generated by small groups of professional astronomers: \citet[][hereafter NA10]{nai10} and EFIGI \citep{bai11}. The fact that GZ2 and both expert catalogues used data from the same survey allows for direct comparison of the results.

The catalogue of NA10 is based on images of 14,034 galaxies from SDSS DR4. Galaxies were selected from a redshift range of $0.01<z<0.1$, with an extinction-corrected apparent magnitude limit of $g<16$. In comparison, the GZ2 sample is deeper, spans a larger redshift range, and contains a more recent data release. 12,480 galaxies were classified in both GZ2 and NA10 -- this comprises nearly all (89.9\%) of the NA10 catalogue, but only 4.5\% of GZ2. 

NA10 is based on visual classifications of monochrome $g$-band images by a single astronomer (P.~Nair). The data include RC3 T-types \citep[a numerical index of a galaxy's stage along the Hubble sequence;][]{dev91} as well as measurements of bars, rings, lenses, pairs, interactions, and tails. The NA10 data does not contain information on the likelihood or uncertainty associated with morphological features, although it does measure some features by their relative strengths (dividing barred galaxies into strong, medium, and weak classes, for example). 


EFIGI consists of classifications of 4,458 galaxies, which are a subset of the RC3 catalogue with 5-colour imaging in SDSS DR4. Almost all galaxies in EFIGI are at $0.0001<z<0.08$. Classifications on composite $gri$ images were performed by a group of 11 professional astronomers, each of whom classified a subset of 445 galaxies. A training set of 100 galaxies was also completed by all 11 astronomers to adjust for biases among individual classifiers. 3,411 galaxies are in both EFIGI and GZ2. This constitutes 77\% of EFIGI and 1.2\% of the GZ2 sample. 

T-types in EFIGI were assigned using a slightly modified version of the RC3 Hubble classifications. Peculiar galaxies were not considered a separate type, and ellipticals were subdivided into various types: compact, elongated (standard elliptical), cD (giant elliptical), and dwarf spheroidals. The remaining morphological information, dubbed ``attributes'', is divided between six groups:

\begin{itemize}
	\item appearance: {\tt inclination/elongation }
	\item environment: {\tt multiplicity, contamination}
	\item bulge: {\tt B/T ratio}
	\item spiral arms: {\tt arm strength, arm curvature, rotation}
	\item texture: {\tt visible dust, dust dispersion, flocculence, hot spots}
	\item dynamics: {\tt bar length, inner ring, outer ring, pseudo-ring, perturbation}
\end{itemize}

\noindent EFIGI attributes were measured on a five-step scale from 0 to 1 (0, 0.25, 0.50, 0.75, 1). For some attributes (e.g., arm strength, rings), the scale is set by the fraction of the flux contribution of the feature relative to that of the entire galaxy. For others (e.g., inclination or multiplicity), it ranges between the extrema of possible values. 

The EFIGI and NA10 catalogues were compared in detail by \citet{bai11}. T-type classifications for the two catalogues strongly agree; EFIGI lenticular and early spirals have slightly later average classifications in NA10, while later EFIGI galaxies have slightly earlier NA10 T-types. EFIGI has a major fraction of galaxies with slight-to-moderate perturbations with no interaction flags set in the NA10 catalogue, indicating that NA10 is less sensitive toward more benign features (e.g., spiral arm asymmetry). The bar length scale is consistent between the two samples; good agreement is also found for ring classifications. 



\subsubsection{Bars}\label{sssec-bars}

To analyse the overlap between bars detected in expert classifications and GZ2, we restrict comparisons to galaxies identified as possessing disks and being ``not edge-on''. For the rest of this paper, we refer to such ``not edge-on'' disks as {\em oblique} disks (since many of them have inclination angles high enough that ``face-on'' is not an accurate description). Oblique galaxies are selected from the GZ2 data as having $p_\mathrm{features/disk}>0.430, p_\mathrm{not~edgeon}>=0.715$, and $N_\mathrm{not~edgeon}\geq20$ (Table~\ref{tbl-thresholds}). This restricts overlap of GZ2 oblique galaxies in NA10 to 5,526 objects. The ``not edge-on'' cut is similar to a restriction on inclination angle of $\lesssim70^\circ$, based on the average axial ratio from the SDSS exponential profile fits.

NA10 detected 2537 barred galaxies, 18\% of their total. For objects with T-types later than E/S0, this rises to 25\% of the sample. This is consistent with the bar fraction from \citet{mas11c} for oblique disk galaxies (29\%). Of the objects NA10 identify as barred galaxies, 2348 (93\%) are objects in GZ2. \citet{mas11c} analysed bar classifications in NA10 and the RC3 and the GZ2 bar classifications (albeit before the classification bias was applied). They found good agreement, particularly finding that values of $f_\mathrm{bar}>0.5$ identified almost all strongly barred NA10 and RC3 galaxies, and that $f_\mathrm{bar}<0.2$ correlated strongly with galaxies identified as unbarred by NA10 and RC3. 

\begin{figure}
\includegraphics[angle=0,width=3.5in]{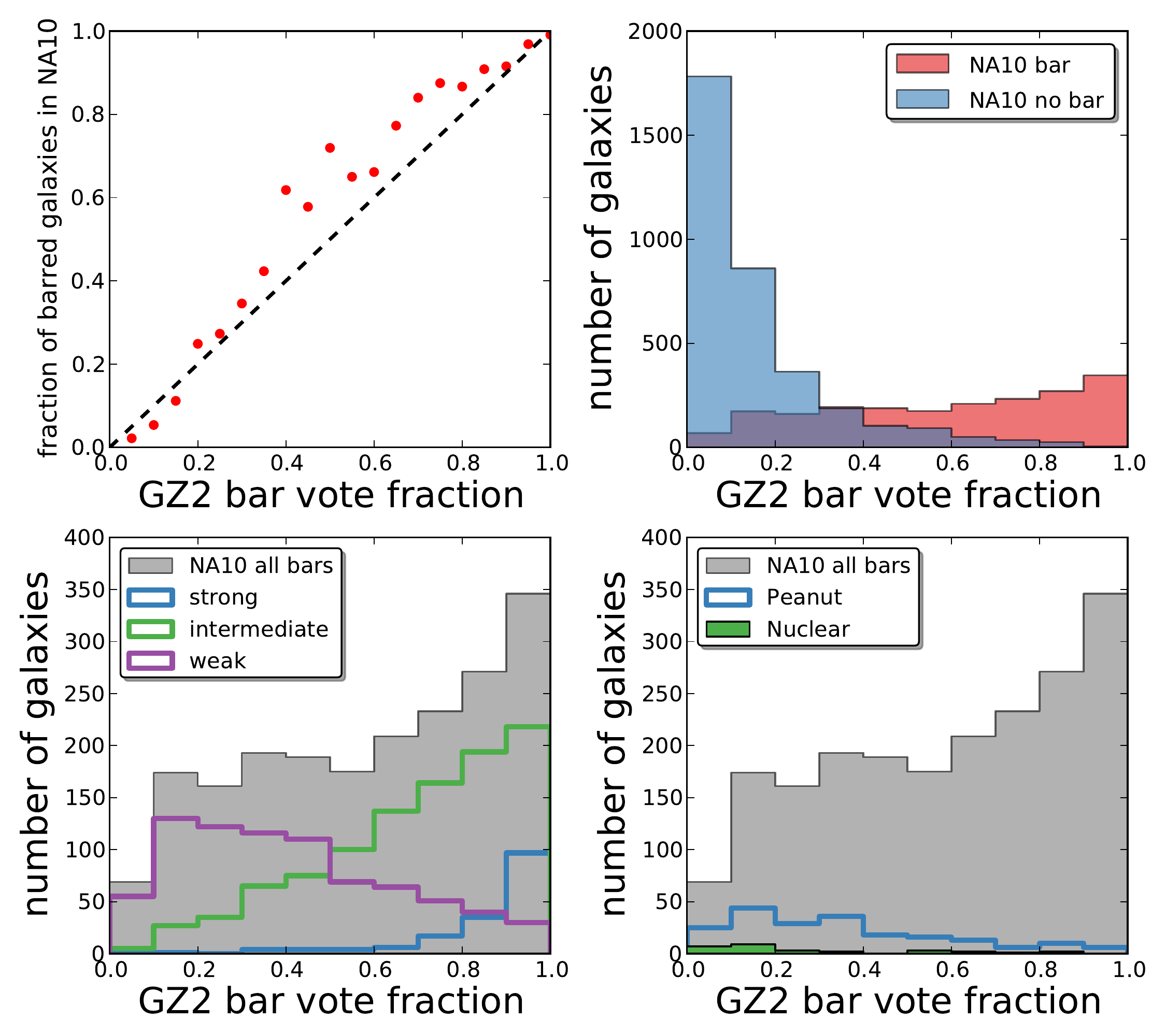}
\caption{Classifications for galactic bars in GZ2 and NA10. Data are for the 5,526 galaxies in both samples classified by GZ2 as not-edge-on disks and with $\geq20$ bar classifications. Top left: mean bar vote fraction per galaxy in GZ2 vs. the ratio of barred to all galaxies in NA10. Dashed line shows the one-to-one relationship. Top right: distribution of the GZ2 debiased bar vote fraction, separated by NA10 classifications. Bottom left: distribution of GZ2 bar vote fraction for the three disk-scale bar categories of NA10. Bottom right: distribution of GZ2 bar vote fraction for peanut and nuclear bars from NA10. 
\label{fig-na_bars}}
\end{figure}

Bars in NA10 are classified according to either bar strength (weak, intermediate, strong) or by other morphological features (ansae, peanuts, or nuclear bar). A galaxy may in rare cases have both a disk-scale (strong, intermediate, or weak) and a nuclear bar. Figure~\ref{fig-na_bars} ({\it top left}) shows that the GZ2 average vote fraction for bars closely agrees with the NA10 fraction of barred galaxies for each GZ2 bin. The two quantities are not identical; the x-axis plots {\it individual classifications} of galaxies with varying vote fractions for the presence of a bar. The y-axis shows the ratio of barred to unbarred galaxies in NA10. The data have a correlation coefficient of $\rho=0.984$, and lie slightly above a linear relationship for $p_\mathrm{bar}>0.4$. For bar identification, the aggregate votes of volunteers closely reproduce overall trends in expert classification. 

The top right panel of Figure~\ref{fig-na_bars} shows the distribution of GZ2 bar votes by simply splitting the NA10 sample in two: galaxies without a bar and galaxies with a bar (of any kind). Both samples show a strong trend toward extrema, with the peak near zero for non-barred galaxies indicating that GZ2 classifiers are very consistent at identifying unbarred disk galaxies. Possession of a bar is less straightforward; while the frequency of NA10 bars does increase with GZ2 fraction, 39\% of barred galaxies from NA10 have a GZ2 $p_\mathrm{bar}<0.5$. Conversely, only 6\% of non-barred NA10 galaxies have GZ2 bar vote fractions above 0.5. 

For galaxies where our identification of a bar ($p_\mathrm{bar}\geq0.5$) disagrees with NA10, inspection shows that almost all are in fact true bars, with some overlap from galaxies with outer rings. Galaxies with GZ2 vote fractions between 0.3 and 0.5 show more of a mix, with some likely bars and some spurious identifications from GZ2. Interestingly, there is also no difference in the average colour, size, or apparent magnitude for galaxies in which the NA10 and GZ2 classifications disagree when compared to those in which they do agree. 



The bottom left panel of Figure~\ref{fig-na_bars} shows the distribution of GZ2 vote fraction split by bar strength from NA10. The distribution for all bars is the same as shown in the top right, increasing with GZ2 vote fraction. There is a clear difference in the GZ2 classifications as a function of NA10 bar strength; all three are statistically highly distinct from each other and from the overall barred sample, according to a two-sided K-S test. The majority of both the strong and intermediate barred population have high GZ2 vote fractions, with 78\% of strong bars and 40\% of intermediate bars at $p_\mathrm{bar}>0.8$. This increases to 94\% and 80\%, respectively, if the majority criterion of 0.5 \citep{mas11c} for the GZ2 vote fraction is used instead. Only 9\% of weakly-barred galaxies have GZ2 vote fractions above 0.8, and 32\% have vote fractions above 0.5. 

The lack of sensitivity to weak bars from NA10 may also be related to the design of the GZ2 interface. When asked if a bar is present, the image shown in the web interface is an icon with two examples of a barred galaxy (Figure~\ref{fig-flowchart}). The example image has the bar extending across the disk's full diameter, fitting the typical definition of a strong bar. With this as the only example (and no continuum of options between the two choices), GZ2 participants may not have looked for bars shorter than the disk diameter, or have been less confident in voting for ``yes'' if they were identified. Results from \citet{hoy11} show that classifiers are fully capable of identifying weak bars in other contexts.

Ansae, peanuts, and nuclear bars as identified by NA10 do not correlate strongly with the GZ2 bar parameter. In fact, the median bar vote fraction for peanuts and nuclear bars (no ansae appear in the oblique sample) is only $p_\mathrm{bar}=0.29$. Nuclear bars are the only feature that overlaps with the NA10 bar strength classifications; out of 283 nuclear bars, 3 galaxies also have strong bars, 44 have intermediate bars, and 166 have weak bars.

\begin{figure}
\includegraphics[angle=0,width=3.5in]{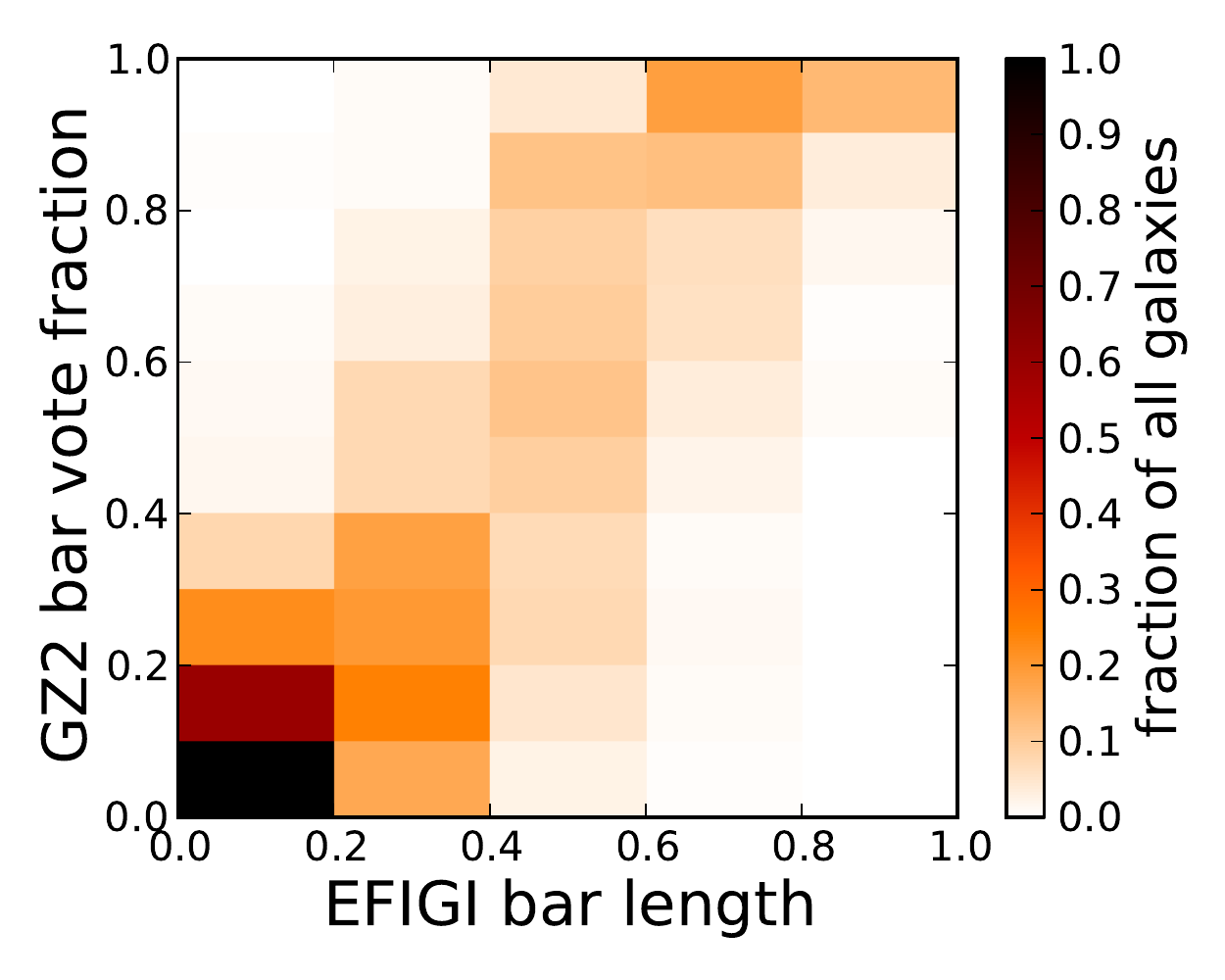}
\caption{EFIGI bar length classifications compared to their GZ2 vote fractions for the presence of a bar. Data are for the 2,232 oblique disk galaxies in both EFIGI and GZ2 with at least 10 bar classifications. 
\label{fig-efigi_bars}}
\end{figure}

The EFIGI bar length attribute is measured with respect to $D_{25}$, the decimal logarithm of the mean isophote diameter at a surface brightness of $\mu_B=25$~mag~arcsec$^{-2}$. A value of 1.0 (the strongest bar) extends more than half the length of $D_{25}$, while the median value of 0.5 would be about one-third the length of $D_{25}$. The overall fraction of barred galaxies in EFIGI is 42\% (1439/3354); this is essentially unchanged if only oblique galaxies are considered (915/2099 = 44\%). This is significantly higher than the mean bar fraction of \citet{mas11c}, at 29.5\%, but consistent with results using automated ellipse-fitting techniques \citep{bar08,agu09}. 

The higher fraction in EFIGI is due to the contributions of galaxies with bar length attributes of 0.25, the majority of which have GZ2 vote fractions below 0.5. If only EFIGI galaxies at 0.5 and above are considered to be barred, then the bar fraction falls to 17\%. Only some of the galaxies in the 0.25 EFIGI bin are being classified by GZ2 as barred, however, \citet{bai11} defines these as ``barely visible'' bars. 

There is a strong correlation between the GZ2 bar vote fractions and the attribute strength from EFIGI (Figure~\ref{fig-efigi_bars}). 65\% of galaxies in both EFIGI and GZ2 sample have no strong evidence for a bar ($p_\mathrm{bar}<0.3$); of those, 77\% had EFIGI bar attributes of zero and 94\% had 0.25 or less. For galaxies where GZ2 $p_\mathrm{bar}>0.8$, the EFIGI attribute lies almost entirely at either 0.75 or 1.0. The correlation coefficient between the EFIGI and GZ2 bar measurements is $\rho=0.75$. 

Using the criteria for oblique galaxies from Table~\ref{tbl-thresholds}, there are 1,543 galaxies with an EFIGI bar classification. Barred galaxies as identified by GZ2 ($p_\mathrm{bar}\geq0.3$) agree very well with EFIGI; less than 5\% of GZ2 barred galaxies have EFIGI attributes of 0, with a mean value of 0.56. This could indicate a selection preference toward medium-length bars (one-third to one-half of $D_{25}$), or could genuinely reflect the fact that medium bars are the most common length in disk galaxies. 

Data from both NA10 and EFIGI can be used to quantify a threshold to identify barred galaxies in GZ2 data. The fraction of non-barred oblique galaxies as identified by both expert catalogues drops to less than $5\%$ at a GZ2 vote fraction $p_\mathrm{bar}=0.3$. This threshold may be changed depending on the specific science needs, but offers a useful trade-off between inclusion of nearly all (97\% from NA10) strong and intermediate bars and most (75\%) of the weak bars. This is a slightly more inclusive threshold than the $f\geq0.5$ used by \citet{mas11c}. We also note that the strong correlation between $p_\mathrm{bar}$ and EFIGI bar strength suggests that $p_\mathrm{bar}$ may be used directly (with caution) as a measure of bar strength in GZ2 galaxies.




\subsubsection{Rings}

NA10 classify three types of ringed galaxies based on criteria from \citet{but96}: inner rings (between the bulge and disk), outer rings (external to the spiral arms), and nuclear rings (lying in the bulge region). In GZ2, rings can be identified only if the user selects ``yes'' for the question {\it ``Anything odd?''} Since the ``odd feature'' task has seven responses, of which only one can be selected, any galaxies with multiple ``odd'' features will have votes split among the features, with only one option achieving a plurality (see \S\ref{ssec-gz1gz2}). While this means that some galaxies with rings may have low vote fractions in the GZ2 classifications, those with high vote fractions are typically strong and distinct.

In the NA10 catalogue, 18.2\% of all galaxies (30\% of disks) have a ring. Of those, 10\% are nuclear rings, 74\% are inner rings, and 32\% are outer rings (the sum is more than 100\% since one-third of ringed galaxies have multiple rings flagged). NA10 and GZ2 ring classifications are compared for the oblique galaxies in both samples. No cut is applied to the vote fraction for the ``anything odd'' question; even a comparatively low cut of $p_\mathrm{odd}>0.2$ eliminates roughly 40\% of the ringed galaxies identified in NA10. 

\begin{figure}
\includegraphics[angle=0,width=3.3in]{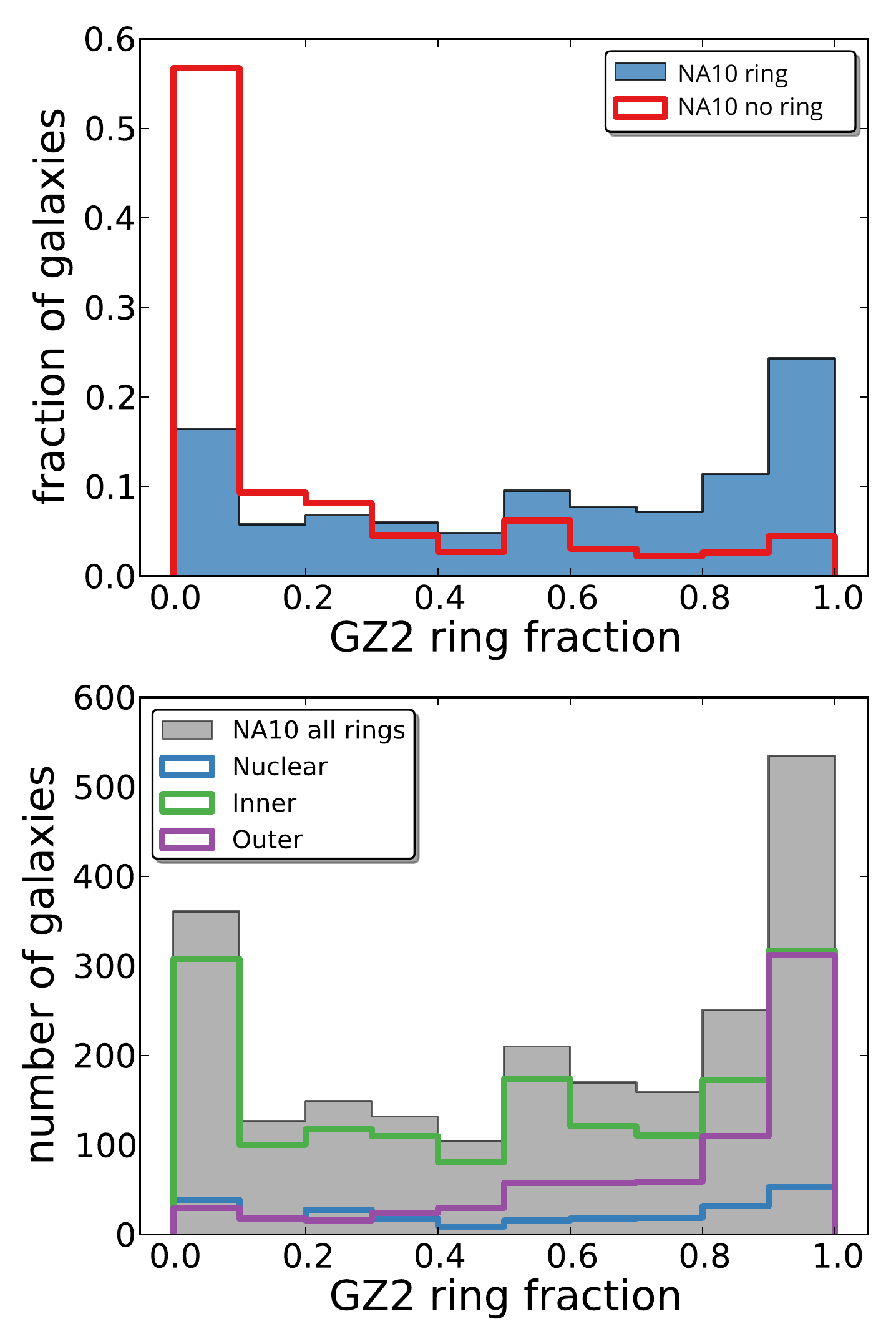} 
\caption{Ring classifications in GZ2 and NA10. Data are for the 7,245 oblique galaxies in both samples. Top: GZ2 vote fraction for rings ($N_\mathrm{ring}/N_\mathrm{odd}$) for all galaxies, split by their NA10 ring identifications. Bottom: GZ2 ring vote fraction for all rings identified by NA10, separated by ring type.
\label{fig-na_rings}}
\end{figure}

Figure~\ref{fig-na_rings} shows the distribution of the GZ2 ring vote fraction ($p_\mathrm{ring}$) in the oblique sample, split by the identification of a ring in NA10. While there is a marked increase in the fraction of ringed galaxies at $p_\mathrm{ring}>0.5$, more than a third of these galaxies are identified by NA10 as ringless. The agreement is significantly better if a limit is placed on the number of votes. Setting $N_\mathrm{ring} > 5$, for example, increases the agreement to $\sim75\%$. 

The distribution of $p_\mathrm{ring}$ is strongly affected by the ring type. Among galaxies that NA10 identifies as rings for which GZ2 strongly disagrees ($p_\mathrm{ring}<0.5$), the majority are classified as inner rings. There are 308 ringed galaxies from NA10 that have no ring votes at all in GZ2; 84\% of these are inner rings. For galaxies on which the NA10 and GZ2 ring classifications agree, the percentage of outer ringed galaxies is much higher. In the absence of specific instructions on different types of ring (the icon in Figure~\ref{fig-flowchart} does not indicate the size of the disk relative to the ring), GZ2 classifiers are much more likely to identify outer rings. The flat distribution of $p_\mathrm{ring}$ for nuclear rings indicates that there is also no strong correlation between GZ2 classifications and ring structures in the bulge. 

Most galaxies with $p_\mathrm{ring}>0.5$ are classified as outer rings in NA10, especially if constraints on $N_\mathrm{odd}$ and/or $p_\mathrm{odd}$ are added. Part of the reason for the remaining disagreements may relate to the placement of the ring classification in GZ2 at the end of the tree, and only as a result of the user identifying something ``odd''. Without having seen examples of ringed galaxies (especially as their structures connect to spiral arms), users may have been less likely to characterize the galaxy as odd and thus will not address the ring question. 

In EFIGI, rings are classified as inner, outer, and pseudo types. Both outer and pseudo ringed galaxies show reasonably strong correlations with GZ2 ring classifications, with a mean ring vote fraction of 0.69 for outer ringed galaxies and 0.71 for pseudo-ringed galaxies. The mean GZ2 ring vote fraction for inner rings is only 0.41. For galaxies in both EFIGI and GZ2, a high GZ2 ring vote fraction agrees significantly with the expert classification of a ring. 89\% of galaxies with $p_\mathrm{ring}>0.5$ and having at least 10 votes for ``Anything odd?'' were classified as rings in EFIGI. 

\begin{figure}
\includegraphics[angle=0,width=3.5in]{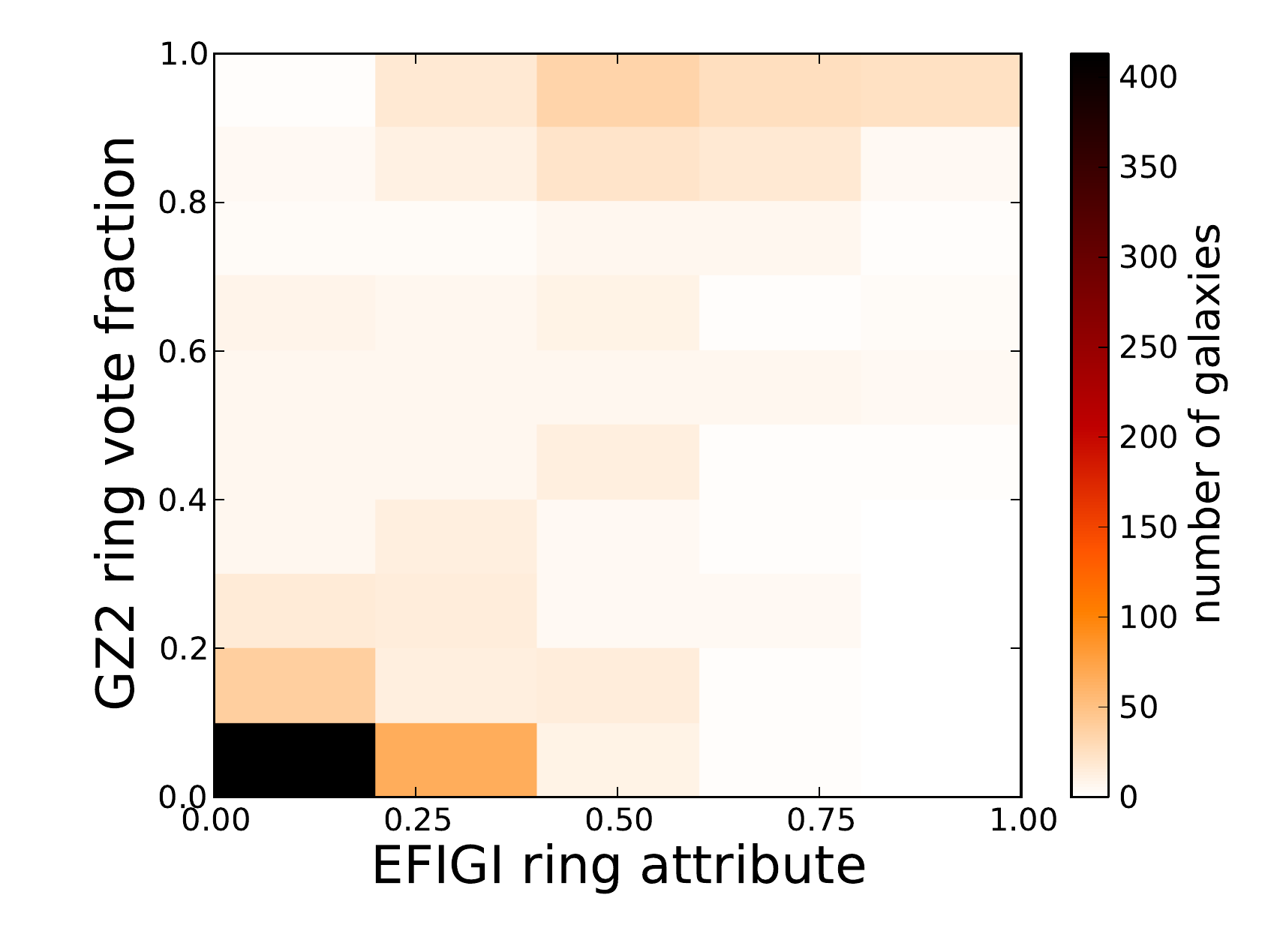}
\caption{EFIGI ring classifications compared to their GZ2 ring vote fractions. The EFIGI data is the strongest attribute among the combined inner, outer, and pseudo-ring categories. Data are for the 1,080 galaxies in both EFIGI and GZ2 with at least 10 responses to Task 08 (odd feature). 
\label{fig-efigi_rings}}
\end{figure}

Figure~\ref{fig-efigi_rings} shows a moderate correlation between the EFIGI ring attributes and the GZ2 ring vote fractions. The relationship is dominated by galaxies for which the methods agree strongly on either no ring or a ring with high contributions to the total galaxy flux. For intermediate (between 0.25 and 0.75) values of the EFIGI ring attribute, the GZ2 vote fraction has relatively little predictive power. 

\subsubsection{Mergers and interacting galaxies}

Galaxies in GZ2 are classified as mergers in Task~08 {\it ``anything odd?''} NA10 classify possible mergers in two ways: both as pairs of objects and as galaxies with visible interaction signatures. The paired objects are sorted by relative separation (close, projected, apparent, or overlapping pairs), and interacting galaxies by morphology (disturbed, warp, shells, tails, or bridges). 

In NA10, 22.3\% of galaxies are paired with another object; of these, 72\% are close pairs. Interacting galaxies are a much smaller subset, comprising only 7\% of the NA10 sample. In GZ2, only 252 galaxies have $p_\mathrm{odd}>0.8$ and $p_\mathrm{merger}>0.8$. 3\% of the NA10 paired galaxies have at least 10~GZ2 votes for a merger. 

\begin{figure}
\includegraphics[angle=0,width=3.3in]{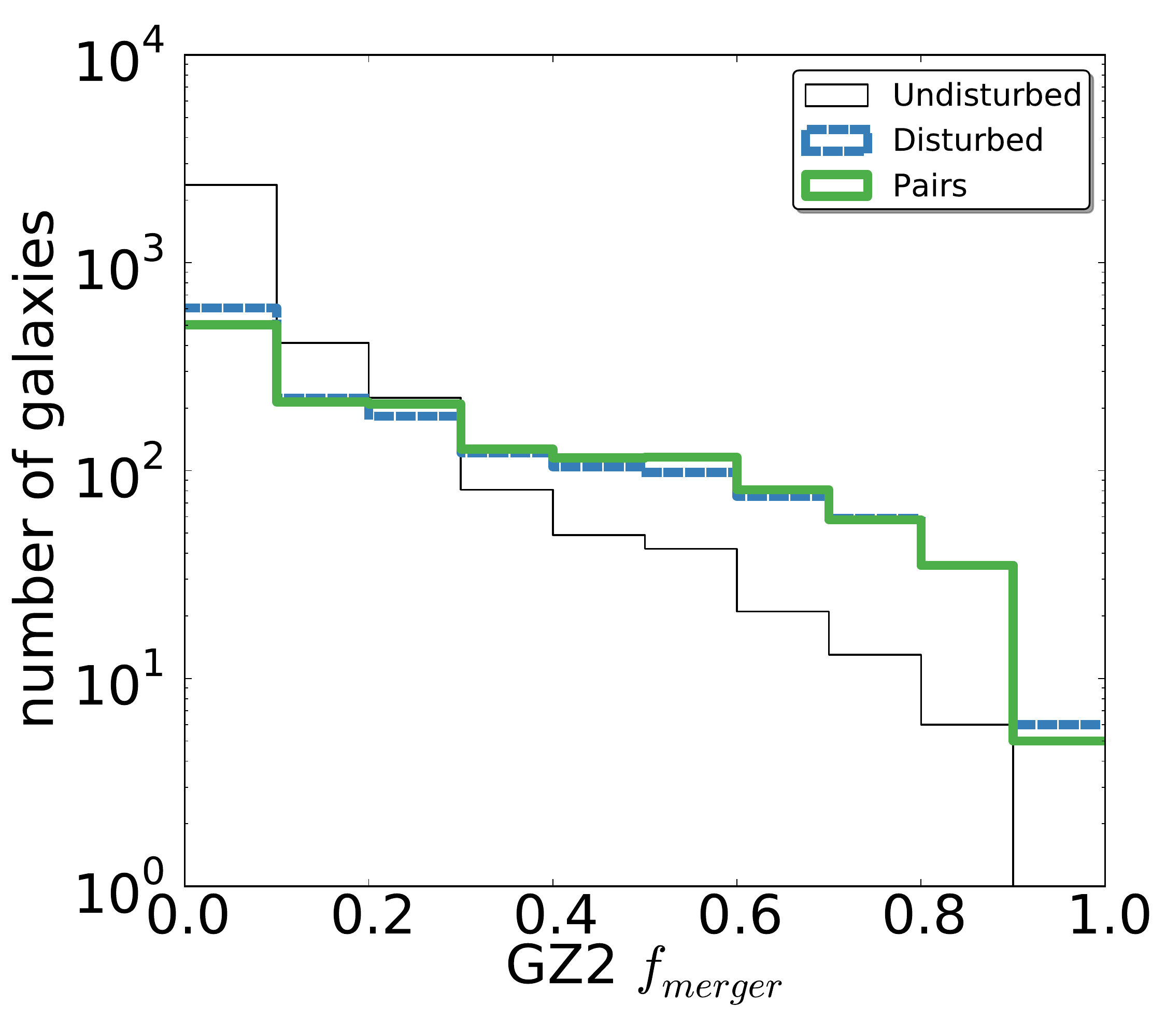}
\caption{Merger classifications in GZ2 and NA10. Data are for the 3,878 galaxies in both samples with $p_\mathrm{odd}>0.223$, showing the distribution of the vote fraction for the ``merger'' response to Task~08 in GZ2. The majority of galaxies have $p_\mathrm{merger}<0.1$. Galaxies classified by NA10 both as disturbed and in pairs dominate at $p_\mathrm{merger}>0.5$, but there remains a significant population of undisturbed galaxies even at the highest GZ2 vote fractions.
\label{fig-na_pairs}}
\end{figure}

Figure~\ref{fig-na_pairs} shows the distributions of NA10 paired and interacting galaxies with at least 10 votes for ``yes'' (something odd) for Task~06. Most galaxies have no votes for a merger, with only 6\% of galaxies having $N_\mathrm{merger}\ge5$. The numbers of both paired and interacting galaxies identified by NA10 begin to exceed the non-interacting population at a merger fraction above $p_\mathrm{merger}>0.25$. There is a significant population of non-interacting galaxies up to very high GZ2 vote fractions, however, which means that a simple cutoff is insufficient to produce a pure merger population by this criterion. 

We visually examined galaxies that have high GZ2 merger fractions ($p_\mathrm{merger}>0.5$) but are classified by NA10 as non-interacting. The majority of these galaxies show obvious nearby companions, many of which appear to be tidally stripped or otherwise deformed. Some of these galaxies are likely the result of projection effects and are not truly interacting pairs -- however, a significant fraction may be true interactions not identified in NA10. The contrary case (galaxies identified as interacting by NA10, but $p_\mathrm{merger}<0.1$ in GZ2), generally show faint extended features -- mostly shells and tidal tails -- that are clear signs of interacting. Most of these galaxies have no apparent companion visible in the image, however.

%
%

\begin{figure}
\includegraphics[angle=0,width=3.5in]{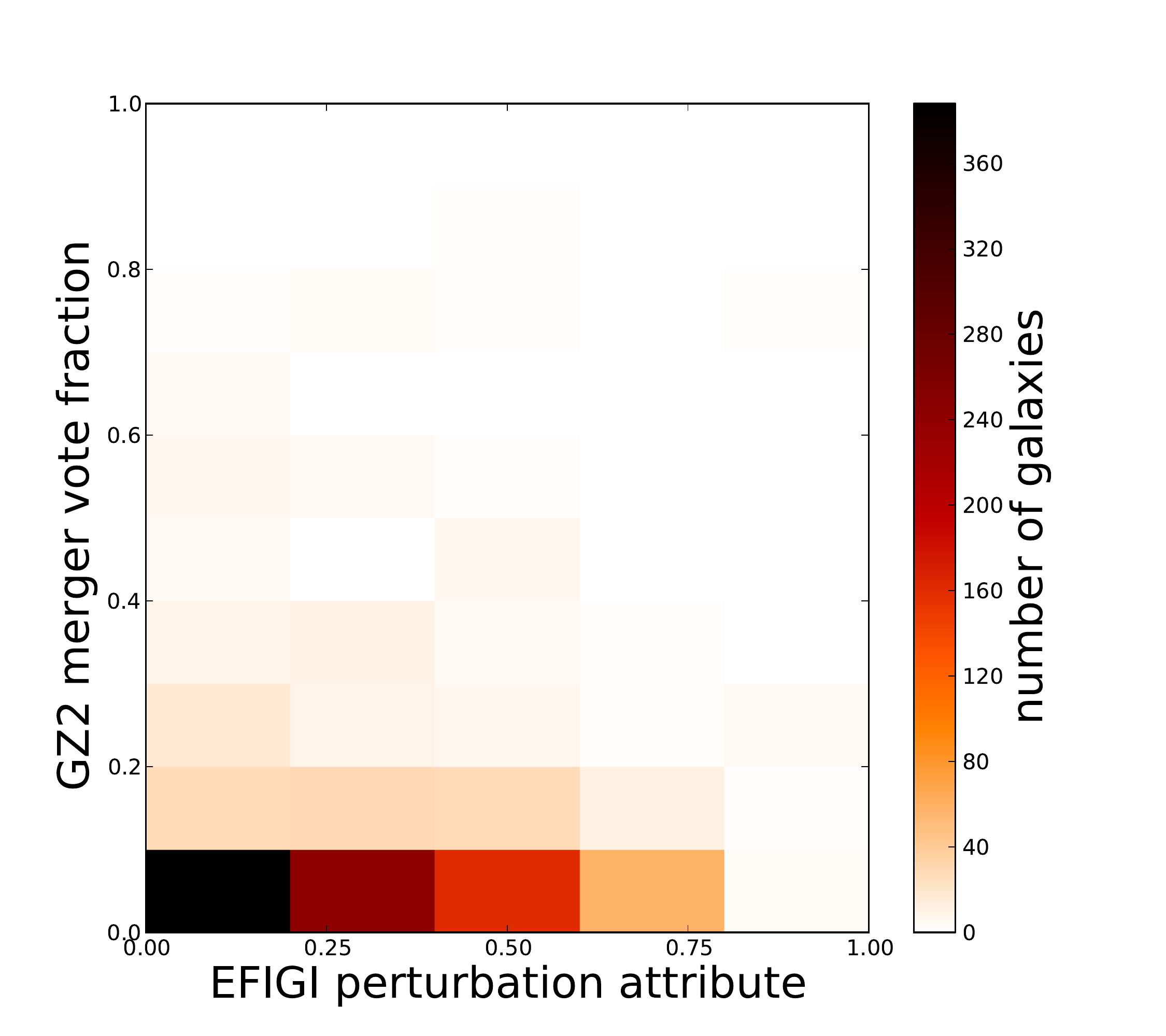}
\caption{EFIGI perturbation classifications compared to GZ2 merger vote fractions. Data are for the 1,080 galaxies in both EFIGI and GZ2 with at least 10 responses to Task 08 (odd feature). 
\label{fig-efigi_mergers}}
\end{figure}

EFIGI has no dedicated category for mergers; galaxies are classified on whether they have any close companions (``contamination'') or distortions in the galaxy profile (``perturbation''), which may or may not be merger-related. Galaxies cleanly classified by GZ2 as mergers are only weakly correlated with both attributes; the mean EFIGI value in GZ2 mergers is 0.31 for the perturbation attribute and 0.48 for contamination. Figure~\ref{fig-efigi_mergers} shows only a very weak correlation ($\rho=0.14$) between EFIGI perturbation and GZ2 merger vote fraction. Highly-perturbed galaxies with low GZ2 $p_\mathrm{merger}$ are mostly dwarf peculiar and irregular galaxies with no sign of tidal features or an interacting companion. 

Results from both expert catalogues are consistent with \citet{cas13}, who found that the mean vote fraction for mergers increases with decreasing projected separations ($r_p$), but then drops off significantly for the closest pairs at $r_p < 10$~kpc. At these separations, the GZ2 votes for Task~08 go instead to the ``irregular'' and ``disturbed'' responses. 

\subsubsection{T-types}

One of the primary challenges for morphological classification in GZ2 is matching the classification tree to T-types, which are not a category in the decision tree. The classifications from expert catalogues are thus extremely valuable as a calibration sample. 

\begin{figure}
\includegraphics[angle=0,width=3.5in]{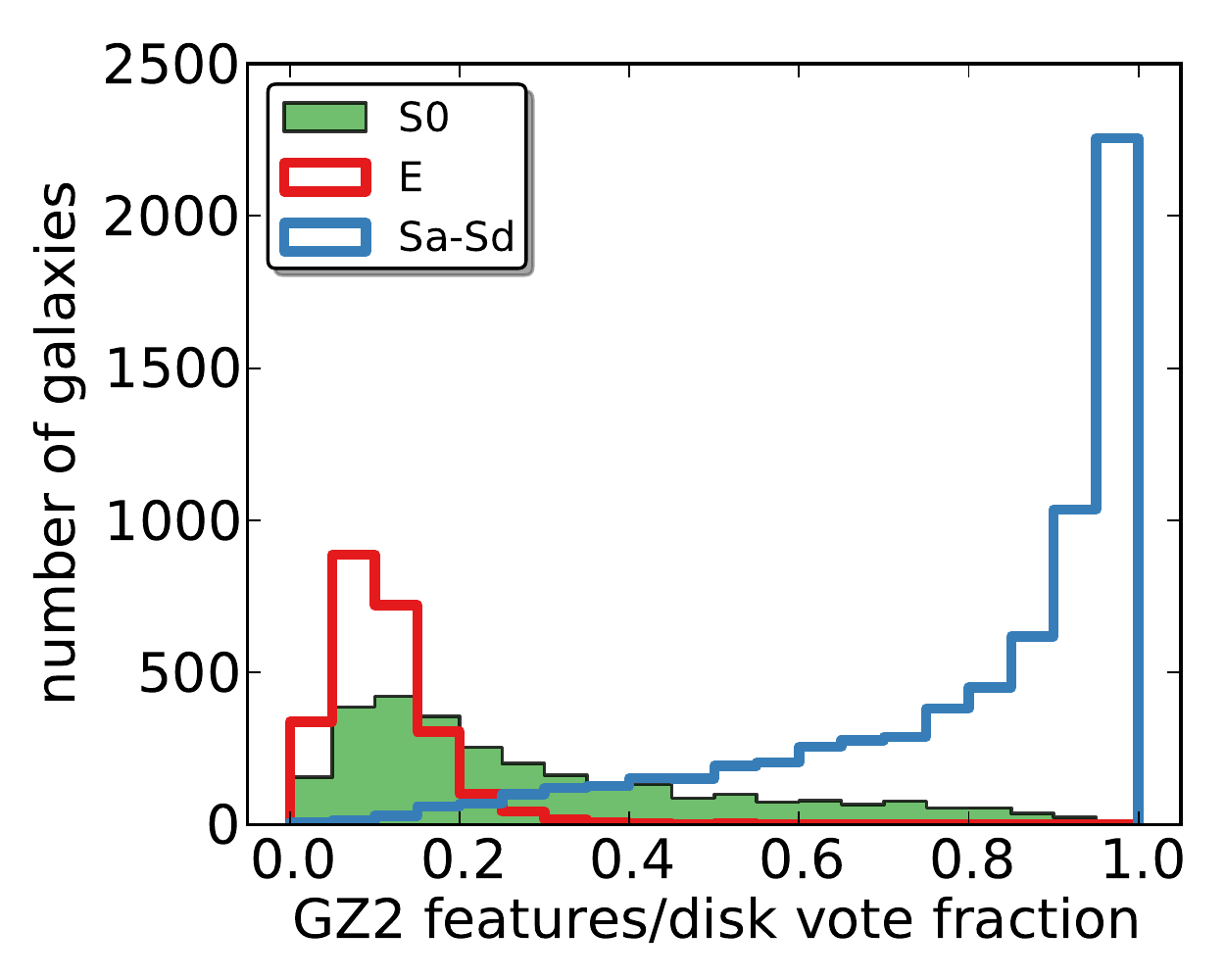}
\caption{T-type classifications for NA10 and GZ2. Data are for the 12,480 galaxies found in both samples. The distribution of GZ2 vote fractions is separated by their T-type classification from NA10. Both elliptical and late-type spirals are strongly correlated with their GZ2 vote fraction. S0 galaxies are more commonly classified as ellipticals, but have a significant tail of high GZ2 features/disk vote fractions. 
\label{fig-na_ttype}}
\end{figure}

Figure~\ref{fig-na_ttype} shows the percentage of galaxies identified as having either a disk or features from the first question in the GZ2 tree, colour-coded by their NA10 T-types. There is a clear separation in the GZ2 fractions for galaxies classified as E versus Sa--Sd. Disk galaxies, including S0, have a median fraction for the GZ2 ``features or disk'' question of 0.80, with a standard deviation of 0.29. Disks with few GZ2 votes for ``feature'' are found to be primarily lenticular (S0) galaxies. If only galaxies with T-types Sa or later are considered, the peak at lower GZ2 vote fractions disappears. The median GZ2 vote fraction for these galaxies is 0.88, with a standard deviation of 0.23. The highest GZ2 vote fraction for an elliptical galaxy in NA10 is 0.741; therefore, any cut above this includes galaxies {\it exclusively} identified by NA10 as late-type. 


Since few objects are identified as stars or artifacts in GZ2 Task~01, the vote fraction for smooth galaxies is approximately $p_\mathrm{smooth} = (1 - p_\mathrm{features/disk})$. Elliptical galaxies have a median vote fraction for the GZ2 ``smooth'' question of $0.86\pm0.07$. The GZ2 votes for the NA10 ellipticals are more sharply peaked than NA10 late-types, lacking the long tail seen even for the very late types. A cut on GZ2 votes for smooth galaxies at 0.8, for example, includes only 4\% late-type galaxies (20\% if S0 galaxies are defined as ``late-type''). 

For galaxies identified as oblique disks, GZ2 users vote if the galaxy has visible spiral structure (Task~04). For the few NA10 elliptical galaxies that have votes for this question, 85\% have GZ2 vote fractions of zero, with the remainder weakly clustered around $p_\mathrm{spiral}\sim0.3$. For NA10 late-type galaxies, the majority of disk/feature objects have high GZ2 spiral structure vote fractions. For galaxies with at least 10 votes on Task~04, 70\% of Sa or later-types have $p_\mathrm{spiral}>0.8$ from GZ2. This drops to 60\% if S0 galaxies are included as late-type. The missing population is thus made up of galaxies that NA10 classify as having significant spiral structure, but for which GZ2 does not distinguish the arms. One might expect these galaxies to have lower magnitudes or surface brightnesses compared to the rest of the sample, thus lowering the confidence of GZ2 votes (there is no analog parameter associated with NA10 classifications). However, the apparent $g$ and $r$ magnitudes, as well as the absolute $g$-band magnitude, show no difference between galaxies above and below the 80\% cutoff. Changing the value for the GZ2 vote fraction does not affect the results, so it appears that lower GZ2 vote fractions for spirals indicate intrinsically weaker (or less clearly-defined) spiral arms.

\begin{figure*}
\includegraphics[angle=0,width=7.0in]{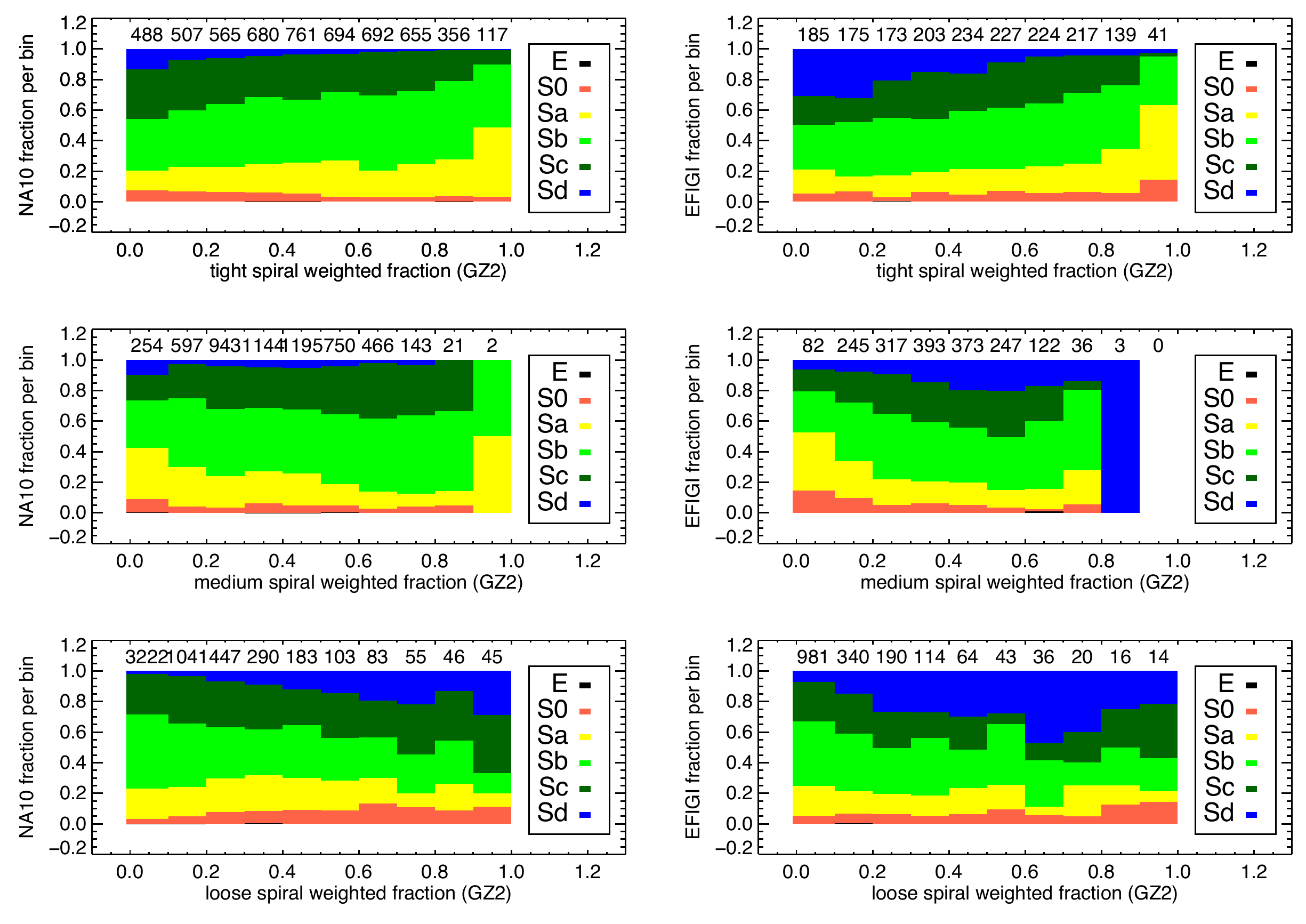}
\caption{T-type classifications compared to the GZ2 vote fractions for spiral tightness (Task~10). Left side is separated by NA10 T-types; right side is EFIGI T-types. Data are for the 5,515 (NA10) and 1,907 (EFIGI) galaxies, respectively, with at least 10 GZ2 votes for Task~10. The number of galaxies per vote fraction bin is given along the top of each panel. 
\label{fig-spiraltightness}}
\end{figure*}

For disk galaxies with spiral structure, Task~10 in GZ2 asked users to classify the ``tightness'' of the arms. This had three options: tight, medium, or loose, accompanied with icons illustrating example pitch angles (Figure~\ref{fig-flowchart}). This allows investigation of the parameters which contribute to the Hubble classification of late-type galaxies which depends on both spiral arm and bulge morphology; tight spirals are presumed to be Sa/Sb, medium spirals Sb/Sc, and loose spirals Sc/Sd. 

The left side of Figure~\ref{fig-spiraltightness} shows the distribution of NA10 T-types for galaxies based on their GZ2 vote fractions for winding arms. Vote fractions for both tight and medium winding arms are relatively normally distributed, with the mean $p_\mathrm{tight}=0.46$ and $p_\mathrm{medium}=0.37$. Strongly-classified loose spirals are much rarer, with 75\% of galaxies having $p_\mathrm{loose}<0.2$. Almost no elliptical galaxies from the NA10 catalogue are included in the oblique disk sample, although there are significant numbers of S0 galaxies. 

For tight spirals, the category of galaxies with the highest vote fractions has more earlier-type spirals than galaxies with a low vote for tight spiral winding arms. For a tight spiral vote fraction above 0.9, 85\% of galaxies are Sb or earlier. Medium-wound spirals with high vote fractions tend to be Sb and Sc -- the proportion of both types increases as a function of $p_\mathrm{medium}$, and constitute 84\% of galaxies when $p_\mathrm{medium}>0.6$. Galaxies classified as strongly medium-wound are rare, however, with only 23 galaxies having $p_\mathrm{medium}>0.8$.  Loose spirals are dominated by Sc and Sd galaxies at high vote fractions, comprising more than 50\% of galaxies with $p_\mathrm{loose}>0.7$. \citet{cas13} found that galaxies with high $p_\mathrm{loose}$ often show tidal features and host a significant proportion of interacting galaxies. This distribution may reflect the experimental design of GZ2, with volunteers preferring extreme ends of a distribution rather than an indistinct `central' option. 

There are less than 30 galaxies classified by GZ2 as smooth and as Sa or later-type by NA10. Individual inspection reveals that these galaxies show no evidence of a disk, and so their NA10 classification is purely bulge-related. There also exist $\sim700$ galaxies classified by GZ2 as smooth but as S0 or S0/a by NA10; these are mostly smooth, face-on galaxies with prominent bulges. 

EFIGI T-types (Figure~\ref{fig-spiraltightness}) show similar trends with respect to GZ2 spiral arm classifications. Late-type spirals (Sc--Sd) constitute about half of disk galaxies with $p_\mathrm{loose}>0.5$, with early-type spirals (Sa--Sb) occupying a similar distribution at $p_\mathrm{tight}>0.5$. S0 galaxies show nearly a flat distribution of GZ2 spiral tightness vote fractions; this is unsurprising, since by definition there is no pitch angle without the presence of spiral arms.

Overall, a clear trend is demonstrated for looser GZ2 spiral arms to correspond with later spiral T-types from expert classifications. High vote fractions are mostly Sa/Sb galaxies for tight winding, Sb/Sc galaxies for medium winding, and Sc/Sd galaxies for loose winding. Individual galaxies, however, can show significant scatter in their GZ2 vote fractions and do not always separate the morphologies on the level of the Hubble T-types. Classifications of spiral galaxies into subcategories (Sa, Sb and Sc) by experts have been shown to be dominated by bulge classification, and to pay little attention to the arm pitch angle, despite the original definition of the late-type categories. 

\begin{figure*}
\includegraphics[angle=0,width=7.0in]{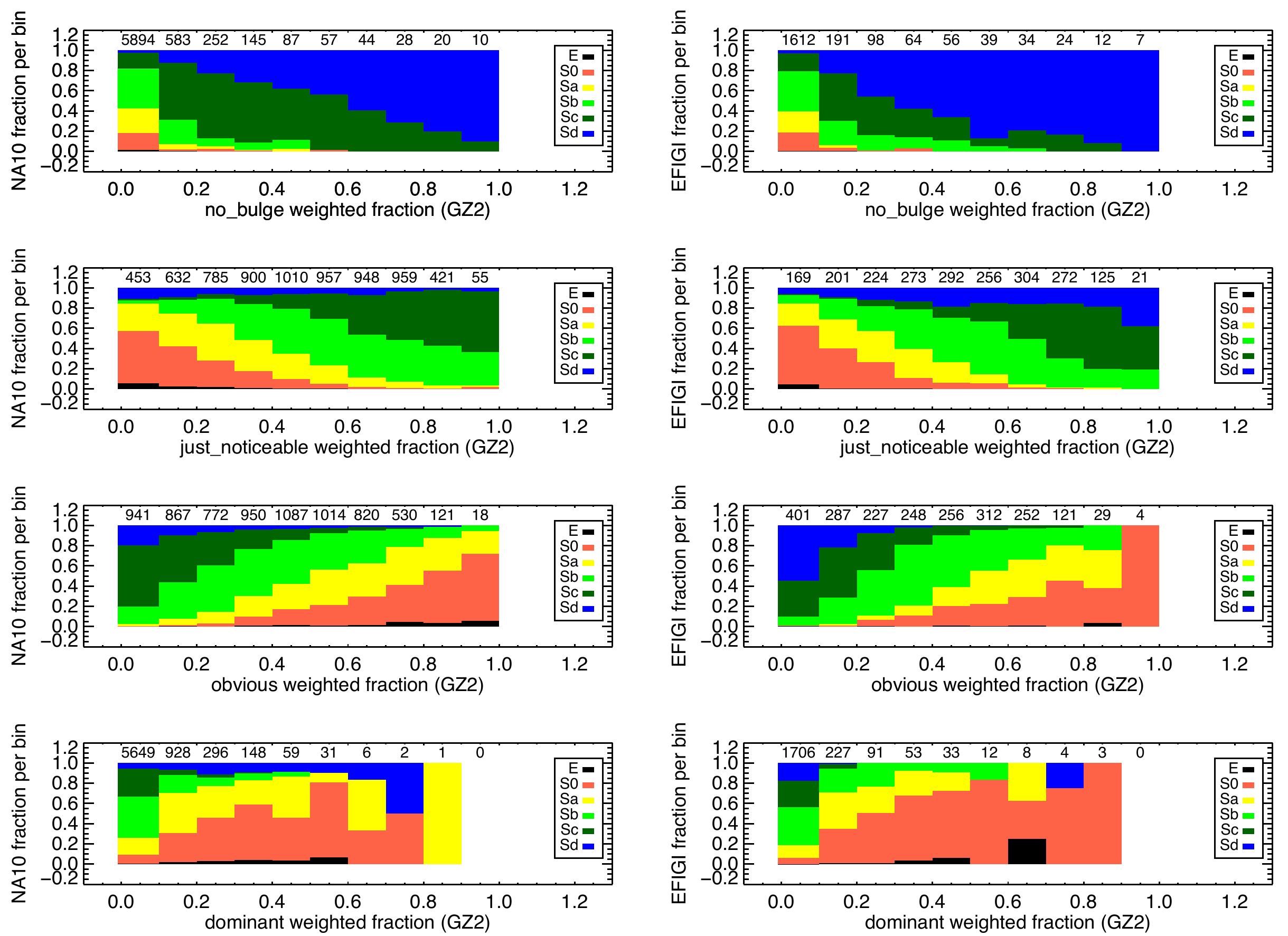}
\caption{T-type classifications compared to the GZ2 vote fractions for bulge prominence (Task~05). Left side is separated by NA10 T-types; right side is EFIGI T-types. Data are for the 7,120 (NA10) and 2,321 (EFIGI) galaxies, respectively, with at least 10 GZ2 votes for Task~05. The number of galaxies per vote fraction bin is given along the top of each panel. 
\label{fig-bulgeprominence}}
\end{figure*}

Having considered the effect of spiral arm tightness, we examine the relationship between bulge morphology and T-type. Disk galaxies in GZ2 are also classified by the visible level of bulge dominance (Task~05), irrespective of whether spiral structure is also identified. This task has four options: ``no bulge'', ``just noticeable'', ``obvious'', and ``dominant'' (Figure~\ref{fig-flowchart}). 

The left side of Figure~\ref{fig-bulgeprominence} shows the distribution of NA10 T-types for galaxies based on their GZ2 vote fractions for bulge prominence, including only galaxies with at least 10 votes for Task~05. Vote fractions for both the ``no bulge'' and ``dominant'' responses peak strongly near zero and tail off as the vote fraction increases. The responses to the middle options (``just noticeable'' and ``obvious'') are both symmetrically distributed around a peak near 0.5. 

``No bulge'' galaxies in GZ2 are dominated by Sc and Sd spirals. For vote fractions above 0.1, 81\% of galaxies are Sc or later; this rises to 100\% for vote fractions higher than 0.6. ``Just noticeable'' galaxies show a smooth change in T-type distribution; galaxies with low $p_\mathrm{just~noticeable}$ are mostly S0 and Sa, while high vote fractions are Sb--Sd. ``Obvious'' bulge galaxies are almost a mirror image of the ``just noticeable'' data; low vote fractions are Sb--Sd galaxies, and high vote fractions are S0--Sa galaxies. Inspection of the few Sa galaxies with $p_\mathrm{obvious}<0.2$ reveals that these are universally very tightly-wound spirals with point-source like bulges. Among galaxies classified as ``dominant'', less than 10 galaxies have vote fractions above 0.6 (which are a diverse mix of S0, Sa, and Sd). Most remaining galaxies have dominant vote fractions of less than 0.1; the T-types of the remaining galaxies between 0.1 and 0.6 mostly contain S0 and Sa spirals. There are also no Sc galaxies with a dominant bulge marked in GZ2. 

The link to T-type is more sharply defined for GZ2 bulge prominence than for spiral tightness, according to expert classifications. Very clean samples of late-type (Sb--Sd) spirals can be selected using only the ``no bulge'' parameter; additional samples with $\sim10$\% contamination can be selected with the ``just noticeable'' and ``obvious'' distributions. Elliptical galaxies that have bulge prominence classified in GZ2 are most often ``dominant'', but there is no obvious separation of ellipticals from disk galaxies based on this task alone. 

EFIGI T-types also correlate strongly with GZ2 bulge dominance. More than 90\% of galaxies with $p_\mathrm{no bulge}>0.5$ are late-type spirals, with the bulk of these Sd galaxies. Both $p_\mathrm{just noticeable}$ and $p_\mathrm{obvious}$ show a continuum of T-types as the vote fractions increase, with Sc and Sd galaxies having high vote fractions for the former and S0, Sa, and Sb galaxies in the latter. Galaxies with high vote fractions for $p_\mathrm{dominant}$ are primarily S0s, along with a few elliptical galaxies that had enough votes as disk galaxies in GZ2 to answer the bulge classification question. 


Since Hubble types are based on both the relative size of the bulge and the extent to which arms are unwound \citep{hub36}, we explored whether the combination of Tasks~05 and 10 from GZ2 can be mapped directly to T-types. The numerical T-types from NA10 were fit with a linear combination of the GZ2 vote fractions for the bulge dominance and arms winding tasks. The best-fit result using symbolic regression \citep{sch09c}, however, depends {\em only} on parameters relating to bulge dominance:

\begin{eqnarray}
\label{eqn-eureqa}
\text{T-type} = 4.63 + 4.17\times p_\mathrm{no bulge} - 2.27\times p_\mathrm{obvious} \\ \nonumber
- 8.38\times p_\mathrm{dominant}.
\end{eqnarray}

\noindent Note that the $p_\mathrm{just noticeable}$ is implicitly included in this equation since the vote fractions for Task~05 must sum to 1. Inclusion of any vote fractions for arms winding responses made no significant difference in the $r^2$ goodness-of-fit metric.

\begin{figure}
\includegraphics[angle=0,width=3.3in]{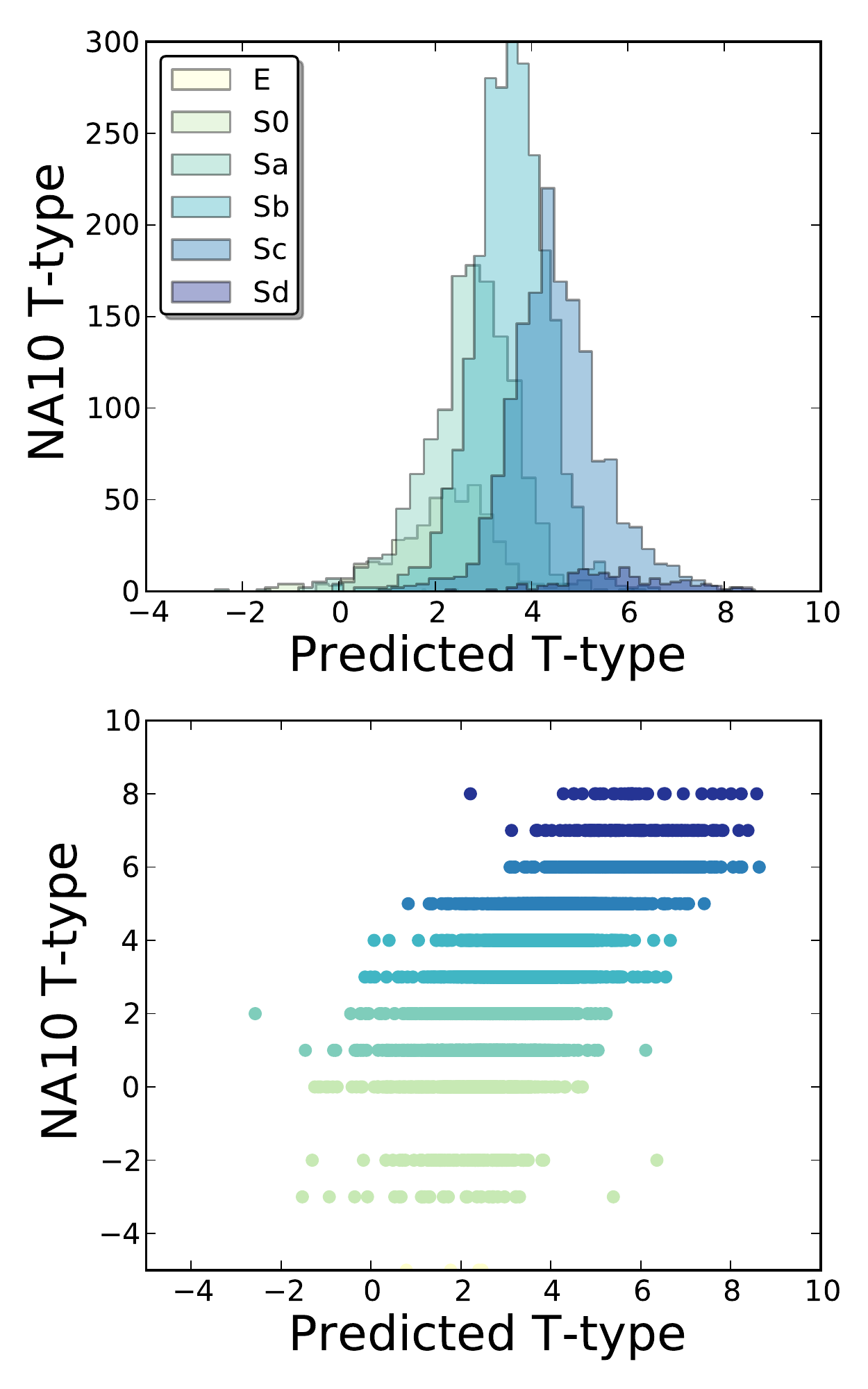}
\caption{Predicted T-type classifications as fit by symbolic regression to the GZ2 data. Galaxies are colour-coded by their morphologies as identified by NA10. The top panel shows the histogram of predicted T-type based on Equation~\ref{eqn-eureqa}. The bottom shows the predicted T-types plotted against their NA10 values. Galaxies shown are only those with sufficient answers to characterize the arms winding and arms number GZ2 tasks, which selects heavily for late-type galaxies. This explains the lack of ellipticals in the plot, but highlights the fact that S0 galaxies do not agree well with the linear sequence. 
\label{fig-eureqa}}
\end{figure}

This technique assumes that the difference in morphology is well-defined by mapping T-types to a linear scale, which is far from being justified. Figure~\ref{fig-eureqa} shows the distribution of the GZ2-derived T-type from Equation~\ref{eqn-eureqa} compared to the NA10 values. The large amounts of overlap between adjoining T-types show that this clearly does not serve as a clean discriminator. One could make a cut between the earliest (Sa) and latest (Sd) spiral types based only on the vote fractions. Alternatively, the relative numbers of galaxies could be used as the weights to construct the {\em probability} of a given T-type. This has yet to be conclusively tested. 

The distributions in Figure~\ref{fig-eureqa} also show that S0~galaxies in particular would typically be mistakenly judged as later types (overlapping strongly with Sa) on average using only this metric. This is consistent with the ``parallel-sequence'' model of \citet{van76} and later revised by several groups \citep[including][]{cap11,lau11,kor12}. 

Finally, we note that \citet{sim13} identified a significant effect in which nuclear point sources, such as AGN, can mimic bulges in the GZ2 classifications. This has not been accounted for in this analysis, but could potentially be addressed by separating the sample into AGN and quiescent galaxies (via BPT line ratios) and looking for systematic differences between the two samples. 

%
%
%
%

%

\subsubsection{Bulge prominence}

\begin{figure}
\includegraphics[angle=0,width=3.5in]{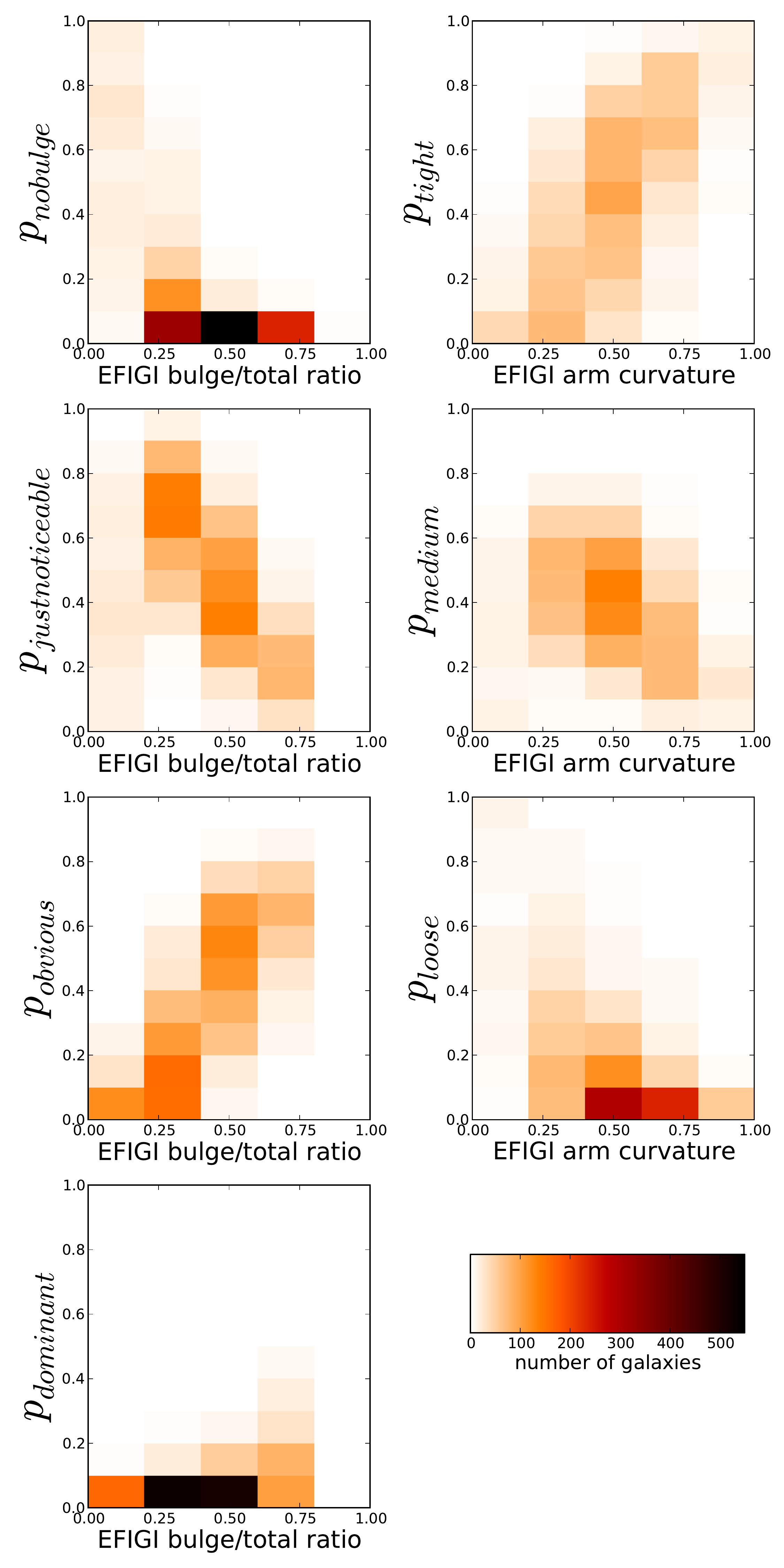}
\caption{Left: EFIGI bulge/total ratio attributes compared to GZ2 vote fractions for ``bulge prominence''. Right: EFIGI arm curvature attributes compared to GZ2 vote fractions for ``arms winding''. Data are for the 1,544 oblique disk galaxies in both samples.  
\label{fig-efigi_bulgearms}}
\end{figure}

EFIGI measures the bulge/total light ratio $(B/T)$ in each galaxy, with the attribute strength corresponding to the relative contribution of the bulge. Elliptical galaxies have $B/T=1$ and irregular galaxies $B/T=0$. \citet{bai11} show that \bt~is correlated with arm curvature and anti-correlated with the presence of flocculent structure and hot~spots, consistent with movement along the Hubble sequence.  

Figure~\ref{fig-efigi_bulgearms} (left panels) show the relationship between \bt~and the GZ2 bulge dominance vote fractions for oblique disk galaxies. $p_\mathrm{obvious}$ is strongly correlated ($\rho=0.65$) with \bt, while $p_\mathrm{just~noticeable}$ has a nearly equal and opposite anti-correlation. Very few galaxies in the sample have either $p_\mathrm{no~bulge}>0$ or $p_\mathrm{dominant}>0$, but those that do show corresponding changes in the EFIGI \bt. In particular, the number of galaxies with $B/T=0$ and $p_\mathrm{just~noticeable}>0$ reinforces the results of \citet{sim13}, who showed that GZ2 bulge prominences increase with the presence of central point sources in the image (such as AGN). 

\subsubsection{Arm curvature}

EFIGI also measures the arm curvature of each galaxy, with classifications very similar to the ``tightness of spiral arms'' question (Task~10) in GZ2. If both expert and citizen science classifiers agree, one would expect galaxies with high GZ2 vote fractions for tight spirals to have EFIGI classifications at 0.75--1.0; GZ2 galaxies classified as medium spirals to be centered around 0.5; and loose spirals to have arm curvatures of 0.0--0.25. 

The EFIGI arm curvature classifications broadly follow the trends expected from matching targets with GZ2. $p_\mathrm{tight}$ is the most strongly correlated with the EFIGI arm curvature parameter (Figure~\ref{fig-efigi_bulgearms}, right panels). The Spearman's correlation coefficient for tight spirals is $\rho=0.62$. The medium spiral vote fraction is clustered in the middle of the EFIGI values, where galaxies with the highest GZ2 vote fraction have EFIGI values of 0.25--0.50, with $\rho=-0.26$. Loose spirals shows an anti-correlation ($\rho=-0.54$); very few galaxies have GZ2 vote fractions above 0.5, but those which do have low EFIGI arm curvature values at 0.25 or below. 

\subsection{Automated classifications}

\begin{figure}
\includegraphics[angle=0,width=3.5in]{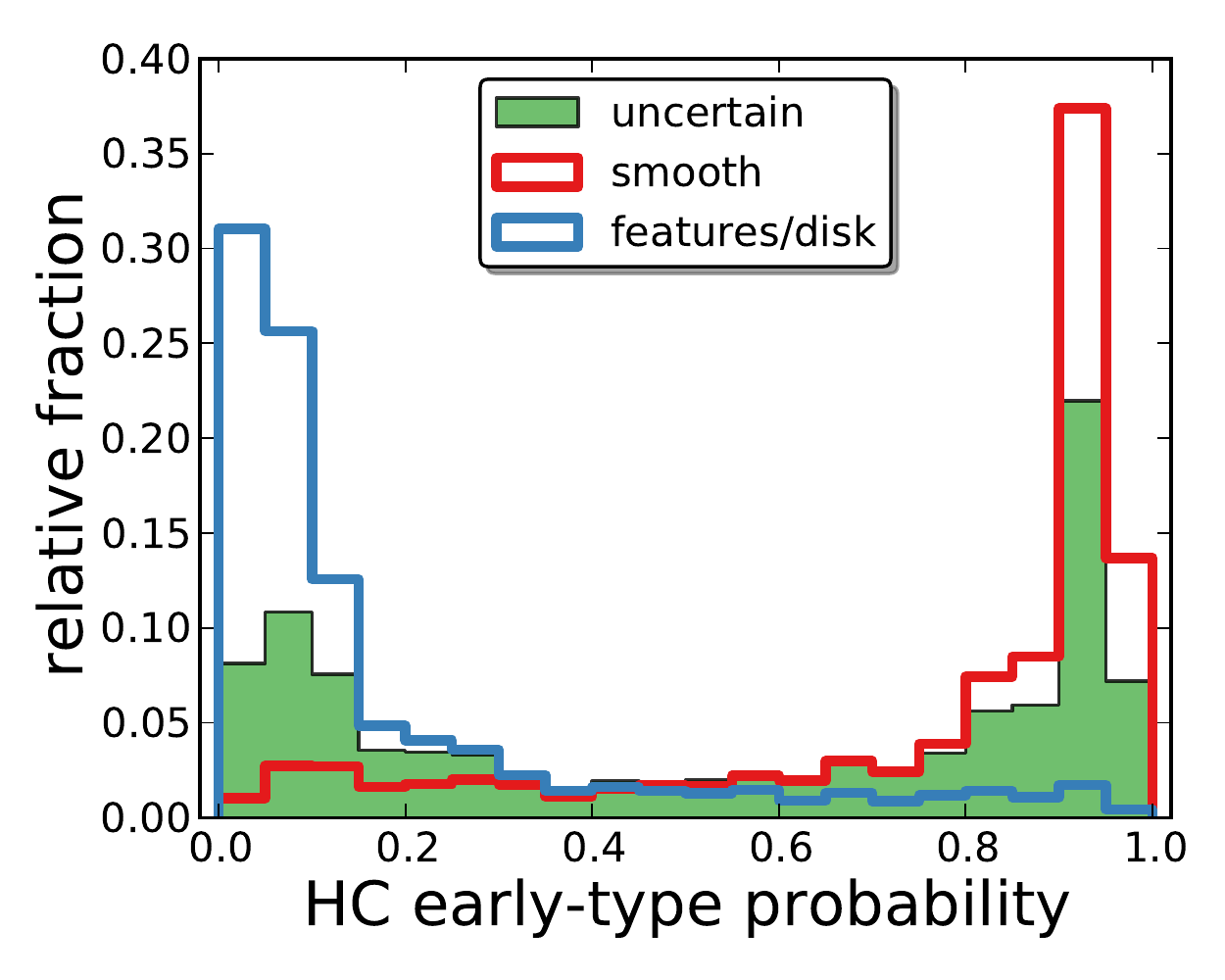}
\caption{Distribution of HC11 early-type probabilities for galaxies split by their GZ2 classification. Data for smooth and features/disk are for galaxies with ``clean'' flags in Table~\ref{tbl-mainclass}; the uncertain classifications comprise galaxies with no flags set for Task~01. 
\label{fig-hc_histogram}}
\end{figure}

\citet[][HC11]{hue11} have generated a large set of morphological classifications for the SDSS spectroscopic sample using an automated Bayesian approach. The broad nature of their probabilities (four broad morphological categories), do not directly relate to the majority of the GZ2 fine structure questions, such as bar or spiral arm structure. Comparison between the two samples, however, is useful to demonstrate the effect that smaller-scale features (as classified by GZ2) may have on automatically-assigned morphologies.

The sample classified by HC11 is limited to galaxies with $z<0.25$ that have both good photometric data and clean spectra. Their total of 698,420 galaxies is approximately twice the size of GZ2. The HC11 sample goes to fainter magnitudes, with more than 400,000 galaxies below the GZ2 limit of $m_r>17$. Their morphological classification algorithm is implemented with support vector machine (SVM) software that tries to find boundaries between regions in $N$-dimensional space, where $N$ is determined by criteria including morphology, luminosity, colour, and redshift \citep{hue08}. The training set is the 2,253 galaxies in \citet{fuk07}, which are already classified by T-type. Each galaxy is assigned a probability of being in one of four subclasses: E, S0, Sab, and Scd (the latter two combining their respective late-type categories). 

We note that the inclusion of colour means that HC11 classifications are not purely morphological, but include information about present-day star formation as well as the dynamical history which determines morphology. Studies of red spiral \citep{mas10a} and blue elliptical galaxies \citep{sch09}, for example, demonstrate the advantages of keeping these criteria separate. 

\citet{hue11} directly compared their results to the GZ1 sample from \citet{lin11}. They found that robust classifications in GZ1 (flagged as either confirmed ellipticals or spirals) have median probabilities of 0.92 according to their algorithm, indicating that sure GZ1 classifications are also sure in their catalogue. They also showed a near-linear relationship between the GZ1 debiased vote fraction and the HC11 probabilities. This is one of the first independent confirmations that the vote fractions may be related to the actual {\em probability} of a galaxy displaying a morphological feature. 


Figure~\ref{fig-hc_histogram} shows the distributions of the HC11 early- and late-type probabilities for GZ2 galaxies robustly identified ($p>0.8$) as either smooth or having features/disks. The median HC11 early-type probability for GZ2 ellipticals is 0.85, and the late-type probability for GZ2 spirals is 0.95. This confirms the result that robust classifications in Galaxy~Zoo agree with the automated algorithm for broad morphological categories. 

An exception to this is a population of galaxies classified as ``smooth'' by GZ2, but which have very low early-type probabilities from HC11 (Figure~\ref{fig-hc_histogram}). The mean GZ2 vote fraction for these galaxies is consistent with those with high early-type probabilities -- these galaxies are not marginally classified as ellipticals in GZ2. The roundness of the galaxy (Task~07 in GZ2) seems to play some role, as the low-HC11 smooth galaxies have fewer round galaxies and many more ``cigar-shaped'' galaxies in this sample. A high axial ratio might train the HC11 algorithm to infer the existence of a disk; the absence of any obvious spiral features or bulge/disk separation (verified by eye in a small subsample of the images) lead GZ2 to categorise these as ``smooth''. There is a clear dependence on apparent magnitude; the lower peak disappears if only galaxies with $r<16$ are included. Early-type galaxies that disagree with the HC11 classification are also significantly bluer, with respective colours of $(g-r)=0.67$ and $(g-r)=0.97$. Since the SVM method does include SDSS colours as a parameter, we conjecture that the low HC11 early-type probability is in part due to the fact that they are blue, in addition to morphological features such as shape and concentration. 

Figure~\ref{fig-hc_histogram} also shows the distribution of ``uncertain'' galaxies, for which none of the responses for Task~01 had a vote fraction $>0.8$. The HC11 probability for these galaxies is bimodal, with the larger fraction classified as HC11 late-type and a smaller fraction as HC11 early-type. 

\begin{figure}
\includegraphics[angle=0,width=3.3in]{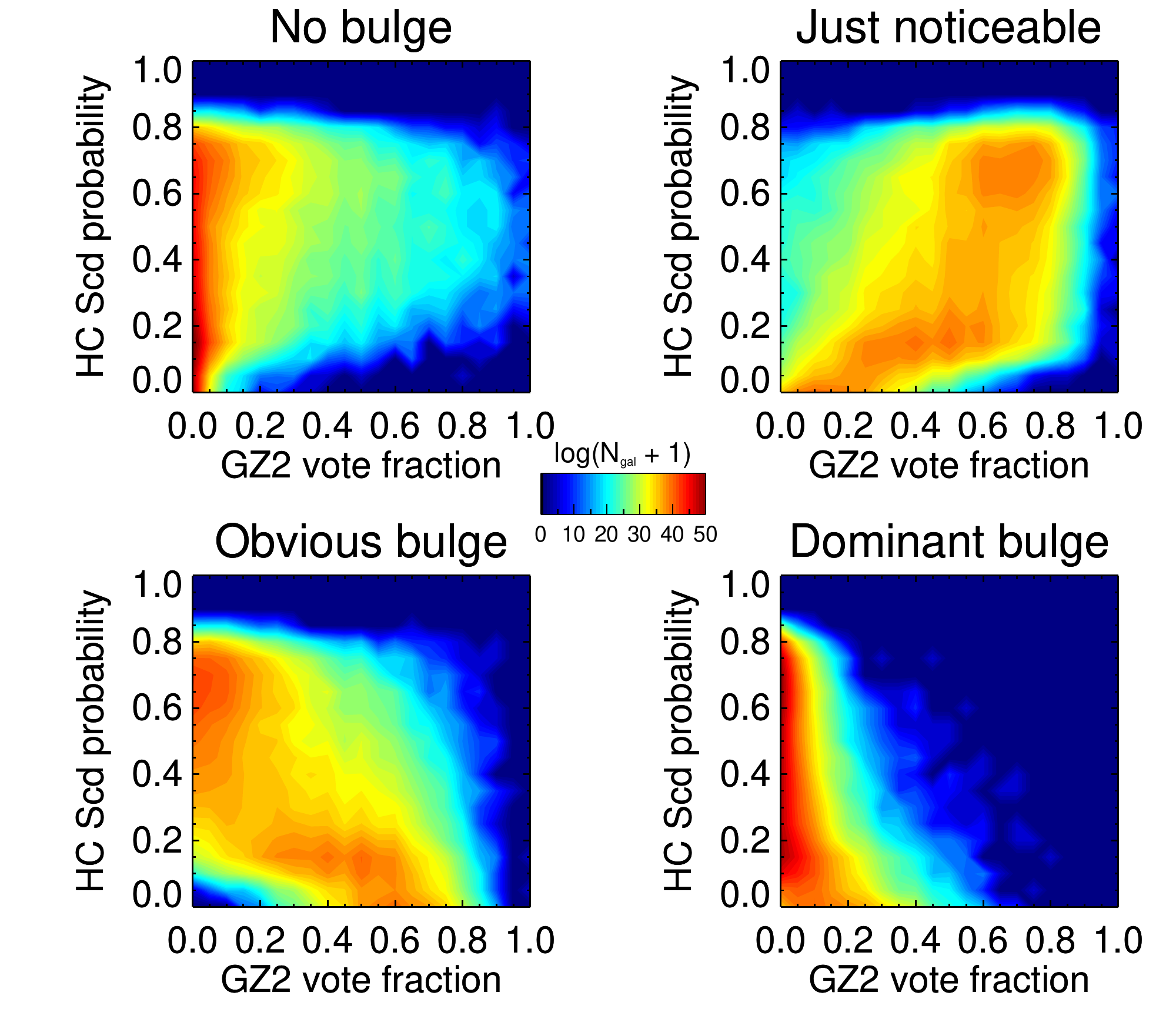}
\caption{\citet{hue11} late-type spiral probability as a function of the GZ2 vote fraction for bulge dominance. The colour of the contours is log~$(N_\mathrm{gal} + 1)$, where $N_\mathrm{gal}$ ranges from 0 to $1.5\times10^3$. Data are the 54,987 oblique disk galaxies appearing in both GZ2 and HC11.
\label{fig-hc_gz2_bulge_contour}}
\end{figure}

Similar to the results from expert visual classifications, morphology in HC11 has a strong dependence on bulge dominance (as measured from GZ2). Figure~\ref{fig-hc_gz2_bulge_contour} shows the HC11 late-type spiral probability for disk galaxies as a function of the GZ2 vote fraction for bulge dominance. Since the majority of galaxies have both low $p_\mathrm{no bulge}$ and $p_\mathrm{dominant}$, the automated probabilities are primarily flat. There is a slight correlation between no bulge and later-type galaxies -- even at $p_\mathrm{no bulge}\simeq0.8$, though, the HC11 algorithm gives galaxies roughly equivalent probabilities between 0.2 and 0.8. 

The relationship between bulge dominance and late-type probability is much stronger for the two intermediate responses for GZ2. Galaxies for which $p_\mathrm{just~noticeable}>0.6$ have a rapid increase in their late-type probabilities, with a sharp transition from the constant late-type probability between 0.25 and 0.6. As expected, the opposite effect occurs for obvious bulges; a vote fraction of $p_\mathrm{obvious}<0.2$ gives a very strong probability of being an Scd galaxy, while galaxies with $p_\mathrm{obvious}>0.5$ are favored to be classified as Sb or earlier. 

\begin{figure}
\includegraphics[angle=0,width=3.3in]{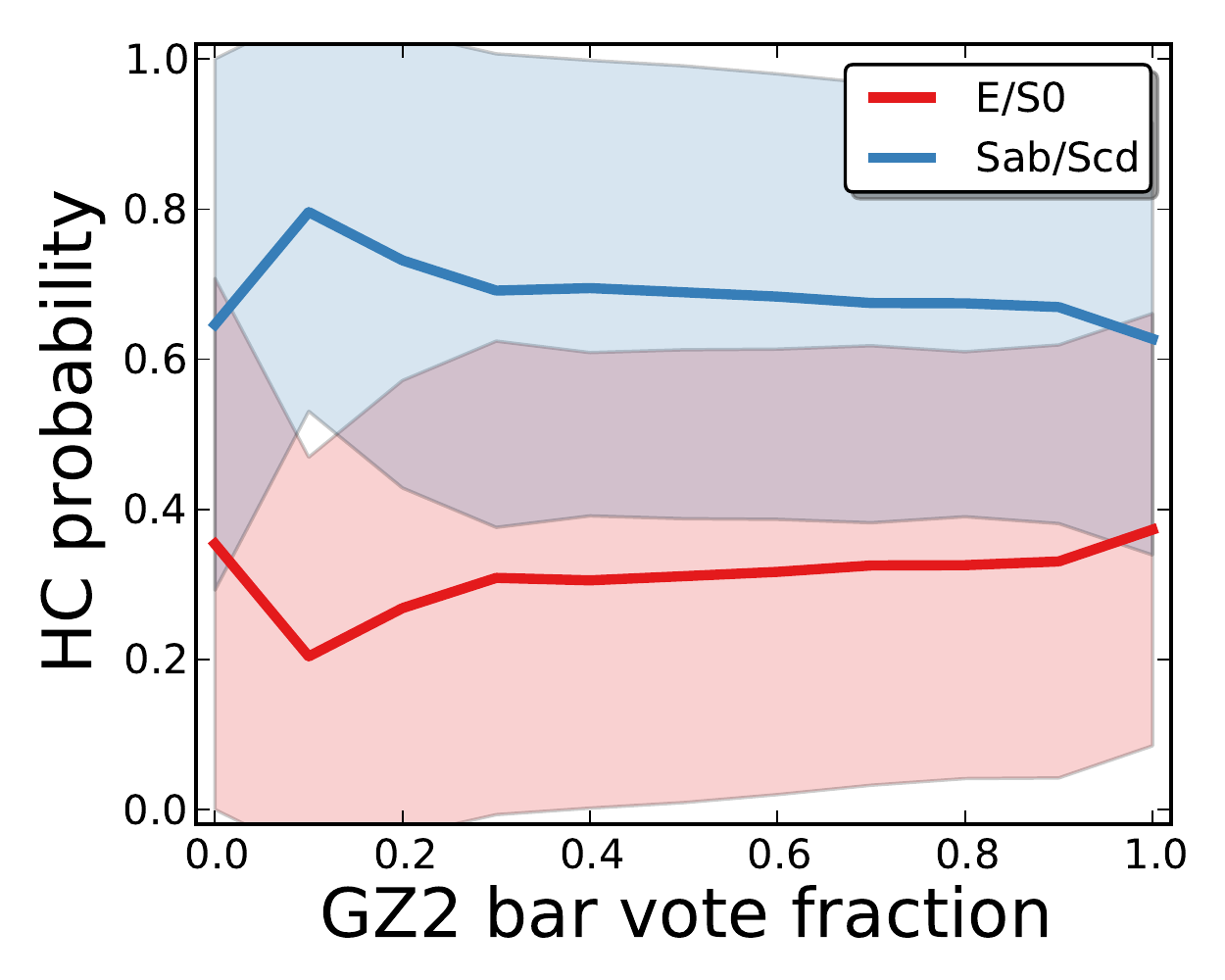}
\caption{HC11 probabilities as a function of GZ2 bar vote fraction for 54,987 oblique disk galaxies. Points give the mean probability in each bin of 0.1 width; shaded areas give the measured $1\sigma$ standard deviation. 
\label{fig-hc_gz2_bar}}
\end{figure}

Finally, we examined the potential effect of bars on the automated classifications. Figure~\ref{fig-hc_gz2_bar} shows the average HC11 probability as a function of GZ2 $p_\mathrm{bar}$ for oblique disk galaxies. The relative proportions of galaxies as classified by HC11 is flat as a function of GZ2 $p_\mathrm{bar}$, with 31\% for early-type and 69\% late-type. The presence of a bar thus does not strongly affect automated classifications, at least as far as distinguishing early- from late-type galaxies. 

%


%


\section{Conclusions}\label{sec-conclusion}

We present the data release for the Galaxy~Zoo~2 (GZ2) project. GZ2 uses crowd-sourced votes from citizen scientist classifiers to characterize morphology of more than 300,000 galaxies from the SDSS DR7. GZ2 classified $gri$ colour composite images selected on the basis of magnitude ($m_r<17$), angular size ($r_{90}>3\arcsec$), and redshift ($0.0005<z<0.25$) criteria. Deeper images from Stripe~82 are also included at both normal and coadded image depths. 

GZ2 expands on the original Galaxy~Zoo results by classifying a large array of fine morphological structures. In addition to previous distinctions between elliptical and spiral galaxies, GZ2 identifies the presence of bars, spiral structure, dust lanes, mergers, disturbed/interacting morphologies, and gravitational lenses. It also quantifies the relative strengths of galactic bulges (both edge-on and face-on), the tightness and multiplicity of spiral arms, and the relative roundness of elliptical galaxies. Classification was done via a multi-step decision tree presented to users in a web-based interface. The final catalogue is the result of nearly 60 million individual classifications of images.

Data reduction for the catalogue begins by weighting individual classifiers. Repeat classifications of objects by the same user are omitted from the catalogue, and then an iterative weighting scheme is applied to users for each task based on their overall consistency. Votes for each galaxy are combined to generate the overall classification; the strength of a particular feature is measured by the fraction of votes for a particular response (among all possible responses). The nature of the GZ2 classification scheme means that these vote fractions are akin to conditional probabilities, however -- for example, a galaxy must first be identified both as possessing a disk {\em and} as being ``not edge-on'' to measure $p_\mathrm{bar}$.

Vote fractions for each response are also adjusted for classification bias, the effect of fine morphological features being more difficult to detect in smaller and fainter galaxies. Corrections to determine the debiased vote fractions are derived directly from the GZ2 data itself. 

The final catalogue consists of five tables, comprising morphological classifications for the GZ2 main sample (separated into galaxies with spectroscopic and photometric redshifts) and galaxies from Stripe~82 (for normal-depth and two sets of coadded images with spectroscopic redshifts). Data for each galaxy includes (for each response) the raw and weighted number of votes, the raw and weighted vote fractions, the debiased vote fraction, and an optional flag which indicates if a feature has been robustly identified. Portions of the data are presented in Tables~\ref{tbl-mainclass}--\ref{tbl-stripe82_coadd2}; full machine-readable tables are available at \url{http://data.galaxyzoo.org} and in SDSS Data Release~10. 

We have compared the GZ2 classifications in detail to several other morphological catalogues. Early and late-type classifications are consistent with results from the original Galaxy~Zoo, especially for galaxies in the clean samples. Expert catalogues \citep{nai10,bai11} show good agreement for galaxies with medium to strong bars; GZ2 is less confident in identifying expert-classified weak and/or nuclear bars. In ringed galaxies, GZ2 recovers the majority of outer rings, but relatively few inner or nuclear rings due to the design of the GZ2 question. Pairs and interacting galaxies are more difficult to reliably cross-match in a clean sample, although \citet{cas13} have already shown that the GZ2 ``loose winding arms'' parameter is a reliable proxy for interaction. The GZ2 bulge dominance parameter strongly correlates with the Hubble T-type from both expert catalogues. Adding GZ2 measurements of the spiral arm tightness, though, does not increase the T-type classification accuracy. Automated classifications from \citet{hue11} agree well with GZ2 in separating elliptical and late-type spirals, although identification of S0 galaxies still represents a challenge. 

GZ2 contains more than an order of magnitude more galaxies than the largest comparable expert-classified catalogues (NA10, EFIGI) while still classifying detailed morphological features not replicable by automated classifications. GZ2 data have already been used to demonstrate a relationship between bar fraction and the colour, gas fractions, and bulge size of disk galaxies \citep{mas11c,mas12a}, as well as studies of the bar colour and length itself \citep{hoy11}. The size of the catalogues has allowed for the discovery and study of comparatively rare objects, such as early-type dust lane galaxies \citep{kav12a} and bulgeless AGN hosts \citep{sim13}. Direct use of the GZ2 likelihoods has also been used to quantify the environmental dependence on morphology, showing a correlation for barred and bulge-dominated galaxies \citep{ski12} and identifying reliable signatures of interaction from GZ2 data \citep{cas13}. 

The scientific productivity of the Galaxy Zoo project has already shown that the use of multiple independent volunteer classifications is a robust method for the analysis of large datasets of galaxy images. This public release of the Galaxy~Zoo~2 catalogue intends to build on this success, by demonstrating the reliability and benefit of these classifications over both expert and automated classifications. We publicly release these Galaxy Zoo 2 classifications both as a rich dataset that can be used to study galaxy evolution, and as training sets for refining future automated classification techniques.


\section*{Acknowledgments}
The data in this paper are the result of the efforts of the Galaxy~Zoo~2 volunteers, without whom none of this work would be possible. Their efforts are individually acknowledged at \url{http://authors.galaxyzoo.org}. 

The development of Galaxy Zoo 2 was supported by The Leverhulme Trust. KWW and LFF would like to acknowledge support from the US National Science Foundation under grant DRL-0941610. CJL acknowledges support from the Science and Technology Facilities Council (STFC) Science in Society program. KS gratefully acknowledges support from Swiss National Science Foundation Grant PP00P2\_138979/1. TM acknowledges funding from the STFC ST/J500665/1. RCN was partially supported by STFC grant ST/K00090X/1. BDS acknowledges support from Worcester College, Oxford, and from the Oxford Martin School program on computational cosmology. Please contact the author(s) to request access to research materials discussed in this paper. 

This research made use of Montage, funded by the National Aeronautics and Space Administration's Earth Science Technology Office, Computation Technologies Project, under Cooperative Agreement Number NCC5-626 between NASA and the California Institute of Technology. Montage is maintained by the NASA/IPAC Infrared Science Archive. It also made extensive use of the Tool for OPerations on Catalogues And Tables (TOPCAT), which can be found at \url{http://www.starlink.ac.uk/topcat/} \citep{tay05,tay11}. 

We thank the referee for useful comments that improved the content and structure of this paper.

Funding for the SDSS and SDSS-II has been provided by the Alfred P. Sloan Foundation, the Participating Institutions, the National Science Foundation, the U.S. Department of Energy, the National Aeronautics and Space Administration, the Japanese Monbukagakusho, the Max Planck Society, and the Higher Education Funding Council for England. The SDSS website is \url{http://www.sdss.org/}.

The SDSS is managed by the Astrophysical Research Consortium for the Participating Institutions. The Participating Institutions are the American Museum of Natural History, Astrophysical Institute Potsdam, University of Basel, University of Cambridge, Case Western Reserve University, University of Chicago, Drexel University, Fermilab, the Institute for Advanced Study, the Japan Participation Group, Johns Hopkins University, the Joint Institute for Nuclear Astrophysics, the Kavli Institute for Particle Astrophysics and Cosmology, the Korean Scientist Group, the Chinese Academy of Sciences (LAMOST), Los Alamos National Laboratory, the Max-Planck-Institute for Astronomy (MPIA), the Max-Planck-Institute for Astrophysics (MPA), New Mexico State University, Ohio State University, University of Pittsburgh, University of Portsmouth, Princeton University, the United States Naval Observatory, and the University of Washington.



\bibliography{kwrefs}

\begin{thebibliography}{59}
\expandafter\ifx\csname natexlab\endcsname\relax\def\natexlab#1{#1}\fi

\bibitem[{{Abazajian} {et~al}\mbox{.}(2009){Abazajian}, {Adelman-McCarthy},
  {Ag{\"u}eros}, {Allam}, {Allende Prieto}, {An}, {Anderson}, {Anderson},
  {Annis}, {Bahcall}, \& et~al.}]{aba09}
{Abazajian} K.~N. {et~al.}, 2009, \apjs, 182, 543

\bibitem[{{Aguerri}, {M{\'e}ndez-Abreu} \& {Corsini}(2009){Aguerri},
  {M{\'e}ndez-Abreu}, \& {Corsini}}]{agu09}
{Aguerri} J.~A.~L., {M{\'e}ndez-Abreu} J., {Corsini} E.~M., 2009, \aap, 495,
  491

\bibitem[{{Ahn} {et~al}\mbox{.}(2013){Ahn}, {Alexandroff}, {Allende Prieto},
  {Anders}, {Anderson}, {Anderton}, {Andrews}, {Aubourg}, {Bailey}, {Bastien},
  \& et~al.}]{ahn13}
{Ahn} C.~P. {et~al.}, 2013, ArXiv e-prints

\bibitem[{{Annis} {et~al}\mbox{.}(2011){Annis}, {Soares-Santos}, {Strauss},
  {Becker}, {Dodelson}, {Fan}, {Gunn}, {Hao}, {Ivezic}, {Jester}, {Jiang},
  {Johnston}, {Kubo}, {Lampeitl}, {Lin}, {Lupton}, {Miknaitis}, {Seo}, {Simet},
  \& {Yanny}}]{ann11}
{Annis} J. {et~al.}, 2011, ArXiv e-prints

\bibitem[{{Baillard} {et~al}\mbox{.}(2011){Baillard}, {Bertin}, {de Lapparent},
  {Fouqu{\'e}}, {Arnouts}, {Mellier}, {Pell{\'o}}, {Leborgne}, {Prugniel},
  {Makarov}, {Makarova}, {McCracken}, {Bijaoui}, \& {Tasca}}]{bai11}
{Baillard} A. {et~al.}, 2011, \aap, 532, A74

\bibitem[{{Bamford} {et~al}\mbox{.}(2009){Bamford}, {Nichol}, {Baldry}, {Land},
  {Lintott}, {Schawinski}, {Slosar}, {Szalay}, {Thomas}, {Torki}, {Andreescu},
  {Edmondson}, {Miller}, {Murray}, {Raddick}, \& {Vandenberg}}]{bam09}
{Bamford} S.~P. {et~al.}, 2009, \mnras, 393, 1324

\bibitem[{{Banerji} {et~al}\mbox{.}(2010){Banerji}, {Lahav}, {Lintott},
  {Abdalla}, {Schawinski}, {Bamford}, {Andreescu}, {Murray}, {Raddick},
  {Slosar}, {Szalay}, {Thomas}, \& {Vandenberg}}]{ban10}
{Banerji} M. {et~al.}, 2010, \mnras, 406, 342

\bibitem[{{Barazza}, {Jogee} \& {Marinova}(2008){Barazza}, {Jogee}, \&
  {Marinova}}]{bar08}
{Barazza} F.~D., {Jogee} S., {Marinova} I., 2008, \apj, 675, 1194

\bibitem[{{Blanton} {et~al}\mbox{.}(2003){Blanton}, {Hogg}, {Bahcall},
  {Brinkmann}, {Britton}, {Connolly}, {Csabai}, {Fukugita}, {Loveday},
  {Meiksin}, {Munn}, {Nichol}, {Okamura}, {Quinn}, {Schneider}, {Shimasaku},
  {Strauss}, {Tegmark}, {Vogeley}, \& {Weinberg}}]{bla03a}
{Blanton} M.~R. {et~al.}, 2003, \apj, 592, 819

\bibitem[{{Buta}(1995)}]{but95}
{Buta} R., 1995, \apjs, 96, 39

\bibitem[{{Buta} \& {Combes}(1996)}]{but96}
{Buta} R., {Combes} F., 1996, \fcp, 17, 95

\bibitem[{{Buta}, {Corwin} \& {Odewahn}(2002){Buta}, {Corwin}, \&
  {Odewahn}}]{but02}
{Buta} R., {Corwin}, Jr. H.~G., {Odewahn} S.~C., 2002, in Astronomical Society
  of the Pacific Conference Series, Vol. 275, Disks of Galaxies: Kinematics,
  Dynamics and Peturbations, {Athanassoula} E., {Bosma} A., {Mujica} R., eds.,
  p. 102

\bibitem[{{Buta}(2013)}]{but13}
{Buta} R.~J., 2013, {Galaxy Morphology}, {Oswalt} T.~D., {Keel} W.~C., eds.,
  Springer

\bibitem[{{Cappellari} {et~al}\mbox{.}(2011){Cappellari}, {Emsellem},
  {Krajnovi{\'c}}, {McDermid}, {Serra}, {Alatalo}, {Blitz}, {Bois}, {Bournaud},
  {Bureau}, {Davies}, {Davis}, {de Zeeuw}, {Khochfar}, {Kuntschner},
  {Lablanche}, {Morganti}, {Naab}, {Oosterloo}, {Sarzi}, {Scott}, {Weijmans},
  \& {Young}}]{cap11}
{Cappellari} M. {et~al.}, 2011, \mnras, 416, 1680

\bibitem[{{Casteels} {et~al}\mbox{.}(2013){Casteels}, {Bamford}, {Skibba},
  {Masters}, {Lintott}, {Keel}, {Schawinski}, {Nichol}, \& {Smith}}]{cas13}
{Casteels} K.~R.~V. {et~al.}, 2013, \mnras, 429, 1051

\bibitem[{{Csabai} {et~al}\mbox{.}(2003){Csabai}, {Budav{\'a}ri}, {Connolly},
  {Szalay}, {Gy{\H o}ry}, {Ben{\'{\i}}tez}, {Annis}, {Brinkmann}, {Eisenstein},
  {Fukugita}, {Gunn}, {Kent}, {Lupton}, {Nichol}, \& {Stoughton}}]{csa03}
{Csabai} I. {et~al.}, 2003, \aj, 125, 580

\bibitem[{{Darg} {et~al}\mbox{.}(2010){Darg}, {Kaviraj}, {Lintott},
  {Schawinski}, {Sarzi}, {Bamford}, {Silk}, {Proctor}, {Andreescu}, {Murray},
  {Nichol}, {Raddick}, {Slosar}, {Szalay}, {Thomas}, \& {Vandenberg}}]{dar10a}
{Darg} D.~W. {et~al.}, 2010, \mnras, 401, 1043

\bibitem[{{Davis} \& {Hayes}(2013)}]{dav13}
{Davis} D., {Hayes} W., 2013, ArXiv e-prints

\bibitem[{{de~Vaucouleurs} {et~al}\mbox{.}(1991){de~Vaucouleurs},
  {de~Vaucouleurs}, {Corwin}, {Buta}, {Paturel}, \& {Fouqu{\'e}}}]{dev91}
{de~Vaucouleurs} G., {de~Vaucouleurs} A., {Corwin}, Jr. H.~G., {Buta} R.~J.,
  {Paturel} G., {Fouqu{\'e}} P., 1991, {Third Reference Catalogue of Bright
  Galaxies}. Springer-Verlag

\bibitem[{{Fischer} {et~al}\mbox{.}(2012){Fischer}, {Schwamb}, {Schawinski},
  {Lintott}, {Brewer}, {Giguere}, {Lynn}, {Parrish}, {Sartori}, {Simpson},
  {Smith}, {Spronck}, {Batalha}, {Rowe}, {Jenkins}, {Bryson}, {Prsa},
  {Tenenbaum}, {Crepp}, {Morton}, {Howard}, {Beleu}, {Kaplan}, {Vannispen},
  {Sharzer}, {Defouw}, {Hajduk}, {Neal}, {Nemec}, {Schuepbach}, \&
  {Zimmermann}}]{fis12}
{Fischer} D.~A. {et~al.}, 2012, \mnras, 419, 2900

\bibitem[{{Fortson} {et~al}\mbox{.}(2012){Fortson}, {Masters}, {Nichol},
  {Borne}, {Edmondson}, {Lintott}, {Raddick}, {Schawinski}, \&
  {Wallin}}]{for12}
{Fortson} L. {et~al.}, 2012, {Galaxy Zoo: Morphological Classification and
  Citizen Science}, {Way} M.~J., {Scargle} J.~D., {Ali} K.~M., {Srivastava}
  A.~N., eds., CRC Press, Taylor \& Francis Group, pp. 213--236

\bibitem[{{Fukugita} {et~al}\mbox{.}(2007){Fukugita}, {Nakamura}, {Okamura},
  {Yasuda}, {Barentine}, {Brinkmann}, {Gunn}, {Harvanek}, {Ichikawa}, {Lupton},
  {Schneider}, {Strauss}, \& {York}}]{fuk07}
{Fukugita} M. {et~al.}, 2007, \aj, 134, 579

\bibitem[{{Hinshaw} {et~al}\mbox{.}(2012){Hinshaw}, {Larson}, {Komatsu},
  {Spergel}, {Bennett}, {Dunkley}, {Nolta}, {Halpern}, {Hill}, {Odegard},
  {Page}, {Smith}, {Weiland}, {Gold}, {Jarosik}, {Kogut}, {Limon}, {Meyer},
  {Tucker}, {Wollack}, \& {Wright}}]{hin12}
{Hinshaw} G. {et~al.}, 2012, ArXiv e-prints

\bibitem[{{Hoyle} {et~al}\mbox{.}(2011){Hoyle}, {Masters}, {Nichol},
  {Edmondson}, {Smith}, {Lintott}, {Scranton}, {Bamford}, {Schawinski}, \&
  {Thomas}}]{hoy11}
{Hoyle} B. {et~al.}, 2011, \mnras, 415, 3627

\bibitem[{{Hubble}(1926)}]{hub26}
{Hubble} E.~P., 1926, \apj, 64, 321

\bibitem[{{Hubble}(1936)}]{hub36}
{Hubble} E.~P., 1936, {Realm of the Nebulae}. Yale University Press

\bibitem[{{Huertas-Company} {et~al}\mbox{.}(2011){Huertas-Company}, {Aguerri},
  {Bernardi}, {Mei}, \& {S{\'a}nchez Almeida}}]{hue11}
{Huertas-Company} M., {Aguerri} J.~A.~L., {Bernardi} M., {Mei} S., {S{\'a}nchez
  Almeida} J., 2011, \aap, 525, A157+

\bibitem[{{Huertas-Company} {et~al}\mbox{.}(2008){Huertas-Company}, {Rouan},
  {Tasca}, {Soucail}, \& {Le F{\`e}vre}}]{hue08}
{Huertas-Company} M., {Rouan} D., {Tasca} L., {Soucail} G., {Le F{\`e}vre} O.,
  2008, \aap, 478, 971

\bibitem[{{Jacob} {et~al}\mbox{.}(2010){Jacob}, {Katz}, {Berriman}, {Good},
  {Laity}, {Deelman}, {Kesselman}, {Singh}, {Su}, {Prince}, \&
  {Williams}}]{jac10}
{Jacob} J.~C. {et~al.}, 2010, {Montage: An Astronomical Image Mosaicking
  Toolkit}. Astrophysics Source Code Library

\bibitem[{{Kaviraj} {et~al}\mbox{.}(2012){Kaviraj}, {Ting}, {Bureau},
  {Shabala}, {Crockett}, {Silk}, {Lintott}, {Smith}, {Keel}, {Masters},
  {Schawinski}, \& {Bamford}}]{kav12a}
{Kaviraj} S. {et~al.}, 2012, \mnras, 423, 49

\bibitem[{{Kendrew} {et~al}\mbox{.}(2012){Kendrew}, {Simpson}, {Bressert},
  {Povich}, {Sherman}, {Lintott}, {Robitaille}, {Schawinski}, \&
  {Wolf-Chase}}]{ken12}
{Kendrew} S. {et~al.}, 2012, \apj, 755, 71

\bibitem[{{Kormendy} \& {Bender}(2012)}]{kor12}
{Kormendy} J., {Bender} R., 2012, \apjs, 198, 2

\bibitem[{{Kormendy} \& {Kennicutt}(2004)}]{kor04}
{Kormendy} J., {Kennicutt}, Jr. R.~C., 2004, \araa, 42, 603

\bibitem[{{Laurikainen} {et~al}\mbox{.}(2011){Laurikainen}, {Salo}, {Buta}, \&
  {Knapen}}]{lau11}
{Laurikainen} E., {Salo} H., {Buta} R., {Knapen} J.~H., 2011, \mnras, 418, 1452

\bibitem[{{Lintott} {et~al}\mbox{.}(2011){Lintott}, {Schawinski}, {Bamford},
  {Slosar}, {Land}, {Thomas}, {Edmondson}, {Masters}, {Nichol}, {Raddick},
  {Szalay}, {Andreescu}, {Murray}, \& {Vandenberg}}]{lin11}
{Lintott} C. {et~al.}, 2011, \mnras, 410, 166

\bibitem[{{Lintott} {et~al}\mbox{.}(2008){Lintott}, {Schawinski}, {Slosar},
  {Land}, {Bamford}, {Thomas}, {Raddick}, {Nichol}, {Szalay}, {Andreescu},
  {Murray}, \& {Vandenberg}}]{lin08}
{Lintott} C.~J. {et~al.}, 2008, \mnras, 389, 1179

\bibitem[{{Lupton} {et~al}\mbox{.}(2004){Lupton}, {Blanton}, {Fekete}, {Hogg},
  {O'Mullane}, {Szalay}, \& {Wherry}}]{lup04}
{Lupton} R., {Blanton} M.~R., {Fekete} G., {Hogg} D.~W., {O'Mullane} W.,
  {Szalay} A., {Wherry} N., 2004, \pasp, 116, 133

\bibitem[{{Martig} {et~al}\mbox{.}(2012){Martig}, {Bournaud}, {Croton},
  {Dekel}, \& {Teyssier}}]{mar12}
{Martig} M., {Bournaud} F., {Croton} D.~J., {Dekel} A., {Teyssier} R., 2012,
  \apj, 756, 26

\bibitem[{{Masters} {et~al}\mbox{.}(2010){Masters}, {Mosleh}, {Romer},
  {Nichol}, {Bamford}, {Schawinski}, {Lintott}, {Andreescu}, {Campbell},
  {Crowcroft}, {Doyle}, {Edmondson}, {Murray}, {Raddick}, {Slosar}, {Szalay},
  \& {Vandenberg}}]{mas10a}
{Masters} K.~L. {et~al.}, 2010, \mnras, 405, 783

\bibitem[{{Masters} {et~al}\mbox{.}(2012){Masters}, {Nichol}, {Haynes}, {Keel},
  {Lintott}, {Simmons}, {Skibba}, {Bamford}, {Giovanelli}, \&
  {Schawinski}}]{mas12a}
{Masters} K.~L. {et~al.}, 2012, \mnras, 424, 2180

\bibitem[{{Masters} {et~al}\mbox{.}(2011){Masters}, {Nichol}, {Hoyle},
  {Lintott}, {Bamford}, {Edmondson}, {Fortson}, {Keel}, {Schawinski}, {Smith},
  \& {Thomas}}]{mas11c}
{Masters} K.~L. {et~al.}, 2011, \mnras, 411, 2026

\bibitem[{{Nair} \& {Abraham}(2010)}]{nai10}
{Nair} P.~B., {Abraham} R.~G., 2010, \apjs, 186, 427

\bibitem[{{Nieto-Santisteban}, {Szalay} \& {Gray}(2004){Nieto-Santisteban},
  {Szalay}, \& {Gray}}]{nie04}
{Nieto-Santisteban} M.~A., {Szalay} A.~S., {Gray} J., 2004, in Astronomical
  Society of the Pacific Conference Series, Vol. 314, Astronomical Data
  Analysis Software and Systems (ADASS) XIII, {Ochsenbein} F., {Allen} M.~G.,
  {Egret} D., eds., p. 666

\bibitem[{{Roberts} \& {Haynes}(1994)}]{rob94}
{Roberts} M.~S., {Haynes} M.~P., 1994, \araa, 32, 115

\bibitem[{{Sandage}(1961)}]{san61}
{Sandage} A., 1961, {The Hubble atlas of galaxies}. {Carnegie Institute of
  Washington}

\bibitem[{{Sandage} \& {Bedke}(1994)}]{san94}
{Sandage} A., {Bedke} J., 1994, {The Carnegie Atlas of Galaxies. Volumes I,
  II.} {Carnegie Institute of Washington}

\bibitem[{{Schawinski} {et~al}\mbox{.}(2009){Schawinski}, {Lintott}, {Thomas},
  {Sarzi}, {Andreescu}, {Bamford}, {Kaviraj}, {Khochfar}, {Land}, {Murray},
  {Nichol}, {Raddick}, {Slosar}, {Szalay}, {Vandenberg}, \& {Yi}}]{sch09}
{Schawinski} K. {et~al.}, 2009, \mnras, 396, 818

\bibitem[{{Schmidt} \& {Lipson}(2009)}]{sch09c}
{Schmidt} M., {Lipson} H., 2009, Science, 324, 81

\bibitem[{{Schwamb} {et~al}\mbox{.}(2012){Schwamb}, {Lintott}, {Fischer},
  {Giguere}, {Lynn}, {Smith}, {Brewer}, {Parrish}, {Schawinski}, \&
  {Simpson}}]{sch12}
{Schwamb} M.~E. {et~al.}, 2012, \apj, 754, 129

\bibitem[{{Simmons} {et~al}\mbox{.}(2013){Simmons}, {Lintott}, {Schawinski},
  {Moran}, {Han}, {Kaviraj}, {Masters}, {Urry}, {Willett}, {Bamford}, \&
  {Nichol}}]{sim13}
{Simmons} B.~D. {et~al.}, 2013, \mnras, 429, 2199

\bibitem[{{Simpson} {et~al}\mbox{.}(2012){Simpson}, {Povich}, {Kendrew},
  {Lintott}, {Bressert}, {Arvidsson}, {Cyganowski}, {Maddison}, {Schawinski},
  {Sherman}, {Smith}, \& {Wolf-Chase}}]{sim12a}
{Simpson} R.~J. {et~al.}, 2012, \mnras, 424, 2442

\bibitem[{{Skibba} {et~al}\mbox{.}(2012){Skibba}, {Masters}, {Nichol},
  {Zehavi}, {Hoyle}, {Edmondson}, {Bamford}, {Cardamone}, {Keel}, {Lintott}, \&
  {Schawinski}}]{ski12}
{Skibba} R.~A. {et~al.}, 2012, \mnras, 423, 1485

\bibitem[{{Smith} {et~al}\mbox{.}(2011){Smith}, {Lynn}, {Sullivan}, {Lintott},
  {Nugent}, {Botyanszki}, {Kasliwal}, {Quimby}, {Bamford}, {Fortson},
  {Schawinski}, {Hook}, {Blake}, {Podsiadlowski}, {J{\"o}nsson}, {Gal-Yam},
  {Arcavi}, {Howell}, {Bloom}, {Jacobsen}, {Kulkarni}, {Law}, {Ofek}, \&
  {Walters}}]{smi11}
{Smith} A.~M. {et~al.}, 2011, \mnras, 412, 1309

\bibitem[{{Stoughton} {et~al}\mbox{.}(2002){Stoughton}, {Lupton}, {Bernardi},
  {Blanton}, {Burles}, {Castander}, {Connolly}, {Eisenstein}, {Frieman},
  {Hennessy}, {Hindsley}, {Ivezi{\'c}}, {Kent}, {Kunszt}, {Lee}, {Meiksin},
  {Munn}, {Newberg}, {Nichol}, {Nicinski}, {Pier}, {Richards}, {Richmond},
  {Schlegel}, {Smith}, {Strauss}, {SubbaRao}, {Szalay}, {Thakar}, {Tucker},
  {Vanden Berk}, {Yanny}, {Adelman}, {Anderson}, {Anderson}, {Annis},
  {Bahcall}, {Bakken}, {Bartelmann}, {Bastian}, {Bauer}, {Berman},
  {B{\"o}hringer}, {Boroski}, {Bracker}, {Briegel}, {Briggs}, {Brinkmann},
  {Brunner}, {Carey}, {Carr}, {Chen}, {Christian}, {Colestock}, {Crocker},
  {Csabai}, {Czarapata}, {Dalcanton}, {Davidsen}, {Davis}, {Dehnen},
  {Dodelson}, {Doi}, {Dombeck}, {Donahue}, {Ellman}, {Elms}, {Evans}, {Eyer},
  {Fan}, {Federwitz}, {Friedman}, {Fukugita}, {Gal}, {Gillespie}, {Glazebrook},
  {Gray}, {Grebel}, {Greenawalt}, {Greene}, {Gunn}, {de Haas}, {Haiman},
  {Haldeman}, {Hall}, {Hamabe}, {Hansen}, {Harris}, {Harris}, {Harvanek},
  {Hawley}, {Hayes}, {Heckman}, {Helmi}, {Henden}, {Hogan}, {Hogg}, {Holmgren},
  {Holtzman}, {Huang}, {Hull}, {Ichikawa}, {Ichikawa}, {Johnston}, {Kauffmann},
  {Kim}, {Kimball}, {Kinney}, {Klaene}, {Kleinman}, {Klypin}, {Knapp},
  {Korienek}, {Krolik}, {Kron}, {Krzesi{\'n}ski}, {Lamb}, {Leger},
  {Limmongkol}, {Lindenmeyer}, {Long}, {Loomis}, {Loveday}, {MacKinnon},
  {Mannery}, {Mantsch}, {Margon}, {McGehee}, {McKay}, {McLean}, {Menou},
  {Merelli}, {Mo}, {Monet}, {Nakamura}, {Narayanan}, {Nash}, {Neilsen},
  {Newman}, {Nitta}, {Odenkirchen}, {Okada}, {Okamura}, {Ostriker}, {Owen},
  {Pauls}, {Peoples}, {Peterson}, {Petravick}, {Pope}, {Pordes}, {Postman},
  {Prosapio}, {Quinn}, {Rechenmacher}, {Rivetta}, {Rix}, {Rockosi}, {Rosner},
  {Ruthmansdorfer}, {Sandford}, {Schneider}, {Scranton}, {Sekiguchi}, {Sergey},
  {Sheth}, {Shimasaku}, {Smee}, {Snedden}, {Stebbins}, {Stubbs}, {Szapudi},
  {Szkody}, {Szokoly}, {Tabachnik}, {Tsvetanov}, {Uomoto}, {Vogeley}, {Voges},
  {Waddell}, {Walterbos}, {Wang}, {Watanabe}, {Weinberg}, {White}, {White},
  {Wilhite}, {Wolfe}, {Yasuda}, {York}, {Zehavi}, \& {Zheng}}]{sto02}
{Stoughton} C. {et~al.}, 2002, \aj, 123, 485

\bibitem[{{Strauss} {et~al}\mbox{.}(2002){Strauss}, {Weinberg}, {Lupton},
  {Narayanan}, {Annis}, {Bernardi}, {Blanton}, {Burles}, {Connolly},
  {Dalcanton}, {Doi}, {Eisenstein}, {Frieman}, {Fukugita}, {Gunn},
  {Ivezi{\'c}}, {Kent}, {Kim}, {Knapp}, {Kron}, {Munn}, {Newberg}, {Nichol},
  {Okamura}, {Quinn}, {Richmond}, {Schlegel}, {Shimasaku}, {SubbaRao},
  {Szalay}, {Vanden Berk}, {Vogeley}, {Yanny}, {Yasuda}, {York}, \&
  {Zehavi}}]{str02}
{Strauss} M.~A. {et~al.}, 2002, \aj, 124, 1810

\bibitem[{{Taylor}(2011)}]{tay11}
{Taylor} M., 2011, {TOPCAT: Tool for OPerations on Catalogues And Tables}.
  Astrophysics Source Code Library

\bibitem[{{Taylor}(2005)}]{tay05}
{Taylor} M.~B., 2005, in Astronomical Society of the Pacific Conference Series,
  Vol. 347, Astronomical Data Analysis Software and Systems XIV, {Shopbell} P.,
  {Britton} M., {Ebert} R., eds., p.~29

\bibitem[{{van den Bergh}(1976)}]{van76}
{van den Bergh} S., 1976, \apj, 206, 883

\bibitem[{{York} {et~al}\mbox{.}(2000){York}, {Adelman}, {Anderson},
  {Anderson}, {Annis}, {Bahcall}, {Bakken}, {Barkhouser}, {Bastian}, {Berman},
  {Boroski}, {Bracker}, {Briegel}, {Briggs}, {Brinkmann}, {Brunner}, {Burles},
  {Carey}, {Carr}, {Castander}, {Chen}, {Colestock}, {Connolly}, {Crocker},
  {Csabai}, {Czarapata}, {Davis}, {Doi}, {Dombeck}, {Eisenstein}, {Ellman},
  {Elms}, {Evans}, {Fan}, {Federwitz}, {Fiscelli}, {Friedman}, {Frieman},
  {Fukugita}, {Gillespie}, {Gunn}, {Gurbani}, {de Haas}, {Haldeman}, {Harris},
  {Hayes}, {Heckman}, {Hennessy}, {Hindsley}, {Holm}, {Holmgren}, {Huang},
  {Hull}, {Husby}, {Ichikawa}, {Ichikawa}, {Ivezi{\'c}}, {Kent}, {Kim},
  {Kinney}, {Klaene}, {Kleinman}, {Kleinman}, {Knapp}, {Korienek}, {Kron},
  {Kunszt}, {Lamb}, {Lee}, {Leger}, {Limmongkol}, {Lindenmeyer}, {Long},
  {Loomis}, {Loveday}, {Lucinio}, {Lupton}, {MacKinnon}, {Mannery}, {Mantsch},
  {Margon}, {McGehee}, {McKay}, {Meiksin}, {Merelli}, {Monet}, {Munn},
  {Narayanan}, {Nash}, {Neilsen}, {Neswold}, {Newberg}, {Nichol}, {Nicinski},
  {Nonino}, {Okada}, {Okamura}, {Ostriker}, {Owen}, {Pauls}, {Peoples},
  {Peterson}, {Petravick}, {Pier}, {Pope}, {Pordes}, {Prosapio},
  {Rechenmacher}, {Quinn}, {Richards}, {Richmond}, {Rivetta}, {Rockosi},
  {Ruthmansdorfer}, {Sandford}, {Schlegel}, {Schneider}, {Sekiguchi}, {Sergey},
  {Shimasaku}, {Siegmund}, {Smee}, {Smith}, {Snedden}, {Stone}, {Stoughton},
  {Strauss}, {Stubbs}, {SubbaRao}, {Szalay}, {Szapudi}, {Szokoly}, {Thakar},
  {Tremonti}, {Tucker}, {Uomoto}, {Vanden Berk}, {Vogeley}, {Waddell}, {Wang},
  {Watanabe}, {Weinberg}, {Yanny}, {Yasuda}, \& {SDSS Collaboration}}]{yor00}
{York} D.~G. {et~al.}, 2000, \aj, 120, 1579

\end{thebibliography}

\bsp

\tabletypesize{\scriptsize}
\begin{deluxetable}{lcclcc|rrrrrr|c}
\rotate
\tablecolumns{12}
\tablewidth{0pc}
\tablecaption{Morphological classifications of GZ2 main sample galaxies with spectra \label{tbl-mainclass}}
\tabletypesize{\scriptsize}
\tablehead{
 &  &   &  & & & 
\multicolumn{6}{c}{\underline{t01\_smooth\_or\_features\_a01\_smooth\_}} &
\colhead{$\dots$}
\\
\colhead{Stripe82 objID} & 
\colhead{RA} & 
\colhead{dec} & 
\colhead{$gz2\_class$} & 
\colhead{$N_\mathrm{class}$} & 
\colhead{$N_\mathrm{votes}$} & 
\colhead{count} & 
\colhead{wt\_count} & 
\colhead{fraction} & 
\colhead{wt\_fraction} & 
\colhead{debiased} &
\colhead{flag} &
\colhead{}
}
\small
\startdata
588017703996096547 & 10:43:57.70 & $+$11:42:13.6 &      SBc?t &  44 & 349 &   1 &   0.1 & 0.023 & 0.002 & 0.002 & 0 \\
587738569780428805 & 12:49:38.60 & $+$15:09:51.1 &        Ser &  45 & 185 &   5 &   5.0 & 0.111 & 0.115 & 0.115 & 0 \\
587735695913320507 & 14:03:12.53 & $+$54:20:56.2 &       Sb+t &  46 & 372 &   0 &   0.0 & 0.000 & 0.000 & 0.000 & 0 \\
587742775634624545 & 12:21:12.82 & $+$18:22:57.7 &     SBb(r) &  45 & 289 &   8 &   8.0 & 0.178 & 0.178 & 0.178 & 0 \\
587732769983889439 & 12:29:28.03 & $+$08:44:59.7 &        Ser &  49 & 210 &  12 &  12.0 & 0.245 & 0.249 & 0.454 & 0 \\
588017725475782665 & 12:34:05.41 & $+$07:41:35.8 &         Ec &  42 & 149 &  27 &  27.0 & 0.643 & 0.686 & 0.771 & 0 \\
588017702391578633 & 11:40:58.75 & $+$11:28:16.1 &       Sc+t &  45 & 356 &   0 &   0.0 & 0.000 & 0.000 & 0.000 & 0 \\
588297864730181658 & 11:46:07.80 & $+$47:29:41.1 &        Sen &  45 & 206 &   4 &   4.0 & 0.089 & 0.091 & 0.091 & 0 \\
588017704545812500 & 12:43:56.58 & $+$13:07:36.0 &       Sb?t &  43 & 360 &   0 &   0.0 & 0.000 & 0.000 & 0.000 & 0 \\
588017566564155399 & 12:25:46.72 & $+$12:39:42.7 &    Sc?t(u) &  43 & 244 &   6 &   6.0 & 0.140 & 0.143 & 0.143 & 0 \\
\enddata
\tablecomments{The full, machine-readable version of this table is available at http://data.galaxyzoo.org. A portion is shown here for guidance on form and content, which is described in Section~\ref{sec-catalogue} and Appendix~\ref{app-gzstring} (for $gz2\_class$). The full table contains 252,750 rows (one for every galaxy in the sample), and 226 columns, with six variables for each of the 37 GZ2 morphology classifications. }
\end{deluxetable}

\tabletypesize{\scriptsize}
\begin{deluxetable}{lcclcc|rrrrrr|c}
\rotate
\tablecolumns{12}
\tablewidth{0pc}
\tablecaption{Morphological classifications of GZ2 main sample galaxies with photo-$z$ \label{tbl-mainclass_photoz}}
\tabletypesize{\scriptsize}
\tablehead{
 &  &   &  & & & 
\multicolumn{6}{c}{\underline{t01\_smooth\_or\_features\_a01\_smooth\_}} &
\colhead{$\dots$}
\\
\colhead{Stripe82 objID} & 
\colhead{RA} & 
\colhead{dec} & 
\colhead{$gz2\_class$} & 
\colhead{$N_\mathrm{class}$} & 
\colhead{$N_\mathrm{votes}$} & 
\colhead{count} & 
\colhead{wt\_count} & 
\colhead{fraction} & 
\colhead{wt\_fraction} & 
\colhead{debiased} &
\colhead{flag} &
\colhead{}
}
\small
\startdata
587722981736579107 & 11:27:57.82 & $-$01:12:50.4 &         Ec &  43 & 181 &  27 &  27.0 & 0.628 & 0.648 & 0.648 & 0 \\
587722981741691055 & 12:14:60.00 & $-$01:08:39.8 &         Er &  44 & 133 &  40 &  40.0 & 0.909 & 0.909 & 0.909 & 1 \\
587722981745819655 & 12:52:22.07 & $-$01:11:58.4 &      Sc(o) &  46 & 221 &  17 &  17.0 & 0.370 & 0.378 & 0.378 & 0 \\
587722981746082020 & 12:55:09.43 & $-$01:05:29.6 &      Sc(o) &  44 & 172 &  31 &  31.0 & 0.705 & 0.771 & 0.363 & 0 \\
587722981746344092 & 12:57:22.12 & $-$01:03:28.7 &      SBb2m &  43 & 358 &   0 &   0.0 & 0.000 & 0.000 & 0.000 & 0 \\
587722981747982511 & 13:12:25.88 & $-$01:06:13.1 &      Ei(o) &  45 & 156 &  37 &  37.0 & 0.822 & 0.850 & 0.547 & 0 \\
587722981748375814 & 13:16:06.59 & $-$01:12:39.7 &         Er &  52 & 198 &  44 &  44.0 & 0.846 & 0.846 & 0.626 & 0 \\
587722981748768914 & 13:19:22.53 & $-$01:07:45.9 &      Sb(r) &  46 & 350 &   3 &   3.0 & 0.065 & 0.065 & 0.097 & 0 \\
587722981748768984 & 13:19:34.20 & $-$01:05:20.0 &         Ei &  42 & 140 &  37 &  36.2 & 0.881 & 0.900 & 0.678 & 0 \\
587722981749031027 & 13:21:46.41 & $-$01:09:37.8 &         Ei &  50 & 158 &  46 &  45.8 & 0.920 & 0.932 & 0.682 & 0 \\
\enddata
\tablecomments{The full, machine-readable version of this table is available at http://data.galaxyzoo.org. A portion is shown here for guidance on form and content, which are identical to those in Table~\ref{tbl-mainclass}.}
\end{deluxetable}

\tabletypesize{\scriptsize}
\begin{deluxetable}{lcclcc|rrrrrr|c}
\rotate
\tablecolumns{12}
\tablewidth{0pc}
\tablecaption{GZ2 morphological classifications of normal-depth images of Stripe~82 galaxies \label{tbl-stripe82_normal}}
\tabletypesize{\scriptsize}
\tablehead{
 &  &   &  & & & 
\multicolumn{6}{c}{\underline{t01\_smooth\_or\_features\_a01\_smooth\_}} &
\colhead{$\dots$}
\\
\colhead{Stripe82 objID} & 
\colhead{RA} & 
\colhead{dec} & 
\colhead{$gz2\_class$} & 
\colhead{$N_\mathrm{class}$} & 
\colhead{$N_\mathrm{votes}$} & 
\colhead{count} & 
\colhead{wt\_count} & 
\colhead{fraction} & 
\colhead{wt\_fraction} & 
\colhead{debiased} &
\colhead{flag} &
\colhead{}
}
\small
\startdata
587730845812064684 & 20:37:42.95 & $-$01:10:10.7 &         Ei &  46 & 135 &  38 &  38.0 & 0.826 & 0.851 & 0.842 & 1 \\
587730845812065247 & 20:37:35.59 & $-$01:05:15.4 &         Ei &  49 & 230 &  26 &  26.0 & 0.531 & 0.551 & 0.551 & 0 \\
587730845812196092 & 20:38:42.14 & $-$01:13:05.5 &       Sb2m &  48 & 368 &   2 &   2.0 & 0.042 & 0.042 & 0.042 & 0 \\
587730845812196825 & 20:39:10.35 & $-$01:10:49.5 &         Ei &  42 & 177 &  26 &  26.0 & 0.619 & 0.633 & 0.633 & 0 \\
587730845812524122 & 20:42:07.88 & $-$01:07:06.4 &         Er &  51 & 149 &  48 &  48.0 & 0.941 & 0.961 & 0.961 & 1 \\
587730845812654984 & 20:42:57.20 & $-$01:05:11.9 &         Ei &  49 & 201 &  33 &  33.0 & 0.673 & 0.673 & 0.638 & 0 \\
587730845812655541 & 20:43:15.93 & $-$01:09:24.6 &      Ei(i) &  46 & 193 &  25 &  25.0 & 0.543 & 0.543 & 0.543 & 0 \\
587730845812720365 & 20:43:25.63 & $-$01:05:13.8 &         Ei &  43 & 152 &  34 &  33.8 & 0.791 & 0.790 & 0.721 & 0 \\
587730845812720699 & 20:43:46.54 & $-$01:12:32.1 &         Ei &  46 & 233 &  24 &  22.5 & 0.522 & 0.506 & 0.506 & 0 \\
587730845812851385 & 20:44:45.87 & $-$01:04:58.8 &         Er &  45 & 147 &  39 &  39.0 & 0.867 & 0.884 & 0.837 & 1 \\

\enddata
\tablecomments{The full, machine-readable version of this table is available at http://data.galaxyzoo.org. A portion is shown here for guidance on form and content, which are identical to those in Table~\ref{tbl-mainclass}. Classifications here are for normal-depth images from Stripe~82, which goes to a deeper magnitude limit ($m_r > 17.7$) galaxies in the main sample.}
\end{deluxetable}

\tabletypesize{\scriptsize}
\begin{deluxetable}{lcclcc|rrrrrr|c}
\rotate
\tablecolumns{12}
\tablewidth{0pc}
\tablecaption{GZ2 morphological classifications of coadded images (set 1) of Stripe~82 galaxies \label{tbl-stripe82_coadd1}}
\tabletypesize{\scriptsize}
\tablehead{
 &  &   &  & & & 
\multicolumn{6}{c}{\underline{t01\_smooth\_or\_features\_a01\_smooth\_}} &
\colhead{$\dots$}
\\
\colhead{Stripe82 objID} & 
\colhead{RA} & 
\colhead{dec} & 
\colhead{$gz2\_class$} & 
\colhead{$N_\mathrm{class}$} & 
\colhead{$N_\mathrm{votes}$} & 
\colhead{count} & 
\colhead{wt\_count} & 
\colhead{fraction} & 
\colhead{wt\_fraction} & 
\colhead{debiased} &
\colhead{flag} &
\colhead{}
}
\small
\startdata
8647474690312307154 & 20:37:16.90 & $-$01:08:54.1 &      Ei(m) &  20 &  74 &  15 &  14.4 & 0.750 & 0.742 & 0.749 & 0 \\
8647474690312307877 & 20:37:35.59 & $-$01:05:15.4 &         Ei &  17 &  54 &  13 &  13.0 & 0.765 & 0.765 & 0.765 & 0 \\
8647474690312308880 & 20:37:27.68 & $-$01:12:39.9 &         Ei &  12 &  32 &  10 &  10.0 & 0.833 & 0.833 & 0.833 & 1 \\
8647474690312373464 & 20:37:52.27 & $-$01:08:17.8 &         Er &  22 &  75 &  18 &  18.0 & 0.818 & 0.829 & 0.829 & 1 \\
8647474690312438284 & 20:38:42.14 & $-$01:13:05.5 &       Sc2m &  23 & 149 &   3 &   3.0 & 0.130 & 0.136 & 0.136 & 0 \\
8647474690312505086 & 20:39:10.33 & $-$01:10:49.5 &         Ec &  15 &  58 &  11 &  11.0 & 0.733 & 0.748 & 0.748 & 0 \\
8647474690312832559 & 20:42:07.88 & $-$01:07:06.3 &         Er &  20 &  77 &  14 &  14.0 & 0.700 & 0.700 & 0.781 & 0 \\
8647474690312898532 & 20:42:57.20 & $-$01:05:11.8 &         Ei &  14 &  68 &   9 &   9.0 & 0.643 & 0.643 & 0.643 & 0 \\
8647474690312962734 & 20:43:25.63 & $-$01:05:13.8 &         Ei &  21 &  77 &  15 &  15.0 & 0.714 & 0.714 & 0.679 & 0 \\
8647474690312963665 & 20:43:15.93 & $-$01:09:24.7 &         Ei &  12 &  43 &  11 &  11.0 & 0.917 & 0.917 & 0.917 & 1 \\
\enddata
\tablecomments{The full, machine-readable version of this table is available at http://data.galaxyzoo.org. A portion is shown here for guidance on form and content, which are identical to those in Table~\ref{tbl-mainclass}. Classifications here are for the coadded images (set~1; see \S\ref{ssec-imagecreation}) from Stripe~82, which goes to a deeper magnitude limit and has a better angular resolution than galaxies in the main sample. There is no colour desaturation for background sky pixels in this set of images.}
\end{deluxetable}

\tabletypesize{\scriptsize}
\begin{deluxetable}{lcclcc|rrrrrr|c}
\rotate
\tablecolumns{12}
\tablewidth{0pc}
\tablecaption{GZ2 morphological classifications of coadded images (set 2) of Stripe~82 galaxies \label{tbl-stripe82_coadd2}}
\tabletypesize{\scriptsize}
\tablehead{
 &  &   &  & & & 
\multicolumn{6}{c}{\underline{t01\_smooth\_or\_features\_a01\_smooth\_}} &
\colhead{$\dots$}
\\
\colhead{Stripe82 objID} & 
\colhead{RA} & 
\colhead{dec} & 
\colhead{$gz2\_class$} & 
\colhead{$N_\mathrm{class}$} & 
\colhead{$N_\mathrm{votes}$} & 
\colhead{count} & 
\colhead{wt\_count} & 
\colhead{fraction} & 
\colhead{wt\_fraction} & 
\colhead{debiased} &
\colhead{flag} &
\colhead{}
}
\small
\startdata
8647474690312307154 & 20:37:16.90 & $-$01:08:54.1 &      Ei(o) &  16 &  72 &  10 &  10.0 & 0.625 & 0.625 & 0.629 & 0 \\
8647474690312307877 & 20:37:35.59 & $-$01:05:15.4 &         Ei &  21 &  84 &  17 &  17.0 & 0.810 & 0.810 & 0.810 & 1 \\
8647474690312308318 & 20:37:42.94 & $-$01:10:10.7 &      Ei(m) &  23 &  88 &  18 &  18.0 & 0.783 & 0.783 & 0.722 & 0 \\
8647474690312308880 & 20:37:27.68 & $-$01:12:39.9 &         Er &  16 &  48 &  16 &  16.0 & 1.000 & 1.000 & 1.000 & 1 \\
8647474690312373464 & 20:37:52.27 & $-$01:08:17.8 &         Er &  23 &  89 &  17 &  17.0 & 0.739 & 0.739 & 0.739 & 0 \\
8647474690312438284 & 20:38:42.14 & $-$01:13:05.5 &       Sb2m &  11 &  91 &   0 &   0.0 & 0.000 & 0.000 & 0.000 & 0 \\
8647474690312505086 & 20:39:10.33 & $-$01:10:49.5 &       Sb?m &  12 &  65 &   4 &   3.4 & 0.333 & 0.295 & 0.295 & 0 \\
8647474690312832559 & 20:42:07.88 & $-$01:07:06.3 &         Er &  23 &  75 &  14 &  14.0 & 0.609 & 0.629 & 0.666 & 0 \\
8647474690312898532 & 20:42:57.20 & $-$01:05:11.8 &         Ei &  26 & 129 &  12 &  12.0 & 0.462 & 0.462 & 0.492 & 0 \\
8647474690312962734 & 20:43:25.63 & $-$01:05:13.8 &         Ei &  20 &  69 &  18 &  17.0 & 0.900 & 0.895 & 0.840 & 1 \\
\enddata
\tablecomments{The full, machine-readable version of this table is available at http://data.galaxyzoo.org. A portion is shown here for guidance on form and content, which are identical to those in Table~\ref{tbl-mainclass}. Classifications here are for the coadded images (set~2; see \S\ref{ssec-imagecreation}) from Stripe~82, which goes to a deeper magnitude limit and has a better angular resolution than galaxies in the main sample. Pixels in the sky background are colour desaturated in this set of images.}
\end{deluxetable}

\appendix
\section{Generating the abbreviation for a GZ2 morphological classification}\label{app-gzstring}
As part of the GZ2 data release (Tables~\ref{tbl-mainclass}--\ref{tbl-stripe82_coadd2}), we provide a short abbreviation ($gz2\_class$) that indicates the most common consensus classification for the galaxy. We emphasise that the intent is \underline{not} to create a new classification system; rather, this is only a convenient shorthand for interpreting portions of the GZ2 results.

The $gz2\_class$ string is generated for each galaxy by taking the largest debiased vote fraction (beginning with Task~01) and selecting the most common response for each subsequent task in the decision tree. 

Galaxies that are smooth (from Task~01) have $gz2\_class$ strings beginning with `E'. Their degree of roundness (completely round, in-between, and cigar-shaped) is represented by `r',`i', and `c', respectively. 

Galaxies with features/disks have $gz2\_class$ strings beginning with `S'. Edge-on disks follow this with `er', `eb', or `en' (with the second letter classifying the bulge shape as round, boxy, or none). For oblique disks, the letter following `S' is an upper-case `B' if the galaxies have a bar. The bulge prominence (`d' = none, `c' = just noticeable, `b' = obvious, `a' = dominant). Both bars and bulges follow the same general trends as the Hubble sequence, although the correspondence is not exact. If spiral structure was identified, then the string includes two characters indicating the number ($1,2,3,4,+,?$) and relative winding (`t'=tight, `m'=medium, `l'=loose) of the spiral arms. 

Finally, any feature in the galaxy the users identified as ``odd'' appears at the end of the string in parentheses: `(r)'=ring, `(l)'=lens/arc, `(d)'=disturbed, `(i)'=irregular, `(o)'=other, `(m)'=merger, `(u)'=dust lane. 

Objects that are stars or artifacts have the $gz2\_class$ string `A'.
\\
\\
For example: 
\begin{itemize}
\item Er = smooth galaxy, completely round
\item SBc2m = barred disk galaxy with a just noticeable bulge and two medium-wound spiral arms
\item Seb = edge-on disk galaxy with a boxy bulge
\item Sc(I) = disk galaxy with a just noticeable bulge, no spiral structure, and irregular morphology.
\item A = star
\end{itemize}

\begin{figure*}
\includegraphics[angle=0,width=7.0in]{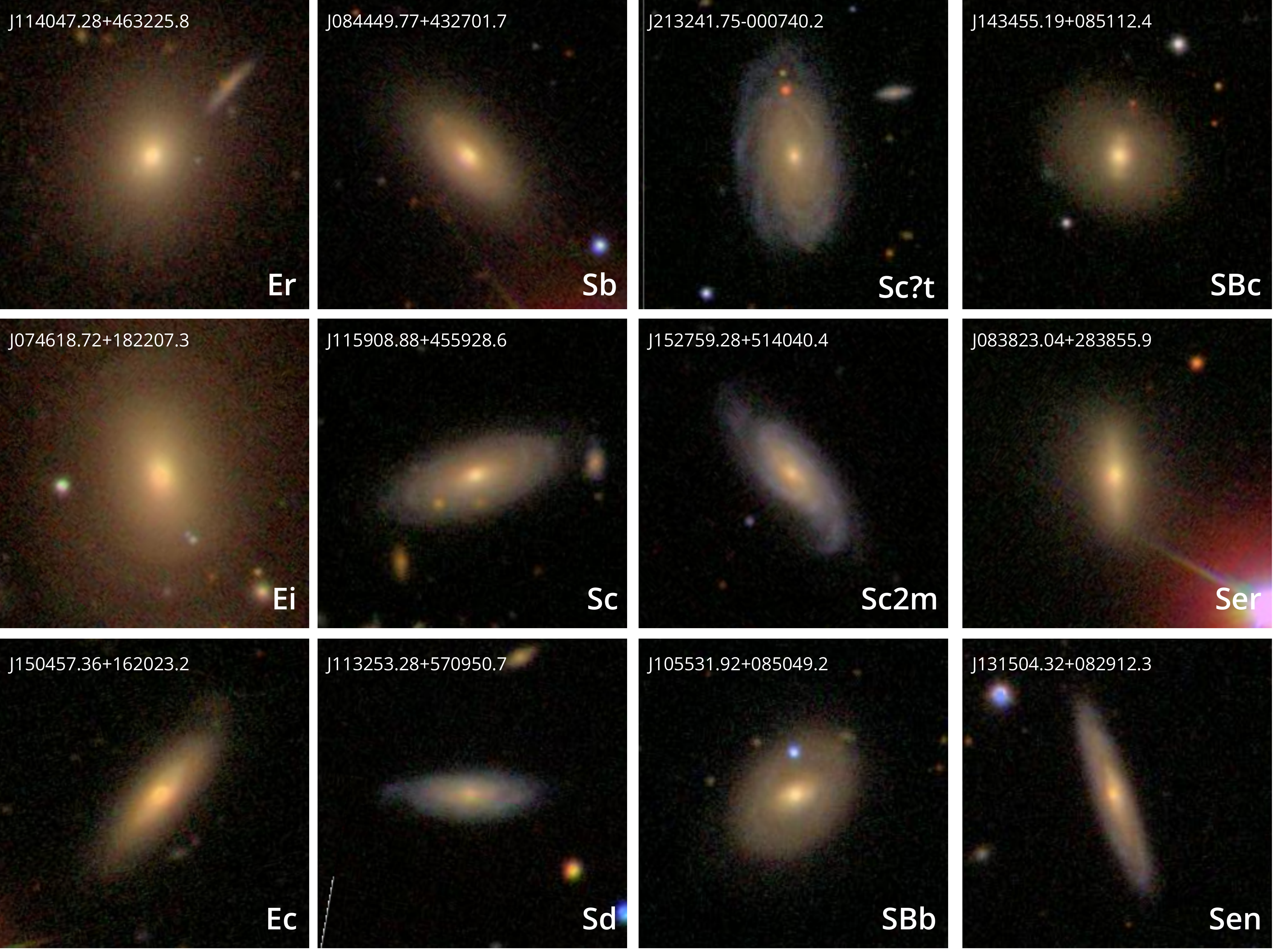}
\caption{Example images with their GZ2 classifications (see Appendix~\ref{app-gzstring}). Galaxies are randomly selected from the GZ2 catalog; all galaxies lie in the redshift range $0.050 < z < 0.055$. Categories shown represent the twelve most common classifications in the GZ2 spectroscopic sample. 
\label{fig-exampleimages}}
\end{figure*}

Sample images of the twelve most common $gz2\_class$ strings are shown in Figure~\ref{fig-exampleimages}. 

\label{lastpage}

\end{document}